\documentclass[11pt,twoside]{article}
\usepackage{atmp}

\newtheorem{theorem}{Theorem}[section]
\newtheorem{corollary}[theorem]{Corollary}
\newtheorem{lemma}[theorem]{Lemma}
\newtheorem{proposition}[theorem]{Proposition}
\newtheorem{definition}[theorem]{Definition}
\newtheorem{remark}[theorem]{Remark}
\newcommand{\epsi}{\varepsilon}
\newcommand{\stackunder}[2]{\stackrel{#1}{#2}}

\newcommand{\Hi}{\mathcal{H}}
\newcommand{\K}{\mathcal{K}}
\newcommand{\Or}{\mathcal{O}}
\newcommand{\R}{\mathbb{R}}
\newcommand{\C}{\mathbb{C}}

\def\1I{\mathchoice{ \hbox{${\rm 1}\hspace{-2.3pt}{\rm l}$} }
                   { \hbox{${\rm 1}\hspace{-2.3pt}{\rm l}$} }
                   { \hbox{$ \scriptstyle  {\rm I}\!{\rm N}$} }
                   { \hbox{$ \scriptscriptstyle  {\rm I}\!{\rm N}$}}}

\copyrightnotice{2003}{7}{145}{204}   %% year, volume, first page, last page.

\begin{document}

\title{Space-Adiabatic Perturbation Theory}

\url{math-ph/0201055}

\author{   Gianluca Panati$^{*}$, Herbert Spohn, Stefan Teufel}
\address{
  Zentrum Mathematik and Physik Department,\\
  Technische Universit\"{a}t M\"{u}nchen,\\ 85747 Garching bei M\"{u}nchen, Germany \smallskip\\
  (*) also Mathematical Physics Sector, SISSA-ISAS, Trieste, Italy}
\addressemail{panati@ma.tum.de, spohn@ma.tum.de, teufel@ma.tum.de}
\markboth{\it Space-Adiabatic Perturbation Theory}{\it G.\ Panati, H.\ Spohn, S.\ Teufel}
%\date{\normalsize January 30, 2002 (Revised on March  25, 2003)}

\begin{abstract} 
  We study approximate solutions to the time-dependent Schr\"odinger equation 
$i\epsi\partial_t\psi_t(x)/\partial t = H(x,-i\epsi\nabla_x)\,\psi_t(x)$
with the Hamiltonian  given as the Weyl quantization of the symbol $H(q,p)$
taking values in the space of bounded operators on
 the Hilbert space $\Hi _{\rm f}$ of fast
``internal'' degrees of freedom.
By assumption $H(q,p)$ has an isolated energy band.
Using a method of Nenciu and Sordoni \cite{NS}  we 
prove that interband transitions are suppressed to any order in $\epsi$.
As a consequence, associated to that energy band there exists a subspace of
$L^2(\mathbb{R}^d,\Hi _{\rm f})$ almost invariant under the unitary
time evolution. We develop a systematic perturbation scheme
for the computation of effective Hamiltonians which govern approximately
the intraband time evolution. As examples for the general perturbation scheme
we discuss the Dirac and Born-Oppenheimer type Hamiltonians and we
 reconsider also the time-adiabatic theory.
\end{abstract}

\cutpage

%%%%%%%%%%%%%%%%%%%%%% INTRODUCTION   %%%%%%%%%%%%%%%%%%%%%%%%

\section{Introduction}

Quantum theory has the remarkable feature that certain dynamical
degrees of freedom may become ``slaved'' and thus lose their
autonomous status. The origin of this phenomenon is  a
separation, both in space and time, into slow and fast degrees of freedom.
The fast modes quickly adapt to the slow modes which in turn are governed by a
suitable effective Hamiltonian. This mechanism is called adiabatic decoupling.

As paradigm we mention the motion of nuclei. The electronic degrees
of freedom rapidly adjust to the state of lowest energy at given
positions of the nuclei and the electronic energy band serves as
effective potential in the Hamiltonian for the nuclei. This
Born-Oppenheimer approximation is the basis for the dynamics of
molecules and, as a consequence, also for the microscopic theory of classical
fluids. There are many other examples of a similar structure. A
very widely studied case are electrons moving in the periodic
crystal potential, which defines the short scale. The envelope of
the electronic wave function is governed by an effective
Hamiltonian obtained from the Peierls substitution, in which the
band energy is taken as effective kinetic energy. For an electron
coupled to the quantized radiation field the photons are the fast
degrees of freedom and the dynamics of the electron is governed by
an effective Hamiltonian accounting for spin precession. These and
other systems have been studied extensively by model specific
approximate methods without realizing that they share a common
structure.  
In fact,  all the examples given can be molded into the generic
form
\begin{equation}\label{a}
i\epsi\frac{\partial}{\partial
t}\psi_t(x)=H(x,-i\epsi\nabla_x)\psi_t(x)\,.
\end{equation}
Here $H(q,p)$ is an operator-valued function on the classical
phase space $\Gamma=\mathbb R^{2d}$ of the slow degrees of freedom
with $q$ a position like and $p$ a momentum like variable.
$H(q,p)$ is self-adjoint and acts on the Hilbert space $\mathcal
H_\mathrm f$ of ``internal" fast degrees of freedom. After quantization
$H(x,-i\epsi\nabla_x)$ becomes the Hamiltonian of our system. To
properly define it one has to specify an ordering of the operators
$x,-i\epsi\nabla_x$, for which we will adopt the Weyl quantization
rule as to be explained in full detail in the Appendix. $\psi_t$ is a
wave function on $\mathbb R^d$ with values in $\mathcal H_\mathrm
f$. Thus the quantum mechanical Hilbert space of states is
$\mathcal H=L^2(\mathbb R^d,\mathcal H_\mathrm f)=L^2(\mathbb
R^d)\otimes\mathcal H_\mathrm f$. Finally $\epsi$ is a
dimensionless parameter which controls the scale separation,
$\epsi\ll 1$. 

Examples will be given in due course and we only
remark that, in general, (\ref{a}) is already the result of a
proper identification of the slow degrees of freedom. For example,
in nonrelativistic QED $q$ stands for the position of the
electron, whereas $p$ is the total momentum of the electron and
photons. For an electron in a periodic potential (\ref{a}) is the
Hamiltonian in the crystal momentum basis with $x$ standing for
the Bloch momentum $k$ in the first Brillouin zone, see \cite{PST2}.

Our goal is to construct approximate solutions to (\ref{a}), a task which we
divide  into four steps.

 \noindent {\bf (i) Almost invariant subspace.}
The adiabatic decoupling can be traced to a spectral property of
$H(q,p)$. It is assumed, and it can be proved in many particular
cases, that for each $q,p$ the spectrum, $\sigma(q,p)$, of
$H(q,p)$ can be decomposed into a ``relevant" part $\sigma_\mathrm
r(q,p)$ and a remainder in such a way that the relevant energy band
$\{(q,p,\lambda)\in\mathbb{R}^{2d+1}:\lambda\in\sigma_{\rm
r}(q,p)\}$  is separated from the remainder
$\{(q,p,\lambda)\in\mathbb{R}^{2d+1}:\lambda\in\sigma_{\rm
r}(q,p)^{\rm c}\}$  by a gap, i.e. by a corridor of finite width.
In many cases of interest the relevant part of the spectrum consists of a 
single, possibly
degenerate eigenvalue and the relevant energy band is the
graph of a smooth function. Of course, a given $H(q,p)$ could have
several such relevant energy bands and we suppose one of them to
be singled out, e.g., through the initial condition. To the relevant 
energy band of $H$ one associates a subspace 
$\Pi\Hi$ of $\Hi$ with the property that if $\psi_0\in\Pi\Hi$, then  
$\psi_t\in\Pi\Hi$ up to an error which is smaller than any power of $\epsi$. 
For this reason $\Pi\Hi$ is called an almost invariant subspace. If
$\widehat H$ denotes  the Weyl quantization of $H$, then $[\widehat H,\Pi]=\Or(\epsi^\infty)$,
where $\Pi$ is the orthogonal projector onto the closed subspace $\Pi\Hi$.
Of course, in practice, approximations to $\Pi$ are constructed, for which at order $n$ the error is 
$\Or(\epsi^{n+1})$.

\noindent {\bf (ii) Reference Hilbert space.}
Clearly, on $\Pi\Hi$ the dynamics is generated by the diagonal Hamiltonian $\Pi\widehat H\Pi$,
up to $\Or(\epsi^\infty)$. While properly defined, for a further analysis of the effective dynamics
inside $\Pi\Hi$ a representation in terms of adapted  coordinates for the slow degrees of freedom 
is more powerful. 
We call the  corresponding Hilbert space $\K = L^2(\R^d,\K_{\rm f})$  the reference space.
By definition $\K$ is independent of $\epsi$. The next   task is thus to construct
a linear map $U :\Pi\Hi\to\K$, which is unitary in the sense that $U^*U=\Pi$ and $UU^*={\bf 1}_\K$.

\noindent {\bf (iii) Effective Hamiltonian.}
 The effective Hamiltonian is defined through $h = U\Pi\widehat H\Pi U^*$ as operator on $\K$. It is
  unitarily equivalent to the diagonal Hamiltonian, but acts on a simpler space.
  While $h$ is still a rather abstract object,   it can be expanded in powers of $\epsi$,
  which is the adiabatic perturbation theory of the title. Already the lowest order
  approximations provide a wealth of information on the motion of the slow degrees of freedom.

 \noindent {\bf (iv) Semiclassical limit.}  In many applications of interest the effective Hamiltonian is  of a form which allows for
 a semiclassical analysis. The simplest situation is a relevant band which consist only of a
 single, possibly degenerate, eigenvalue. Then the effective Hamiltonian has a scalar principle symbol
 and the semiclassical analysis is straightforward.

We are certainly not the first  ones to investigate approximate solutions
to (\ref{a}) and we have to explain which parts of the program outlined above have been achieved
before and which parts are our novel contribution. In addition there have been other approaches to (\ref{a}) 
on which we briefly comment below.

To our knowledge, the notion ``almost invariant subspace'' was first coined by Nenciu \cite{NenciuAIS}
in the context of gauge invariant  perturbation theory. In the context of space-adiabatic problems 
Brummelhuis and Nourrigat \cite{BN} construct $\Pi$ for the particular case of the Dirac equation and
Martinez and Sordoni \cite{MS} based on \cite{Sordoni} consider Born-Oppenheimer type Hamiltonians.
The general scheme for the construction of $\Pi$ is sketched in Nenciu and Sordoni \cite{NS} and applied
to the matrix-valued Klein-Gordon equation. Our construction is based on the one in \cite{NS}, but differs 
in a few technical details.

As in the case of $\Pi$ we construct   the unitary $U$ in several steps. First 
we compute  order by order  the formal symbol of $U$, which, after quantization, 
gives rise to an ``almost unitary'' operator. Finally the almost unitary is modified 
to yield a true unitary exactly intertwining $\Pi\Hi$ and $\K$. 
Our method is specifically designed to deal also with problems as the Dirac equation and the Bloch electron with external magnetic fields, 
where the projector $\Pi$
has no limit for $\epsi\to 0$, see Remark~\ref{NagyRem}.  
While the specific application and the proof are new, the general idea
to construct a pseudodifferential operator which is almost unitary and diagonalizes 
a given pseudodifferential operator has a long tradition, \cite{Nirenberg} Section~7 and references
therein, \cite{TaylorArt},\cite{HS}.
The method of successive diagonalization is also prominent in the physics literature,
for example \cite{FW} in the derivation of the Pauli equation
and its corrections, \cite{BlountBloch} for periodic Schr\"odinger operators,
\cite{BlountDirac} for the Dirac equation, \cite{LF,LW} for Born-Oppenheimer
type Hamiltonians.

Our central result is the expansion of the effective Hamiltonian $h$. We provide a scheme
which is applicable in general and work out explicitly 
the expansion as $h=h_0 +\epsi h_1 +\epsi^2 h_2 +\Or(\epsi^3)$. In particular, for a relevant band
consisting of a single eigenvalue, $h_0$
is the Peierls substitution and $h_1$ contains among other things the information on geometric phases.

Since $h$ has, in general,  a matrix-valued symbol, we discuss and apply
some results on the semiclassical limit for matrix-pseudodifferential
operators with scalar principal symbol. We include this part to make the
paper self-contained and to demonstrate that in many cases of interest the motion of the 
slow degrees of freedom can be approximately described through the appropriate classical Hamiltonian flow.

Surely the reader will have noticed that our program makes no mention of the initial conditions for
(\ref{a}). The reason is simply that our estimates are uniform on $\Pi\Hi$, respectively  on $\K$.
Physically this partial independence from the initial conditions is most welcome, because in general they
cannot be controlled so easily. On the other hand, in the approach of Hagedorn and Joye \cite{JH}
one constructs for a specific $\epsi$-dependent $\psi_0^\epsi$ the wave packet $t\mapsto\psi_t^\epsi\in\Hi$, which
solves (\ref{a}) up to $\Or(\epsi^\infty)$ or even $\Or(e^{-c/\epsi})$. Hagedorn and Joye study Born-Oppenheimer 
Hamiltonians. Periodic Schr\"odinger operators  are considered  in \cite{GRT,DGR}, who employ a related multi-scale analysis. 
Whereas in the other approaches the adiabatic and the semiclassical limit are taken simultaneously, 
both limits are clearly separated in our results. We consider this an important conceptual advantage.

To give a brief outline of our paper. In Section~2 we explain the precise assumptions on the Hamiltonian $H$
and construct the almost invariant subspace $\Pi\Hi$. The reference space $\K$ and the unitary intertwining
it with $\Pi\Hi$ are explained Section~3. The central part of our paper is the expansion of the effective Hamiltonian
in Section~4. In particular, we work out the expansion for Born-Oppenheimer Hamiltonians including order $\epsi^2$.
We also show that by the Howland trick the standard time-adiabatic theory can be subsumed under (\ref{a}).
The   semiclassical analysis for the effective Hamiltonian is  summarized in Section~5. Last but not least,
in Section~6 we discuss the Dirac equation with slowly varying external vector potentials, since it is the
simplest Hamiltonian for which the full generality of our approach is needed and yields interesting physical
results. In this case $\Hi_{\rm f} = \C^4$ and the classical symbol has two
two-fold degenerate energy bands,
one for the electron and one for the positron. Thus the reference
Hilbert space e.g.\ for the electron band is $L^2(\mathbb R^3,\mathbb C^2)$. The
effective Hamiltonian is determined including order $\epsi$. The principal part $h_0$
describes the translational motion through the Peierls
substitution and the subprincipal part $h_1$ yields the spin precession as governed by the BMT
equation. Since the external vector potential appears  through
minimal coupling, the projection $P(q,p)$ on the electron subspace
depends nontrivially on both coordinates,  in contrast to Born-Oppenheimer and time-adiabatic theory,
where the relevant projection $P(q,p)$ depends only on one of the canonical coordinates.
We end the paper with some concluding  remarks in Section~7 and with an Appendix reviewing
some results on pseudodifferential calculus with operator-valued symbols.\bigskip

\noindent \textbf{Acknowledgments:} We are grateful to Andr\'{e} Martinez and 
Gheorghe Nenciu for helpful discussions, which in part initiated our work.
We thank Berndt Thaller for his remarks
on the problem of observables for the Dirac
equation and Patrick G\'erard for pointing out to us reference \cite{Nirenberg}.
S.T.\ also thanks Raymond Brummelhuis, George Hagedorn, 
and Alain Joye for valuable comments and their interest in this work. G.P.\ is grateful
to Fabio Pioli for a useful remark about fiber bundles.

%%%%%%%%%%%%%%%%%%%%%%  PRELIMINARIES   %%%%%%%%%%%%%%%%%%%%%%%%%%%%%%%%

%%%%%%%%%%%%%%%%%%%% PROJECTORS %%%%%%%%%%%%%%%%%%%

\section{General setting and construction of the almost-invariant subspace}
\label{SPRO}

Space-adiabatic perturbation theory
deals with quantum systems in which it is
possible to distinguish between fast and slow degrees of freedom.
In particular we assume that the Hilbert
space $\Hi $  admits a
natural decomposition as $\Hi =L^{2}(\mathbb{R}^{d})\otimes$
$\Hi _{\mathrm{f}}$,
where $L^2(\mathbb{R}^d)$ is the state space for the slow degrees of freedom and
$\Hi _{\mathrm{f}}$ is the state space for the fast
degrees of freedom.

As the second structural ingredient we require that the Hamiltonian
 is given as the quantization
of a $\mathcal{B}(\Hi _{\mathrm{f}})$-valued function on
the classical phase space $\mathbb{R}^{2d}$ of the slow degrees of freedom.
Hence we need to consider the generalization of the usual
quantization rules to the case of $\mathcal{B}(\Hi _{\mathrm{f}})$-valued
functions on $\mathbb{R}^{2d}$. This theory is well covered in the
literature, see for example \cite{Hormander,Folland,Ivrii,GMS}.
Still, for the convenience of the reader,
  we  provide a self-contained review of the basic
results in the Appendix, where we also introduce the relevant notation.

We   now   state the general assumptions on which the
adiabatic perturbation theory will be based in the following.
Let $\Hi _{\rm f}$ be a separable Hilbert space, the state space for the fast
degrees of freedom, and $\Hi =L^2(\mathbb{R}^d)\otimes \Hi _{\rm f}$.
The Hamiltonian $\widehat H$ of the full system is given as the Weyl
quantization of a semiclassical symbol $H\in S^m_\rho(\epsi)$. We assume that~$\widehat H$ 
is essentially self-adjoint on $\mathcal{S}$. A point in the classical
phase space $\mathbb{R}^{2d}$ is denoted by $z= (q,p)\in \mathbb{R}^{2d}$.

The adiabatic decoupling relies on a gap condition for the principal symbol $H_0$ of $H$.
\bigskip

\noindent \textbf{Condition (Gap)}$_{\sigma }.$
For any $z\in\mathbb{R}^{2d}$
the spectrum $\sigma (z)$ of $H_{0}(z)\in \mathcal{B}(\Hi _{\mathrm{f}})$
contains a relevant subset
${\sigma_{\rm r}}(z)$ which is uniformly separated from its complement
$ \sigma(z)\setminus {\sigma_{\rm r}}(z)$ by a gap. More precisely
 there are two
continuous functions $\gamma_{j}:\mathbb{R}^{2d}\rightarrow \mathbb{R}$ $(j=\pm)$
(with $\gamma_{-}\leq \gamma_{+}$) such that:

\begin{enumerate}
\item[(G1)]  for every $z\in \mathbb{R}^{2d}$ the spectral component ${\sigma_{\rm r}}(z)$
is entirely contained in the interval $\mathrm{I}(z):= [\gamma
_{-}(z),\gamma_{+}(z)]$;

\item[(G2)]  the distance between $\sigma (z)\setminus {\sigma_{\rm r}}(z)$ and
the interval $\mathrm{I}(z)$ is uniformly bounded away from zero and
increasing for large momenta, i.e.
\begin{equation}
\mathrm{dist}(\sigma (z)\setminus {\sigma_{\rm r}}(z),\mathrm{I}(z))\geq
C_{g}\left\langle p\right\rangle^{\sigma };  \label{Gap}
\end{equation}

\item[(G3)]  the width of the interval $\mathrm{I}(z)$ is uniformly bounded,
i.e. 
\[ \sup_{z\in \mathbb{R}^{2d}} \left| \gamma
_{+}(z)-\gamma_{-}(z)\right| \leq C<\infty\,.
\]
\end{enumerate}

We denote the spectral projector corresponding to ${\sigma_{\rm r}}(z)$ by $\pi_{0}(z)$.
As explained in the Introduction, one expects interband transitions to be
 suppressed for small $\epsi$.  To prove such a property we need either
one of the following  assumptions to be satisfied.\bigskip

\noindent \textbf{Condition of increasing gap (IG)}$_{m}$\textbf{.} \\ Let $H$  be
an \emph{hermitian} symbol in $S_{\rho }^{m}(\varepsilon,\mathcal{B(H_{\rm f})})$
(with $\rho >0$ and $m\geq0$) such that the principal symbol $H_{0}$ satisfies
condition (Gap)$_{\sigma }$ with $\sigma =m$.\bigskip

\noindent \textbf{Condition of constant gap (CG).} \\Let $H$  be an
\emph{hermitian} symbol in $S_{0}^{0}(\varepsilon ,\mathcal{B(H_{\rm f})} )$
such that the principal symbol $H_{0}$ satisfies condition (Gap)$_{\sigma }$
with $\sigma =0$.\bigskip

Note that for the case $H\in S_{1 }^{1}(\varepsilon,\mathcal{B(H_{\rm f})} )$
 one can show that $\widehat{H}$  -- the Weyl
quantization of $H$ -- is essentially self-adjoint on the domain $\mathcal{S}
(\mathbb{R}^{d},\Hi _{\mathrm{f}})\subseteq \Hi $. The proof is
postponed to an appendix to the space-adiabatic theorem.

In analogy with the
usual time-adiabatic theorem of quantum mechanics, see Section~\ref{STA},
we baptize the following result as {\em space-adiabatic theorem}.
It establishes that there are almost invariant subspaces associated with
isolated energy bands.
In spirit the result is not new.
However, to our knowledge it appears
in this explicit form only recently in the literature.
Brummelhuis and Nourrigat \cite{BN} gave a proof for the Dirac equation,
 Martinez and Sordoni \cite{MS} considered  Born-Oppenheimer type Hamiltonians 
(cf.\ Section~\ref{SBO})
based on results from  \cite{Sordoni}, 
and Nenciu and Sordoni \cite{NS}
sketched the general scheme and applied it to a matrix-valued Klein-Gordon
type problem.

\begin{theorem}[Space-adiabatic theorem] 
\hspace{-5pt}
Assume either\hspace{-1pt} (IG)$_{m}$\hspace{-1pt} or \hspace{-1pt}(CG).  Let  $\widehat{H}$ be the Weyl quantization of
$H$.  Then there exists an orthogonal projector $\Pi \in \mathcal{B}(\Hi )$
such that
\begin{equation}\label{TIScom}
\big[\,\widehat{H},\,\Pi\, \big]=\mathcal{O}_{0}(\varepsilon^{\infty })
\end{equation}
and $\Pi =\widehat{\pi}+\mathcal{O}_{0}(\varepsilon^{\infty })$,
where $\widehat{\pi}$ is the Weyl quantization of a semiclassical symbol
\[
\pi \asymp \sum_{j\geq 0}\varepsilon^{j}\pi_{j}\qquad \mbox{in }S_{\rho
}^{0}(\varepsilon )
\]
whose principal part $\pi_{0}(z)$ is the spectral projector of  $H_{0}(z)$
corresponding to ${\sigma_{\rm r}}(z)$.  \label{Th Invariant subspace}
\end{theorem}

The subspace $\mathrm{Ran}\Pi \subseteq \Hi $ is thus an
 almost invariant subspace for the dynamics generated by the
Hamiltonian $\widehat{H}$, i.e.\ 
$[e^{-i\widehat H t},\Pi]=\mathcal{O}_0(\epsi^\infty|t|)$,
 and it is associated with the spectral band ${\sigma_{\rm r}}(z)$.
The terminology was introduced in  \cite{NenciuAIS}. Note, however, that  
Ran$\Pi$ is, in general, {\em not} an almost invariant subspace 
in the sense of    \cite{NenciuAIS}, since $\Pi$ need not have a limit as $\epsi\to 0$.

\begin{remark}\em
Note that the growth condition on the gap in (IG)$_m$ is stronger than one would  
expect from the analysis in \cite{NS} or \cite{general}. Indeed, in both examples a gap 
which is bounded globally over phase space suffices to prove uniform adiabatic decoupling
also in the presence of a Hamiltonian with principal symbol increasing linearly in momentum.
More general,
uniform adiabatic decoupling should hold whenever  (IG)$_m$
is satisfied with $\gamma = m-\rho$. Indeed (\ref{TIScom})
follows from the following proof with slight modifications under
this weaker condition on the growth of the gap. However, this
modified proof does not give $\pi\in S^0_\rho(\epsi)$, a fact we
will make use of in the following. To avoid further
complications in the presentation, we decided to state only the
stronger result for the stronger growth condition. 
\end{remark}

\noindent {\bf Proof.}\,We decompose the proof into
two steps.\smallskip

\noindent {\bf Step I. Construction of the Moyal projector}

\noindent In
general $\pi_{0}$ is not a projector in the Moyal algebra, i.e. $\pi_{0}\
\#\ \pi_{0}\neq \pi_{0}$. The following lemma shows that $\pi_{0}$
can be corrected, order by order in $\varepsilon$, so to obtain a true
Moyal projector $\pi $ which Moyal commutes with $H$.
Similar constructions appeared in the context of the Schr\"odinger equation
 several times in the literature \cite{NS,BN,EW}.
Our proof was strongly influenced by the one in  \cite{NS}, but differs in relevant
details, since we consider different symbol classes. It relies on the
construction of the {\em local} Moyal resolvent of $H_0(z)$. The construction of the
global inverse of an elliptic symbol, often called the parametrix, is well known
\cite{DiSj,Folland,Nirenberg}.

\begin{lemma}\label{Prop Moyal proj}
Assume either (IG)$_{m}$ or (CG). Then there exists a \emph{unique} formal
symbol
\[
\pi =\sum_{j\geq 0}\varepsilon^{j}\pi_{j}\qquad \pi_{j}\in S_{\rho
}^{-j\rho }(\mathcal{B}(\Hi _{\mathrm{f}}))\,,
\]
such that $\pi_{0}(z)$ is the spectral projector of $H_{0}(z)$
corresponding to ${\sigma_{\rm r}}(z),$ with the following properties:

\begin{enumerate}
\item  $\pi\, \#\, \pi =\pi ,$

\item  $\pi^{*}=\pi ,$

\item  $[\,H,\,\pi\,]_{\#}:= H\, \#\, \pi -\pi\, \#\, H=0.$
\end{enumerate}
\end{lemma}

\noindent

\noindent \textbf{Proof.} We give the proof under the assumption (IG)$_{m}$.
The proof under assumption (CG) is simpler, since all the symbols which appear
belong to $S^{0}_0(\epsi)$.

We first provide a constructive scheme for the special case where 
${\sigma_{\rm r}}(z) = \{{E_{\rm r}}(z)\}$
is an eigenvalue, which, at the same time, proves uniqueness of $\pi$ in the general case.
It follows basically the construction as given in \cite{EW}. The reason for including this
scheme is that  the aim of adiabatic perturbation theory is, in particular, to give an as simple
as possible recipe for explicitly computing the relevant quantities.
The inductive scheme for constructing $\pi$ in the  special case
${\sigma_{\rm r}}(z) = \{{E_{\rm r}}(z)\}$ is much better suited for explicit computations
 than the general construction which will follow later on.

Note that $\pi_0\,\#\,\pi_0 - \pi_0 = \mathcal{O}(\epsi)$ and $[H_0,\pi_0]_\#=\mathcal{O}(\epsi)$
and proceed by induction. Assume that we found $\pi^{(n)} = \sum_{j=0}^n \pi_j$ such that
\begin{equation}\label{piCon1}
\pi^{(n)}\,\#\,\pi^{(n)}-\pi^{(n)} = \epsi^{n+1}\,G_{n+1} + \mathcal{O}(\epsi^{n+2})\,,
\end{equation}
where, in particular,  (\ref{piCon1}) defines $G_{n+1}$.
Thus  the next order term in the expansion $\pi_{n+1}$  must
satisfy
\[
\pi_{n+1}\,\pi_0 + \pi_0\,\pi_{n+1}- \pi_{n+1} = -G_{n+1}\,,
\]
which  uniquely determines
the diagonal part of $\pi_{n+1}$ to be
\begin{equation}\label{piCon2}
\pi_{n+1}^{\rm D}
=- \pi_0\,G_{n+1}\,\pi_0 + (1- \pi_0)\,G_{n+1}\,(1-\pi_0)\,.
\end{equation}
Since $G_{n+1} =  \pi_0\,G_{n+1}\,\pi_0 + (1- \pi_0)\,G_{n+1}\,(1-\pi_0)$ follows
from the fact that $G_{n+1}$ is the principal symbol of
$\epsi^{-n-1}(\pi^{(n)}\,\#\,\pi^{(n)}-\pi^{(n)})$, $\omega^{(n)}
:= \pi^{(n)} + \epsi^{n+1}\pi^{\rm D}_{n+1}$
indeed satisfies  (i) up to an error of order $\mathcal{O}(\epsi^{n+2})$.

By induction assumption we also have that $[H,\pi^{(n)}]_\#=\mathcal{O}(\epsi^{n+1})$
and thus
\begin{equation}\label{piCon3}
[H, \omega^{(n)}]_\#=\epsi^{n+1}F_{n+1} + \mathcal{O}(\epsi^{n+2})\,.
\end{equation}
Hence,  the diagonal part of $\pi_{n+1}$ being fixed already, the
off-diagonal part of $\pi_{n+1}$  must satisfy $[H_0,
\pi_{n+1}^{\rm OD}] = - F_{n+1}$. In particular,
\begin{eqnarray}\label{piCon4}\lefteqn{\hspace{-2.5cm}
H_0(z)\, \left(\pi_0(z) \pi_{n+1}(z)(1-\pi_0(z))\right) -
 \left(\pi_0(z) \pi_{n+1}(z)(1-\pi_0(z))\right) \,H_0(z)
}\nonumber \\&&\qquad=\, -\, \pi_0(z)\,F_{n+1}(z)\,(1-\pi_0(z))
\end{eqnarray}
for all $z\in\mathbb{R}^{2d}$. 
We first show that if (\ref{piCon4}) has a solution $\pi_0(z) \pi_{n+1}(z)(1-\pi_0(z)) =:
\pi_{n+1}^{OD1}(z)$, it is unique, i.e.\ that the kernel of the map $\pi_{n+1}^{OD1}(z)
\mapsto [H_0(z),\pi_{n+1}^{OD1}(z)]$ restricted to Ran$(1-\pi_0(z))$ contains only zero.
To see this let $\overline\sigma_{\rm r}(z) := (\sup\sigma_{\rm r}(z)-\inf\sigma_{\rm r}(z))/2$
and note that, due to the gap condition, 
$H_0(z)-\overline\sigma_{\rm r}(z)$ is invertible on Ran$(1-\pi_0(z))$ with
$\|(H_0(z)-\overline\sigma_{\rm r}(z))^{-1}(1-\pi_0(z))\| < 2/{\rm diam}(\sigma_{\rm r}(z))$.
Hence 
\begin{eqnarray*}\lefteqn{
[H_0(z),\pi_{n+1}^{OD1}(z)] = 0 \Leftrightarrow
[H_0(z)-\overline\sigma_{\rm r}(z) ,\pi_{n+1}^{OD1}(z)] = 0} \\ &\Leftrightarrow & 
\pi_{n+1}^{OD1}(z) = (H_0(z)-\overline\sigma_{\rm r}(z)) \pi_{n+1}^{OD1}(z)
 (H_0(z)-\overline\sigma_{\rm r}(z))^{-1}
\end{eqnarray*}
and therefore
\begin{eqnarray*}\lefteqn{
 \|\pi_{n+1}^{OD1}(z)\|}\\& \leq & \|(H_0(z)-\overline\sigma_{\rm r}(z))\pi_0(z)\|\,\|
 \pi_{n+1}^{OD1}(z)\|\,\|
 (H_0(z)-\overline\sigma_{\rm r}(z))^{-1}(1-\pi_0(z))\|\\& = & C\, \| \pi_{n+1}^{OD1}(z)\|
\end{eqnarray*}
with $C<1$. Hence $\pi_{n+1}^{OD1}(z)=0$
 and we conclude
that $\pi_{n+1}$ is unique when it exists.

In the special case that  ${\sigma_{\rm r}}(z) = \{{E_{\rm r}}(z)\}$,  
(\ref{piCon4}) can be solved, and one finds
\begin{equation}\label{piCon5}
\pi_0 \pi_{n+1}(1-\pi_0) =\pi_0\,F_{n+1}\, (H_0-{E_{\rm r}})^{-1}\,(1-\pi_0)\,.
\end{equation}
Using that $F_{n+1}$ is the principal symbol of $\epsi^{-n-1}[H,\omega^{(n)}]_\#$,
that $\pi_0$  is the principal symbol of $\omega^{(n)}$ and that  $\omega^{(n)}$
satisfies (i) up to $\mathcal{O}(\epsi^{n+2})$, one finds that
$\pi_0 F_{n+1} \pi_0 =(1-\pi_0) F_{n+1}(1- \pi_0)=0$ and thus that
$\pi^{(n+1)}$ defined through (\ref{piCon2}) and (\ref{piCon5}) satisfies
(i) and (iii) up to $\mathcal{O}(\epsi^{n+2})$.

We conclude that by induction we have uniqueness of $\pi$ in the general case, and an
explicit construction for $\pi$ when   ${\sigma_{\rm r}}(z) = \{{E_{\rm r}}(z)\}$.
The latter one involves four steps at each order: [a] Evaluation of $G_{n+1}$ as in
(\ref{piCon1}),
[b] computation of $\pi_{n+1}^{\rm D}$ as in  (\ref{piCon2}),
[c] evaluation of $F_{n+1}$ as in (\ref{piCon3}),  [d]
 computation of $\pi_{n+1}^{\rm OD}$ as in  (\ref{piCon5}).

We now turn to the construction of $\pi$ in the general case.
 Since the Moyal product is a local operation
(it depends only on the pointwise value of the symbols and their derivatives)
it suffices to construct $\pi$ locally in phase space and then uniqueness will
liberate us from gluing the local results together.

 Let us fix a point $z_{0}\in \mathbb{R}^{2d}$.  From the continuity of the
map $z\mapsto H_{0}(z)$ and the gap condition it follows that there exists a
neighborhood $\mathcal{U}_{z_{0}}$ of $z_{0}$ such that for every $z\in
\mathcal{U}_{z_{0}}$ the set ${\sigma_{\rm r}}(z)$ can be enclosed in a
positively-oriented complex circle $\Gamma(z_0)$ (independent of $z$) in such a
way that $\Gamma(z_0)$ is symmetric with respect to the real axis,
\begin{equation} \label{Cgcon}
\mathrm{dist}(\Gamma(z_0) ,\sigma (z))\geq \frac{1}{4}C_{g}\left\langle
p \right\rangle^{\sigma }\qquad {\rm for\, all} \,\,z\in\mathcal{U}_{z_0}
\end{equation}
and
\begin{equation}\label{Crcon}
{\rm Radius}(\Gamma(z_0))\leq C_{\rm r}\,\sup_{z\in \mathcal{U}_{z_0}}\,
\langle p\rangle^\sigma\,,
\end{equation}
where $\mathrm{Radius}\left( \Gamma (z_{0})\right) $ is the radius of the
complex circle $\Gamma =\Gamma (z_{0})$.
The constant $C_{g}$ in (\ref{Cgcon}) is the same as
in (\ref{Gap}) and the existence of a constant $C_{\rm r}$ {\em independent} 
of $z_0$ such that
(\ref{Crcon}) is satisfied follows from assumption (G3).
We keep $\sigma $ in the notation
as a bookkeeping device, in order to distinguish the contributions related
to the gap, although $\sigma=m$.

Let us choose any $\zeta \in \Gamma$ and restrict all the following expressions
to $z\in \mathcal{U}_{z_0}$. There exist a formal symbol
$R(\zeta )$ -- the \emph{local Moyal resolvent} of $H$ -- such that
\begin{equation}
R(\zeta )\ \#\ (H-\zeta 1)=1=(H-\zeta 1)\ \#\ R(\zeta )\qquad 
\mbox{on }\mathcal{U}_{z_{0}}.  \label{Moyal resolvent}
\end{equation}
The symbol $R(\zeta )$ can be explicitly constructed. We abbreviate
\[
R_{0}(\zeta )\ =\ (H_{0}-\zeta 1)^{-1}
\]
where the inverse is understood in the $\mathcal{B}(\Hi _{\mathrm{f}})
$-sense and exists according to (\ref{Cgcon}). By induction, suppose that 
$R_{(n)}(\zeta)=\sum_{j=0}^{n}\varepsilon^{j}R_{j}(\zeta )$ satisfies the first equality
in (\ref{Moyal resolvent}) up to $\mathcal{O}(\varepsilon^{n+1})$-terms,
i.e.\
\[
R^{(n)}(\zeta )\ \#\ (H-\zeta 1)=1+\varepsilon^{n+1}E_{n+1}(\zeta)+
\mathcal{O}(\varepsilon^{n+2}).
\]
By choosing $R_{n+1}=-E_{n+1}\ (H_{0}-\zeta 1)^{-1}$, we obtain that
$R^{(n+1)}=R^{(n)}+\varepsilon^{n+1}R_{n+1}$ satisfies the same equality up
to $\mathcal{O}(\varepsilon^{n+2})$-terms. Then the formal symbol $R(\zeta
)=\sum_{j\geq 0}\varepsilon^{j}R_{j}(\zeta )$ satisfies the first equality
in (\ref{Moyal resolvent}) which -- by the associativity of the Moyal
product -- implies the second one.

Equation (\ref{Moyal resolvent}) implies that $R(\zeta )$ satisfies the
\emph{resolvent equation  }
\begin{equation}
R(\zeta )-R(\zeta^{\prime })=(\zeta -\zeta^{\prime })\ R(\zeta )\ \#\
R(\zeta^{\prime })\qquad \mbox{on }\,\,\mathcal{U}_{z_{0}}.
\label{Resolvent eq}
\end{equation}

\noindent From the resolvent equation it follows -- by using an argument
similar to the standard one in operator theory \cite{Kato book} -- that the
symbol $\pi =\sum_{j\geq 0}\varepsilon^{j}\pi_{j}$ defined by
\begin{equation}
\pi_{j}(z):=\frac{i}{2\pi }\int_{\Gamma }R_{j}(\zeta ,z)\ d\zeta ,\qquad
z\in \mathcal{U}_{z_{0}}\,,  \label{Riesz formula}
\end{equation}
is a Moyal projector such that $[H,\pi ]_{\#}=0$ on $\mathcal{U}_{z_{0}}$.
Indeed, for every fixed $z\in \mathcal{U}_{z_{0}}$ and $j\in \mathbb{N}$, the
map $\zeta \mapsto R_{j}(\zeta ,z)$ is holomorphic in a neighborhood of the
circle $\Gamma(z_0)$. Then $\Gamma(z_0)$ can be expanded to a slightly larger circle
$\Gamma^{\prime }$ without changing the left hand side of (\ref{Riesz formula}) and
we obtain
\begin{eqnarray}
\left( \pi\ \#\ \pi \right)_{j} &=&\left( \frac{i}{2\pi }\right)
^{2}\int_{\Gamma^{\prime }}d\zeta^{\prime }\int_{\Gamma }d\zeta \left(
R(\zeta^{\prime })\ \#\ R(\zeta )\right)_{j}  \label{Pi square} \\
&=&\left( \frac{i}{2\pi }\right)^{2}\int_{\Gamma^{\prime }}d\zeta^{\prime
}\int_{\Gamma }d\zeta \left( \zeta^{\prime }-\zeta \right)^{-1}\left[
R(\zeta^{\prime })-R(\zeta )\right]_{j}  \nonumber \\
&=&\frac{i}{2\pi }\int_{\Gamma }R_{j}(\zeta )\ d\zeta =\pi_{j}  \nonumber
\end{eqnarray}
where (\ref{Resolvent eq}) has been used. The first equality in (\ref{Pi
square}) follows by noticing that for every $\gamma \in \mathbb{N}^{2d}$
\[
\partial_{z}^{\gamma }\pi_{j}(z)=\frac{i}{2\pi }\int_{\Gamma }\partial
_{z}^{\gamma }R_{j}(\zeta ,z)\ d\zeta \qquad z\in \mathcal{U}_{z_{0}},
\]
and by expanding the Moyal product order by order in $\varepsilon $.

Since the circle $\Gamma $ is symmetric with
respect to the real axis one immediately concludes that $\pi^{*}=\pi $,
since $R(\zeta )^{*}=R(\bar{\zeta})$ as a consequence of (\ref{Moyal
resolvent}). From (\ref{Riesz formula}) it follows that $\pi $
Moyal-commutes with $R(\lambda )$ for any $\lambda \in \Gamma $. Then, by
multiplying  $\pi\, \#\, R(\lambda )=R(\lambda )\, \#\, \pi $ by $(H-\lambda 1)$
 on both sides, one obtains that $H\, \#\, \pi =\pi\, \#\, H$.

Finally we have to show that $\pi_{j}\in S_{\rho }^{-j\rho }$ for every
$j\in \mathbb{N}$. From the Riesz formula (\ref{Riesz formula}) it follows that
for every $\gamma \in \mathbb{N}^{2d}$ one has
\[%\begin{equation}
\left\| \left( \partial_{z}^{\gamma }\pi_{j}\right) (z)\right\|_{\mathcal{
B}(\Hi _{\mathrm{f}})}\leq 2\pi \mbox{ }\mathrm{Radius}\left( \Gamma
(z_{0})\right) \sup_{\zeta \in \Gamma (z_{0})}\left\| \left( \partial
_{z}^{\gamma }R_{j}\right) (\zeta ,z)\right\|_{\mathcal{B}(\Hi _{
\mathrm{f}})} \,. %\label{Radius resolvent}
\]%\end{equation}
According to (\ref{Crcon}) we are left to prove that
\begin{equation}
\sup_{\zeta \in \Gamma (z_{0})}\left\| \left( \partial_{q}^{\alpha
}\partial_{p}^{\beta }R_{j}\right) (\zeta ,z)\right\|_{\mathcal{B}
(\Hi _{\mathrm{f}})}\leq C_{\alpha\beta j}\left\langle p\right\rangle^{-\sigma
-j\rho -|\beta |\rho }, \quad \alpha,\beta \,\in \mathbb{N}^d\,,\,\,j\in\mathbb{N}\,,
\label{Sufficient cond}
\end{equation}
where $C_{\alpha\beta j}$ must {\em not} depend on $z_0$.
 As for $R_{0}$, we notice that according to (\ref{Cgcon}) one has
\begin{equation}
\left\| (H_{0}(z)-\zeta 1)^{-1}\right\|_{\mathcal{B}(\Hi _{
\mathrm{f}})}\leq \frac{1}{\mathrm{dist}(\zeta ,\sigma (H_{0}(z)))}\leq
\frac{4}{C_{g}}\left\langle p\right\rangle^{-\sigma }\,,  \label{Bound R0}
\end{equation}
 and moreover,
\begin{eqnarray*}
\left\| \nabla_{p}R_{0}(z)\right\|_{\mathcal{B}(\Hi _{\mathrm{f}})} 
&=&\left\| -(R_{0}\nabla_{p}H_{0}R_{0})(z)\right\|_{\mathcal{B}(\Hi _{\mathrm{f}})}
\\&\leq &\left( \frac{4}{C_{g}}\right)
^{2}\left\langle p\right\rangle^{-2\sigma }\left\| \nabla
_{p}H_{0}(z)\right\|_{\mathcal{B}(\Hi _{\mathrm{f}})} \\
&\leq &C\left\langle p\right\rangle^{-2\sigma + m-\rho } = C\left\langle 
p\right\rangle^{-\sigma-\rho }\,,
\end{eqnarray*}
where the last bound follows from the fact that $H_{0}\in S_{\rho }^{m}$
(recall that $\sigma =m$). By
induction one  controls higher order derivatives and
(\ref{Sufficient cond}) follows for $j=0$.

\noindent Again by induction, assume that $R_{0},\ldots ,R_{n}$ satisfy the bound
(\ref{Sufficient cond}). Then, by writing out
\[
E_{n+1}=\left( R_{(n)}(\zeta )\ \#\ (H-\zeta 1)-1\right)_{n+1}
\]
and using (\ref{Composition}), one concludes that
$R_{n+1}=-E_{n+1}R_{0}$ satisfies (\ref{Sufficient cond}) with $\sigma =m$.
\hfill $\qed$ \newline

\noindent \textbf{Step II. Quantization}

\noindent First of all, by resummation (Prop.\ \ref{Prop resummation}) we
obtain a semiclassical symbol $\pi :\mathbb{R}^{2d}\times [0,\varepsilon
_{0})\rightarrow $ $\mathcal{B}(\Hi _{\mathrm{f}})$ whose asymptotic
expansion is given by $\sum_{j\geq 0}\varepsilon^{j}\pi_{j}$. Then, by
Weyl quantization, one gets a bounded operator 
$\widehat{\pi }\in \mathcal{B}(\Hi )$  (see Prop.\ \ref{Prop CaldVaill}) which is an
almost-projector, in the sense that

\begin{enumerate}
\item  $\widehat{\pi }^{2}=\widehat{\pi }+\mathcal{O}_{-\infty }(\varepsilon
^{\infty })\qquad $

\item  $\widehat{\pi }^{*}=\widehat{\pi }$

\item  $[\widehat{H},\widehat{\pi }]=\mathcal{O}_{-\infty }(\varepsilon
^{\infty })$
\end{enumerate}

\noindent Notice that the assumption $\rho >0$ is crucial in order to obtain
(iii) for an unbounded $\widehat{H}$.

In order to get a true projector we follow the idea of \cite{NS} and notice that
$\|\widehat{\pi }^{2}-\widehat{\pi }\|=\mathcal{O}(\varepsilon
^{\infty })$ and the spectral mapping theorem for self-adjoint operators
imply that for each $n\in\mathbb{N}$ there is a $C_n<\infty$ such that
\[
\sigma(\widehat \pi)\subset [-C_n\epsi^n,C_n\epsi^n]\cup[1-C_n\epsi^n,1+C_n\epsi^n]
=: \sigma^\epsi_{0} \cup   \sigma^\epsi_{1}
\,.
\]
Hence one can define for $\epsi\leq 1/(4C_1)$
\[
\Pi := \frac{i}{2\pi} \int_{|\zeta-1|=\frac{1}{2}}\,(\widehat\pi-\zeta)^{-1}\,d\zeta
\,.
\]
 Then $\Pi^{2}=\Pi $ follows and we claim that 
$\Pi =\widehat{\pi}+\mathcal{O}_{0}(\varepsilon^{\infty })$. Indeed,
\[
\widehat \pi = \int_{ \sigma^\epsi_{0} \cup\, \sigma^\epsi_{1}}\lambda
E(d\lambda) = \mathcal{O}_0(\epsi^n) + \int_{ \sigma^\epsi_{1}}E(d\lambda)
= \Pi +  \mathcal{O}_0(\epsi^n)\qquad \mbox{for all}\,\,n\in\mathbb{N}\,,
\]
where $E(\cdot)$ is the projection valued measure of
$\widehat{\pi }$.
Finally notice that
\begin{eqnarray*}
[\widehat H ,\Pi] &=&
\frac{i}{2\pi} \int_{|\zeta-1|=\frac{1}{2}} [\widehat H, (\widehat\pi-\zeta)^{-1}] d\zeta\\
&=& -\,\frac{i}{2\pi} \int_{|\zeta-1|=\frac{1}{2}}  (\widehat\pi-\zeta)^{-1}
[\,\widehat H,\,\widehat \pi\,]
 (\widehat\pi-\zeta)^{-1} d\zeta\,,
\end{eqnarray*}
which implies that
\[
\big\| [\widehat H ,\Pi]\big\|_{\mathcal{B(H)}} \leq C \big\| [\widehat H ,
\widehat\pi]\big\|_{\mathcal{B(H)}}  = \mathcal{O}_0(\epsi^\infty)\,.
\]
This concludes the proof of the theorem.
\hfill$\qed$\bigskip

\noindent \textbf{Essential self-adjointness of }$\widehat{H}$\textbf{. }

\noindent Since $H$ is an \emph{hermitian} symbol its Weyl quantization 
$\widehat{H}$ is symmetric on the invariant domain $\mathcal{S}(\mathbb{R}^{d},
\mathcal{B}(\Hi _{\mathrm{f}}))\subseteq \Hi $. If $H$
belongs to $S^{0}(\varepsilon )$ then $\widehat{H}$ is a bounded
operator, and there is nothing to prove.

In order to prove essential self-adjointness in the case 
$H\in S_{1}^{1}(\varepsilon )$, we use an argument  of \cite{Robert}. 
The proof does not exploit the smallness of $\varepsilon $ 
and we therefore consider any $\varepsilon >0$.
For $s>0$ let
\[
B_{\pm s}(q,p)=(H_{0}(q,p)\pm is1)^{-1}\,,
\]
which, according to 
Proposition \ref{Prop point inversion},
belongs to $S_{1 }^{0}(\mathcal{B}(\Hi _{\mathrm{f}}))$. Moreover
\[
(H\pm is1)\ \tilde{\#}\ B_{\pm s}={\bf 1}+\varepsilon S_{\pm s}\,,
\]
where $S_{\pm s}\in S_{1 }^{0}(\varepsilon )$, since $H\in
S_{1 }^{1}(\varepsilon )$ and $B_{\pm s}\in S_{1 }^{0}(\varepsilon )$.
After Weyl quantization we obtain that
\[
(\widehat{H}\pm is{\bf 1})\ \widehat{B}_{\pm s}={\bf 1}+\varepsilon \widehat{S}_{\pm s}\qquad
\mbox{with}\,\,\,\,\big\| \widehat{S}_{\pm s}\big\|_{\mathcal{B}(
\Hi )}<\frac{C }{|s|},
\]
the latter bound following (for $s$ large enough) from Proposition 
\ref{Prop CaldVaill} and from estimating the Fr\'{e}chet semi-norms of $S_{\pm s}$.
Essential self-adjointness of $\widehat H$ on the domain $\mathcal S$ follows,
if we can show that Ker$(\widehat H^*\pm is)=\{0\}$ for some $s>0$. 
For this let $\varphi \in$ 
Ker$(\widehat H^*\pm is)$ and $\psi\in \mathcal{S}$. Using $\widehat B_\pm\mathcal S
\subset \mathcal S$, we obtain
\[
0= \langle  (\widehat H^*\mp is)\varphi,\widehat B_\pm \psi\rangle
=  \langle \varphi, (\widehat H\pm is)\widehat B_\pm \psi\rangle
=  \langle \varphi,({\bf 1}+ \varepsilon \widehat S_{\pm s}) \psi\rangle\,.
\] 
Since $\|\epsi \widehat{S}_{\pm s}\|<1$ for $s$ large enough,
 $({\bf 1}+ \varepsilon \widehat S_{\pm s})\mathcal{S}$ is dense in $\mathcal H$
and hence $\varphi =0$ follows.

%%%%%%%%%%%%%%%%%%%%%%%   UNITARIES  %%%%%%%%%%%%%%%%%%%%%%%%%%%%%%%%%5

\section{Reference subspace and intertwining unitaries}
\label{SUNI}

The fact that the subspace associated with an isolated energy band
decouples from its orthogonal complement up to small errors in $\epsi$
leads immediately to the following question. Is there a natural way
to describe the dynamics of the system inside the almost invariant
subspace Ran$\Pi$? The main obstruction for such a simple description is the fact that
the subspace Ran$\Pi$ depends on $\epsi$ and is not easily accessible.
Even worse, in general the limit $\lim_{\epsi\to 0}\Pi$ does not exist, meaning that
Ran$\Pi$ is not even close to an $\epsi$-independent subspace.
In order to obtain a useful description of the effective intraband dynamics
 we thus need to map Ran$\Pi$ to an easily accessible
and $\epsi$-independent {\em reference} subspace.

>From the continuity
of $z\mapsto H_{0}(z)$ and the gap condition it follows that there is a
subspace $\mathcal{K}_{\mathrm{f}} \subset \Hi _{\rm f}$ independent of $(q,p)$
such that the subspaces Ran$\pi_0(q,p)$ are all isomorphic to $\mathcal{K}_{\rm f}$.
Let $\pi_{\rm r}$ be the projection on  $\mathcal{K}_{\mathrm{f}}$, then
${\Pi_{\rm r}} := {\bf 1}\otimes \pi_{\rm r}$ $(=\widehat\pi_{\rm r})$
will serve as the projector
on the {\em reference subspace} $\mathcal{K}:=$ Ran${\Pi_{\rm r}}$.
Of course $\mathcal{K}_{\rm f}$ is {\em highly} non-unique and a convenient
choice must be made in concrete applications.

Once the reference Hilbert space is fixed we next chose a
unitary operator valued smooth function $u_{0}(z)$ which pointwise in
phase space intertwines $\pi_0(z)$ and $\pi_{\rm r}$, i.e.\
\begin{equation}
u_{0}(z) \,\pi_{0}(z)\, u_{0}(z)^*=\pi_{\mathrm{r}}\,.  \label{Intertwining}
\end{equation}
The existence of such a \emph{smooth} map follows from a
bundle-theoretic argument given at  the end of this section.
Again $u_0(z)$ is not unique and must be chosen conveniently. We will
see in Section~\ref{SDIRAC} that there is an optimal
choice for $u_0(z)$, which reflects the physics of the problem. 

We cannot prove that it is possible to
choose $u_{0}$ in $S_{\rho }^{0}(\mathcal{B}(\Hi _{\mathrm{f}}))$.
Indeed, relation (\ref{Intertwining}) does not imply any bound at infinity
on the derivatives of $u_{0}$, as can be seen by multiplying $u_{0}$ with
a highly oscillating phase. Hence we \emph{assume} that  $u_{0}$ is
 in $S_{\rho }^{0}(\mathcal{B}(\Hi _{\mathrm{f}}))$, as
will be the case in the physical examples.

In the following $\mathcal{U}(\Hi )$ will denote
the group of  unitary operators over $\Hi $.

\begin{theorem}
\label{Th unitaries} Assume either (IG)$_{m}$ or (CG) and that there exists
a $\mathcal{U}(\Hi _{\mathrm{f}})$-valued map $u_{0}\in S_{\rho
}^{0}(\mathcal{B}(\Hi _{\mathrm{f}}))$ which satisfies (\ref
{Intertwining}). Then there exist a  unitary operator $U\in \mathcal{B}(\Hi )$
 such that
\begin{equation}
U \,\Pi\, U^*={\Pi_{\rm r}}  \label{Intertwines}
\end{equation}
and $U=\widehat{u}+\mathcal{O}_{0}(\varepsilon^{\infty })$, where
 $u\asymp\sum_{j\geq 0}\varepsilon^{j}u_{j}$
in $S_{\rho }^{0}(\varepsilon )$ with
principal symbol $u_{0}$.
\end{theorem}

\begin{remark}\em \label{NagyRem}
In \cite{NS} the Nagy transformation (\ref{Nagy}) is used in order to map Ran$\Pi$
to the $\epsi$-independent subspace Ran$\widehat \pi_0$. This is possible because
in their application the symbol $\pi_0$ depends only on $q$ and, as a consequence, 
$\widehat \pi_0$ is a projector satisfying $\|\Pi-\widehat \pi_0\|=\Or(\epsi)$.
However, in general $\pi_0$ depends on $q$ and $p$, see Section~6 and \cite{PST2} for relevant examples, and
the mapping to the reference space becomes  more subtle.
\end{remark}

\noindent {\bf Proof.} 
\noindent \textbf{Step I. Construction of the Moyal unitaries.}

\noindent Again $u_{0}$ fails to be a Moyal unitary (i.e.\ $u_{0}^{*}\, \#\,
u_{0}\neq 1$) and to intertwine $\pi$ and $\pi_{\mathrm{r}}$. However, the
following lemma shows that $u_{0}$ can be corrected order by order to
reach this goal. 
The idea of constructing a pseudodifferential operator which is almost unitary
and diagonalizes a given pseudodifferential operator  has a long tradition, 
cf.\ \cite{Nirenberg} Section~7 and references
therein, and was applied in different settings many times, e.g.\ \cite{TaylorArt,HS}.

\begin{lemma}
Assume either (IG)$_{m}$ or (CG) and that there exists
a $\mathcal{U}(\Hi _{\mathrm{f}})$-valued map $u_{0}\in S_{\rho
}^{0}(\mathcal{B}(\Hi _{\mathrm{f}}))$ which satisfies (\ref
{Intertwining}). Then there is a formal symbol
$u=\sum_{j\geq 0}\varepsilon^{j}u_{j}$,  with $u_{j}\in S_{\rho }^{-j\rho }(
\mathcal{B}(\Hi _{\mathrm{f}}))$, such that

\begin{enumerate}
\item  $u^{*} \#\, u=1\qquad$ and \qquad $u\, \#\, u^{*}=1$\,,
\item  $u \, \#\,\, \pi\, \#\, u^*=\pi_{\mathrm{r}}$\,,
\end{enumerate}

\noindent where $\pi$ is the Moyal projector constructed in Lemma
\ref{Prop Moyal proj}. \label{Prop Moyal unitaries}\bigskip
\end{lemma}

\begin{remark} \label{NonUniRem}
\emph{ We emphasize that -- as opposed to the Moyal projector $\pi$
appearing in Lemma \ref{Prop Moyal proj} -- the Moyal unitary $u$ is highly
non-unique even for fixed $u_0$. As it will follow from the proof, all the possible choices of
Moyal unitaries intertwining $\pi$ and $\pi_{\mathrm{r}}$ with prescribed principal symbol
$u_0$ are parametrized by the antihermitian Moyal symbols which are diagonal in the
$\pi_{\mathrm{r}}$-splitting.}\bigskip
\end{remark}

\noindent \textbf{Proof of Lemma \ref{Prop Moyal unitaries}.}
Observe that $u_0$ satisfies (i) and (ii) on the principal symbol level. We proceed by
induction and assume that we found $u^{(n)} = \sum_{j=0}^n\epsi^ju_j$ satisfying (i)
and (ii) up to $\mathcal{O}(\epsi^{n+1})$. We will construct $u_{n+1}$
such that $u^{(n+1)} = u^{(n)} + \epsi^{n+1}u_{n+1}$ satisfies (i) and (ii)
up to  $\mathcal{O}(\epsi^{n+2})$. To this end we write without restriction
\[
u_{n+1} =:  (a_{n+1} + b_{n+1})\,u_0\,,
\]
with $a_{n+1}$ hermitian and $b_{n+1}$ anti-hermitian. By induction assumption
we have
\[%\begin{equation}
\begin{array}{rcl}
u^{(n)}\, \#\, u^{(n)*}-1 & = & \varepsilon^{n+1}A_{n+1}+\mathcal{O}
(\varepsilon^{n+2})  \\[1ex]
u^{(n)*}\, \#\, u^{(n)}-1 &=&\varepsilon^{n+1}\tilde{A}_{n+1}+
\mathcal{O}(\varepsilon^{n+2})\,.
\end{array} %\label{Un}
\]%\end{equation}
Thus $u_{n+1}$ has to solve
\begin{equation}
\begin{array}{rcl}
u_{0} \, u_{n+1}^*+u_{n+1}\, u_{0}^* &=&-A_{n+1},  \\[1ex]
u_{n+1}^*\, u_{0} +u_{0}^*\, u_{n+1}  &=&-\tilde{A}_{n+1}\,.
\end{array} \label{Un2}
\end{equation}
The first equation in (\ref{Un2}) fixes $a_{n+1} = -\frac{1}{2}A_{n+1}$, since $A_{n+1}$ is
hermitian as it is the principal symbol of $\epsi^{-n-1}(u^{(n)}\, \#\, u^{(n)*}-1)$.
The second equation in   (\ref{Un2}) is then also satisfied, since
 the compatibility equation $u_{0}\, A_{n+1}=\tilde{A}_{n+1}\,u_{0}$
follows from
\[
\frac{1}{\varepsilon^{n+1}}u^{(n)}\, \#\, (u^{(n)*}\, \#\, u^{(n)}-1)=
\frac{1}{\varepsilon^{n+1}}(u^{(n)}\, \#\, u^{(n)*}-1)\, \#\, u^{(n)}
\]
by noticing that $u_{0}\, \tilde A_{n+1}$ (resp.\ $A_{n+1}\, u_{0}$) is the
principal symbol of the l.h.s (resp.\ r.h.s).

Note that (\ref{Un2}) puts no constraint on $b_{n+1}$ and we are left to determine it
using (ii).
Let $w^{(n)} = u^{(n)} + \epsi^{n+1}\,a_{n+1}\,u_0$, then by induction assumption
\[%\begin{equation}
w^{(n)}\, \#\, \pi\, \#\, w^{(n)*}-\pi_{\mathrm{r}}=\varepsilon^{n+1}B_{n+1}+
\mathcal{O}(\varepsilon^{n+2})  %\label{Un3b}
\]%\end{equation}
and thus
\[
u^{(n+1)}\, \#\, \pi\, \#\, u^{(n+1)*}-\pi_{\mathrm{r}}=\varepsilon
^{n+1}\left( B_{n+1}+[b_{n+1},\pi_{\mathrm{r}}]\right) +\mathcal{O}
(\varepsilon^{n+2})\,.
\]
Hence we need to find an anti-hermitian $b_{n+1}$ satisfying
\[%\begin{equation}
B_{n+1}+[b_{n+1},\pi_{\mathrm{r}}]=0\,,  %\label{Un5}
\]%\end{equation}
which is given by
\begin{equation}
b_{n+1}=[\pi_{\mathrm{r}},\,B_{n+1}]\,,  \label{Un6}
\end{equation}
provided that $B_{n+1}$ is hermitian and off-diagonal in the $\pi_{\mathrm{r}}$-splitting,
i.e.\ $\pi_{\mathrm{r}}\ B_{n+1}\ \pi_{\mathrm{r}}$ and $(1-\pi
_{\mathrm{r}})\ B_{n+1}\ (1-\pi_{\mathrm{r}})$ vanish. This follows
by noticing that $B_{n+1}$ is the principal symbol of $\varepsilon
^{-(n+1)}\left( w^{(n)}\ \#\ \pi\ \#\ w^{(n)*}-\pi_{\mathrm{r}}\right)$
and then
\begin{eqnarray*}
\lefteqn{(1-\pi_{\mathrm{r}})  B_{n+1}\ (1-\pi_{\mathrm{r}})\stackunder{
\varepsilon \rightarrow 0}{\sim }\frac{1}{\varepsilon^{n+1}}(1-\pi_{
\mathrm{r}})\left( w^{(n)} \, \# \, \pi\,  \# \, w^{(n)*}-\pi_{\mathrm{r}}\right)
  (1-\pi_{\mathrm{r}})} \\
&=&\frac{1}{\varepsilon^{n+1}}(1-\pi_{\mathrm{r}})\left( w^{(n)}\ \#\ \pi
\ \#\ w^{(n)*}\right) (1-\pi_{\mathrm{r}}) \\
&=&\frac{1}{\varepsilon^{n+1}}\left( \varepsilon^{2(n+1)}B_{n+1}\left(
w^{(n)}\ \#\ \pi\, \#\ w^{(n)*}\right) B_{n+1}+\mathcal{O}(\varepsilon
^{n+2})\right) \stackunder{\varepsilon \rightarrow 0}{\sim }0,
\end{eqnarray*}
where for the last equality we inserted $1-\pi_{\mathrm{r}}=w^{(n)}\ \#\ (1-\pi )\ \#\
w^{(n)*}+\varepsilon^{n+1}B_{n+1}+\mathcal{O}(\varepsilon^{n+2})$ and used  that
$w^{(n)}$ solves (i) up to $\mathcal{O}(\epsi^{n+2})$ and that $\pi$ is a Moyal
projector. A
similar argument shows that $\pi_{\mathrm{r}}\ B_{n+1}\ \pi_{\mathrm{r}}$
vanishes too. Note also that (\ref{Un6}) fixes only the off-diagonal part of $b_{n+1}$
and one is free to choose the diagonal part of $b_{n+1}$ arbitrarily, which is
exactly the non-uniqueness mentioned in Remark \ref{NonUniRem}.

It remains to show that the assumption $u_{0}\in S_{\rho }^{0}$ implies that
$u_{j}^{{}}$ belongs to $S_{\rho }^{-j\rho }$.
Assume by induction  that $u^{(n)}\in M_{\rho }^{0}(\varepsilon )$. Then the formula
\[
a_{n+1}=-\frac{1}{2} A_{n+1}=-\frac{1}{2} \left(
u^{(n)}\, \#\, u^{(n)*}-1\right)_{n+1}
\]
shows that $a_{n+1}$ belongs to $S_{\rho }^{-(n+1)\rho }$ as it is
the $(n+1)$-th term of an
element of $M_{\rho }^{0}(\varepsilon )$. By Proposition \ref{Prop point product},
$a_{n+1}\,u_0\in S_{\rho }^{-(n+1)\rho }$ as well.
Analogously we have that $B_{n+1}\in  S_{\rho }^{-(n+1)\rho }$ by induction assumption,
therefore $b_{n+1}\in  S_{\rho }^{-(n+1)\rho }$ and thus 
$b_{n+1}\,u_0\in  S_{\rho }^{-(n+1)\rho }$,
which finally gives $u_{n+1}\in  S_{\rho }^{-(n+1)\rho }$.
\hfill $\qed$ \newline

\noindent \textbf{Step II. Quantization\mathstrut \medskip }

\noindent Now let $u$ denote a resummation of the formal power series 
$u=\sum_{j\geq 0}\varepsilon^{j}u_{j}$ in $S_{\rho }^{0}(\varepsilon )$ (see
Prop.\ \ref{Prop resummation}). Then, by Weyl quantization, one gets a
bounded operator $\widehat{u}\in \mathcal{B}(\Hi )$ (see Prop.\ \ref{Prop
CaldVaill}) such that:

\begin{enumerate}
\item  $\widehat{u}^{\,*}\widehat{u}=1+\mathcal{O}_{-\infty }(\varepsilon^{\infty
})\qquad $and\qquad $\widehat{u}\,\widehat{u}^{\,*}
=1+\mathcal{O}_{-\infty }(\varepsilon
^{\infty })$

\item  $\widehat{u}\,\widehat{\pi }\, \widehat{u}^{\,*}={\Pi_{\rm r}}+
\mathcal{O}_{-\infty }(\varepsilon^{\infty })$.
\end{enumerate}

\noindent
As a first step we
modify $\widehat{u}$ by an $\mathcal{O}_{0}(\varepsilon^{\infty
})$-term in order to get a true unitary operator $\widetilde
U\in \mathcal{U}(\Hi )$ (which, in general, does not correspond to the Weyl 
quantization of any
semiclassical symbol).  Let
\begin{equation}
\widetilde U=\widehat{u}\,\left( \widehat{u}^{\,*}\,\widehat{u}\right)^{-\frac{1}{2}}\,.
\label{Unitarization}
\end{equation}
Notice that $\widehat{u}^{\,*}\,\widehat{u}$ is a self-adjoint positive operator which
is $\mathcal{O}_{0}(\varepsilon^{\infty })$-close to the identity operator.
Then $( \widehat{u}^{\,*}\,\widehat{u})^{-\frac{1}{2}}$ is well-defined and again 
$\mathcal{O}_{0}(\varepsilon^{\infty })$-close to the identity operator. \
Hence (\ref{Unitarization}) defines a unitary operator which moreover is 
$\mathcal{O}_{0}(\varepsilon^{\infty })$-close to $\widehat{u}$.

Finally we modify $\widetilde U$ in order to obtain a unitary which exactly
intertwines ${\Pi_{\rm r}}$ and $\Pi$.
Since $\| \widetilde U\,\Pi\, \widetilde U^{\,*} - {\Pi_{\rm r}}\| <1$
for $\varepsilon $ sufficiently small, the Nagy formula as used in \cite{NS}
\begin{equation}\label{Nagy}
W:=\left[ 1-\left( \widetilde U\,\Pi\,\widetilde
U^{\,*}-{\Pi_{\rm r}}\right)^{2}\right]
^{-\frac{1}{2}}\left[ \widetilde U\,\Pi\,\widetilde
 U^{\,*}\,{\Pi_{\rm r}}+(1-\widetilde U\,\Pi\,\widetilde U^{\,*})(1-
{\Pi_{\rm r}})\right]
\end{equation}
defines a unitary operator $W\in \mathcal{U}(\Hi )$ such that
$W\,\widetilde U\,\Pi\,\widetilde U^{\,*} \,W^*=
{\Pi_{\rm r}}$ and $W= {\bf 1} + \mathcal{O}_0(\epsi^\infty)$.
Thus by defining $U=W\,\widetilde U$ one
obtains  (\ref{Intertwines}), with the desired properties.
\hfill
$\qed$ \newline

\begin{remark}
\emph{We sketch how to prove the existence of a smooth map $u_{0}$
satisfying (\ref{Intertwining}). Given
\[
E=\left\{ (z,\psi )\in \mathbb{R}^{2d}\times \Hi _{\mathrm{f}}:\psi \in
\mathrm{Ran}\pi_0 (z)\right\}
\]
the map $\Pi_{E}:E\rightarrow \mathbb{R}^{2d},$ \ $(z,\psi )\mapsto z$
defines a fibration of Hilbert spaces over the base space $\mathbb{R}^{2d}$. }

\emph{The fibration is locally trivial. Indeed for any $z_{0}\in \mathbb{R}^{2d}
$ there exists a neighborhood $\mathcal{U}_{z_{0}}$ such that $\left\| \pi
_{0}(z)-\pi_{0}(z_{0})\right\| <1$ for any $z\in \mathcal{U}_{z_{0}}$, so
that the Nagy formula
\[
w(z)=\left[ 1-\left( \pi_{0}(z)-\pi_{0}(z_{0})\right)^{2}\right]^{-\frac{
1}{2}}\left[ \pi_{0}(z)\pi_{0}(z_{0})+(1-\pi_{0}(z))(1-\pi
_{0}(z_{0}))\right]
\]
locally defines a unitary operator $w(z)$ such that
 $w(z)^{*}\pi_{0}(z) w(z)=\pi_{0}(z_{0})$. A
local trivialization of the fibration is then explicitly given by
\[
\begin{array}{cccccc}
\Theta : & \Pi_{E}^{-1}(\mathcal{U}_{z_{0}}) & \rightarrow  & 
\mathcal{U}_{z_{0}}\times \mathrm{Ran}\pi (z_{0}) & \rightarrow  & 
\mathcal{U}_{z_{0}}\times \mathcal{K}_{\mathrm{f}} \\
& (z,\psi ) & \mapsto  & (z,w(z)\psi ) & \mapsto  & (z,\phi (z_{0})w(z)\psi )
\end{array}
\]
where we use the fact that there exists a unitary operator 
$\phi (z_{0}):\mathrm{Ran}\pi (z_{0})\rightarrow \mathcal{K}_{\mathrm{f}}$.
The existence of $\phi (z_{0})$ follows from the fact that the
dimension of $\mathrm{Ran}\pi (z_{0})$ is independent of $z_{0}$, but the
map $z_{0}\mapsto \phi (z_{0})$ may be \textit{a priori} even discontinuous.
}

\emph{\noindent Moreover one can check that any two such trivializations are
$\mathcal{U(K}_{\mathrm{f}})$-compatible, and the  previous data define
a linear $\mathcal{U(K}_{\mathrm{f}})$-bundle. }

\emph{Since the base space is contractible, the bundle is trivial and 
the associated principal $\mathcal{U(K}_{\mathrm{f}})$-bundle (i.e. the
bundle of the orthonormal frames) admits a global \emph{smooth} section.
This implies the existence of a \emph{smooth} map $u_{0}:$ 
$\mathbb{R}^{2d}\rightarrow \mathcal{U(H}_{\mathrm{f}})$ such that (\ref{Intertwining})
holds true.\label{Rem bundle}}
\end{remark}

%%%%%%%%%%%%%%%%%   EFFECTIVE HAMILTONIAN  %%%%%%%%%%%%%%%%%%%%%%%%%%%%%
\section{Adiabatic perturbation theory}\label{SLOP}

\subsection{The effective Hamiltonian}

In the previous section we constructed a unitary $U$ on $\Hi $ which exactly
intertwines the almost invariant subspace Ran$\Pi $ and the reference subspace
$\mathcal{K}=$ Ran${\Pi_{\rm r}}$. $U$ and $\Pi$ are
$\mathcal{O}_0(\epsi^\infty)$-close to pseudodifferential operators with
symbols $u$ and $\pi$ both in $S^0_\rho(\epsi)$.

We define  the \emph{effective Hamiltonian} $\widehat h$
as the quantization of a resummation $h$ of the formal symbol
\begin{equation}\label{heffDef}
h = u\,\#\,H\,\#\,u^*\,.
\end{equation}
Recall that we do not distinguish semiclassical symbols and formal 
symbols in the notation.
The following theorem is the basis for the adiabatic perturbation theory, as it
relates the unitary time-evolution generated by the original Hamiltonian $\widehat H$ to
the one generated by the effective Hamiltonian $\widehat h$.

\begin{theorem}
\label{hThm}
Under the assumptions of Theorem \ref{Th unitaries}, one has that
$h\in  S^m_\rho(\epsi)$ and $\widehat h$ is essentially self-adjoint on $\mathcal S$.
Furthermore
\begin{equation}\label{hcommu}
  [\,\widehat{h},\,{\Pi_{\rm r}}\,]=0\,,
\end{equation}
\begin{equation}\label{GroupD1}
e^{-i\widehat H t} - \widehat u^{\,*}\,e^{-i\widehat h t}\,\widehat u =
\mathcal{O}_0(\epsi^\infty|t|)
\end{equation}
and
\begin{equation}\label{GroupD2}
e^{-i\widehat H t} - U^*\,e^{-i\widehat h t}\,U = \mathcal{O}_0(\epsi^\infty(1+|t|))\,.
\end{equation}
\end{theorem}

\noindent \textbf{Proof.} Since $u\in S_{\rho }^{0}(\varepsilon )$
and $H\in S_{\rho }^{m}(\varepsilon )$, the composition rule for
semiclassical operators (see Prop.\ \ref{Prop Moyal product})
yields $h\in S_{\rho }^{m}(\varepsilon )$ and thus $h_{j}\in
S_{\rho }^{m-j\rho }$.

Let $\widetilde h := \widehat  u \,\widehat H\,\widehat u^*$.
Since $\widehat u^{\,*}$ is bounded with bounded inverse, 
one finds, by checking definitions, that
 $\widetilde h$ is self-adjoint on $\widehat u^{\,*\,-1}D(\widehat H)$ and that
$\widetilde h$ is essentially self-adjoint on $\widehat u^{\,*\,-1}\mathcal{S}$.
According to Equation (8.10)  
in \cite{DiSj}, which generalizes to $\mathcal{B(H}_{\rm f})$-valued symbols, 
$\widehat u^{\,*\,-1}\in OPS^0(\epsi)$ and thus
$\widehat u^{\,*\,-1}\mathcal{S}=\mathcal{S}$. Hence $\mathcal{S}$ is a core for $\widetilde h$ and,
since $\widehat h-\widetilde h\in\mathcal{B(H)}$, the same conclusions hold for
$\widehat h$.

Next observe that, by construction, $[h_{j},\pi_{\mathrm{r}}]=0$
for all $j\in \mathbb{N}$ and thus $[h_{j},\pi
_{\mathrm{r}}]_{\widetilde{\#}}=0$ because $\pi_{\mathrm{r}}$
does not depend on $(q,p)\in \mathbb{R}^{2d}$. Hence
$[\widehat{h}_{j},{\Pi_{\rm r}}]=0$ and thus (\ref{hcommu})
follows.

For (\ref{GroupD1}) observe that
\[
e^{-i\widehat H t} - \widehat u^{\,*}\,e^{-i\widehat h t}\,\widehat u = -\,
i\,e^{-i\widehat H t}\int_0^t\,ds\,e^{i\widehat H s}\left( \widehat H \, \widehat u^{\,*}
-  \widehat u^{\,*}\,\widehat h\right)e^{-i\widehat h s}\,\widehat u
 = \mathcal{O}_0(\epsi^\infty|t|)\,,
\]
since, by construction, $( \widehat H \, \widehat u^{\,*}
-  \widehat u^{\,*}\,\widehat h) = \mathcal{O}_{-\infty}(\epsi^\infty)$.
Finally  (\ref{GroupD2}) follows from  (\ref{GroupD1}) using
$U-\widehat u=\mathcal{O}_0(\epsi^\infty)$.
\hfill $\qed$

\begin{remark}\em
It might seem more natural to define the effective Hamiltonian as 
\[
H_{\rm eff} = U\,\Pi\,\widehat H\,\Pi\,U^* +  U\,({\bf 1}-\Pi)\,\widehat H\,({\bf 1}-\Pi)\,U^*\,.
\]
Clearly one should have $H_{\rm eff} - \widehat h = \mathcal{O}(\epsi^\infty)$ in some sense.
However, if $\widehat H$ is unbounded, this closeness does not follow in the norm of
 bounded operators from our results, since $U$ need not be a semiclassical operator.
As a consequence no asymptotic expansion of $H_{\rm eff}$ in the norm of bounded operators would
be available.
\end{remark}

In the remainder of this section we will study the finite order
asymptotic approximations
\[
\widehat{h}^{(n)}:=\sum_{j=0}^{n}\,\varepsilon^{j}\,\widehat{h}_{j}
\]
to the effective Hamiltonian $\widehat{h}$.
By virtue of (\ref{hcommu}), we
can, whenever appropriate, restrict our attention to the reduced Hilbert
space $\mathcal{K}=\mathrm{Ran}{\Pi_{\rm r}}$. Furthermore we
define $\widehat{u}^{(n)}=\sum_{j=0}^{n}\,\varepsilon^{j}\,\widehat{u}_{j}$
and obtain a finite order expansion of the unitary $U$ as 
$\Vert \,U-\widehat{u}^{(n)}\,\Vert_{\mathcal{B(H)}}=\mathcal{O}(\varepsilon^{n+1})$.

Our main interest are approximations to the solution of the time-depen\-dent
Schr\"{o}\-din\-ger equation
\[%\begin{equation}
i\,\frac{\partial \psi_{t}}{\partial t}=\widehat{H}\,\psi_{t}  %\label{SEA}
\]%\end{equation}
over times of order $\varepsilon^{-k}\tau $, where $\tau $ does not depend
on $\varepsilon $ and $k\in \mathbb{N}$ is arbitrary.
Starting with (\ref{GroupD1}) on the almost invariant subspace we obtain
\begin{eqnarray}
e^{-i\widehat{H}t}\Pi  &=& \widehat u^{\,*}\,e^{-i\widehat h t}\,{\Pi_{\rm r}}\,\widehat u +
\mathcal{O}_0(\epsi^\infty|t|)
  \nonumber \\
&=&\widehat{u}^{\,(n)*}\,e^{-it\widehat{h}^{(n+k)}}\,{\Pi_{\rm r}}
\,\widehat{u}^{\,(n)}+(1+|\tau|)\mathcal{O}_0 (\varepsilon^{n+1})\,,\quad
|t|\leq \varepsilon^{-k}\tau \,, \label{UA}
\end{eqnarray}
where $\rho(n+k+1)\geq m$ is assumed in order to have
$\widehat h-\widehat h^{(n+k)}\in\mathcal{B(H)}$.
Hence, given the level of precision $\varepsilon^{n}$ and the time scale $\varepsilon^{-k}
$, the expansion of $\widehat{h}$ must be computed up to order $\widehat{h}
_{n+k}$ and the expansion of $U$ up to order $\widehat{u}_{n}$. Put
differently, in order to improve the error, a better approximation to the
unitary transformation is necessary. On the other hand, in order to enlarge
the time-scale of validity for the space-adiabatic approximation, only the
effective Hamiltonian $\widehat{h}$ must be computed to higher orders.

Specializing (\ref{UA}) to $n=0$ and $k=1$, one obtains the leading order
solution of the Schr\"{o}dinger equation as
\begin{equation}
e^{-i\widehat{H}t}\Pi =\widehat{u}_{0}^{*}\,e^{-i(\widehat{h}_{0}+\varepsilon
\widehat{h}_{1})t}\,{\Pi_{\rm r}}\,\widehat{u}_{0}+(1+|\tau |)\mathcal{O}_0(\varepsilon )\,,\quad |t|\leq \varepsilon
^{-1}\tau \,,  \label{U1}
\end{equation}
where $m\leq2\rho$.
Here the choice of $k=1$ corresponds to the macroscopic or semiclassical
time-scale $t/\varepsilon $. On this time-scale the effective dynamics 
$e^{-i\widehat{h}t/\varepsilon }{\Pi_{\rm r}}$ on the reference
subspace is expected to have a nice semiclassical limit, under suitable
conditions on $\widehat{h}$.

Note that one can replace in (\ref{UA}) and analogously in (\ref{U1}) $\tau $
by $\varepsilon^{-\delta }\tau $ and obtains
\begin{equation}
e^{-i\widehat{H}t}\Pi =\widehat{u}^{(n)*}\,e^{-i\widehat{h}^{(n+k)}}\,
{\Pi_{\rm r}}\,\widehat{u}^{(n)}+(1+|\tau |)\mathcal{O}_0
(\varepsilon^{n+1-\delta })\,,\quad |t|\leq \varepsilon^{-(k+\delta )}\tau
\,.  \label{UAdelta}
\end{equation}
Thus one can enlarge the time-span for which the approximation holds without
the need to compute further terms in the expansion. The price to be paid is
 a larger error, of course.

We emphasize that (\ref{UA}) and (\ref{U1}) are purely space-adiabatic
expansions with \emph{no semiclassical approximation} invoked yet.
As a consequence one obtains uniform results and a simple bound on the
growth of the error with time. Note in particular that the space-adiabatic approximation
holds on time-scales far beyond the Ehrenfest time-scale, the maximal time-scale
for which semiclassical approximations are expected to hold.
For some particular cases semiclassical
expansions of the full propagator $e^{-i\widehat{H}t/\varepsilon }$ have
been derived directly, e.g.\ in the context of  the Dirac equation \cite
{Yajima,BK}. These expansions hold, in general, only for short times, in the
sense that they must be modified each time a caustic in the corresponding
classical flow is encountered. More important, the clear separation of the
space-adiabatic and the semiclassical expansion is not maintained, which is
a severe drawback, since in many physical situations the space-adiabatic
approximation is valid to high accuracy, while the semiclassical
approximation is not, cf.\ Section~\ref{SDIRAC}. On the other hand, a
semiclassical expansion of the right hand side of (\ref{U1}) is
straightforward in many interesting cases, as will be discussed in Section~\ref{SSEMI}.

In parentheses we remark that the space-adiabatic approximation can be used
also in the time-independent setting, i.e.\ to estimate spectral properties
of $\widehat{H}$. If one is able to
compute eigenvalues of $\widehat{h}^{(n)}$ up to errors of order $o(\varepsilon^{n})$,
\[
\widehat{h}^{(n)}\,\psi^{(n)}=E^{(n)}\,\psi^{(n)}\,+\,o(\varepsilon
^{n})\,,
\]
it follows that
\[
\widehat{H}\,\widehat u^{\,*}\,\psi^{(n)}=E^{(n)}\,\widehat u^{\,*}\,\psi^{(n)}\,+\,
o(\varepsilon^{n})\,.
\]
If, in addition, one knows from some a priori arguments that $\widehat{H}$ has
pure point spectrum near $E^{(n)}$, it follows that $\widehat{H}$ has an
eigenvalue $o(\varepsilon^{n})$-close to $E^{(n)}$. Otherwise one can at least
conclude that there is a ``resonance'' in the sense of a quasi bound state 
$o(\varepsilon^{n})$-close to $E^{(n)}$. We stress that no explicit
knowledge of $U$ is needed as long as the interest is in approximate
eigenvalues only. For example, the scheme just described can be applied
 to the time-independent Born-Oppenheimer theory, where one is
interested in the low lying spectrum of a molecule. The standard approaches
to the time-independent Born-Oppenheimer approximation \cite{CDS,HagedornTIBO,KMSW} 
yield in some respects mathematically stronger
results. However, our scheme suffices for estimating asymptotic expansions
of eigenvalues and is simpler to handle, in general.

\subsection{Leading order terms in the expansion of the effective Hamiltonian}

\label{SFOE}

We turn to the explicit determination of the leading order terms $h_{j}$ in
the expansion of $\widehat{h}$ using (\ref{heffDef}). Of course, in
concrete applications only $H$ and $u_{0}$ are given
explicitly, while the higher order terms in the expansion of $u$ must be
calculated using the construction from Section~\ref{SUNI}. For a general
Hamiltonian $\widehat{H}$ such a program is feasible only for the terms
$h_{0}$, $h_{1}$ and possibly $h_{2}$, which will be our concern in the
following.

The principal symbol of $h$ is given by
\[%\begin{equation}
h_{0}=u_{0}\,H_{0}\,u_{0}^{*}\,.
\]%\end{equation}
Higher order terms can be obtained using (\ref{heffDef}). The double Moyal
product becomes rather awkward to handle, and alternatively we proceed
inductively by observing that
\begin{equation}
u\,\#\,H-h_0\,\#\,u=\epsi \,h_1\,\#\,u+ \Or(\epsi^2)=\epsi
\,h_1\,u_0+\Or(\epsi^2)\,,
\label{h11}
\end{equation}
with the subprincipal symbol on the left hand side being
\begin{equation}
\left( u\,\#\,H-h_0\,\#\,u\right)_1
=u_1H_0+u_0H_1-h_0u_1+(u_0\,\#\,H_0)_1-(h_0\,\# \,u_0)_1\,. \label{h12}
\end{equation}
Recall the notation $a\,\#\,b=\sum_{j=0}^{\infty }\,\varepsilon
^{j}\,(a\,\#\,b)_{j}$ for the expansion of the Moyal product, see the Appendix. Combining (\ref{h11}) and (\ref{h12}) one obtains
\begin{equation}
h_1=\big( u_1H_0+u_0H_1-h_0u_1+(u_0\,\#
\,H_0)_1-(h_0\,\#\,u_0)_1\,\big)\,u_0^*\,.  \label{h1}
\end{equation}
The expression (\ref{h1}) further simplifies if one specializes to the case
where ${\sigma_{\rm r}}(q,p)=\{{E_{\rm r}}(q,p)\}$ consists of a single eigenvalue of 
$H_{0}(q,p)$ and one projects on the relevant subspace,
\begin{equation}
\pi_{\rm r}h_1\pi_{\rm r}=\pi_{\rm r}\big(
u_0\,H_1\,u_0^*+(u_0\,\#\,H_0)_1\,u_0^*-(E_*\,\#
\,u_0)_1\,u_0^*\big) \pi_{\rm r}\,.  \label{h1d}
\end{equation}
The right hand side  has the nice property to be independent of $u_{1}$ and
thus to depend only on known quantities.

Along the same lines and under the same condition on ${\sigma_{\rm r}}(q,p)$, one
computes
\begin{eqnarray}
\pi_{\rm r}h_2\pi_{\rm r} &=&\pi_{\rm r}\Big(
u_0H_2+u_1H_1-h_1u_1  \label{h21} \\
&&+\,(u_1\,\#\,H_0)_1+(u_0\,\#\,H_1)_1-(E_*\,\#\,
u_1)_1-(h_1\,\#\,u_0)_1  \nonumber \\
&&+\,(u_0\,\#\,H_0)_2-(E_*\,\#\,u_0)_2\Big)u_0^* \pi_{\rm r}\,.
\nonumber
\end{eqnarray}
Again, (\ref{h21}) does not depend on $u_{2}$ for the special case under
consideration, but it does depend on $u_{1}$, which must now be computed
using the construction from Section~\ref{SUNI}.

Although (\ref{h21}) looks still rather innocent, in general, it requires
some work to compute it explicitly. This is partly because the second order
expansion of the Moyal product in (\ref{h21}) tends to become rather tedious
to obtain. But, in general, also the determination of $u_{1}$ is
nontrivial. To convince the reader, we state without details that the
construction from Sections~\ref{SPRO} and \ref{SUNI} yields
\begin{equation} \label{u1}
u_1^*=u_0^*\Big( -\frac{i}{4}\{u_0,u_0^*\}+\big[u_0\,\pi_1^{\rm
OD}\,u_0^*, \pi_{\rm r}\big] +\frac{i}{4}\big[ \big(
\{u_0,\pi_0\}u_0^*+u_0\{\pi_0,u_0^*\}\big),\pi_{\rm
r}\big]\Big)\,,
\end{equation}
with
\[
\pi_{1}^{\mathrm{OD}}:=\pi_{0}\pi_{1}(1-\pi_{0})+(1-\pi_{0})\pi_{1}\pi_{0}\,,
\]
where we used  that $(a\,\#\,b)_{1}=-\frac{i}{2}\{a,b\}$. Recall the definition 
(\ref{Poisson}) of the Poisson bracket $\{\cdot ,\cdot \}$.

To compute $\pi_{1}$ from the given quantities one has to use
the construction explained in Section~\ref{SPRO}. One finds
\begin{eqnarray*}
\pi_1^{\rm OD} & = & \frac{i}{2}\big(
R_0(E_*)(1-\pi_0)\{H_0+E_*,\pi_0\} \pi_0 \\&& +\,
\pi_0\,\{\pi_0,H_0+E_* \}\,R_0(E_*)\,(1-\pi_0)\big)\\&& + \,\pi_0
H_1 R_0(E_*)(1-\pi_0) + R_0(E_*)(1-\pi_0) H_1 \pi_0 \,,
\end{eqnarray*}
where $R_{0}({E_{\rm r}})(1-\pi_{0})=(H_{0}-{E_{\rm r}})^{-1}(1-\pi
_{0}) $ is uniformly bounded because of the gap condition. For sake of
completeness we mention that $\pi_{1}=\pi_{1}^{\mathrm{OD}}+\frac{i}{2}\{\pi_{0},\pi_{0}\}$ 
in this case.

For the higher orders in the expansion of $h$ we only remark
that, in general, $h_{n}$ depends on $u^{(n)}$, $H^{(n)}$ and $h^{(n-1)}$.
In the special, but interesting case of an isolated eigenvalue 
${E_{\rm r}}(q,p)$, $h_{n}$ depends only on $u^{(n-1)}$, $H^{(n)}$ and $h^{(n-1)}$
and is thus considerably easier to obtain.

\begin{remark}
\emph{Note that in the case of ${\sigma_{\rm r}}(q,p)=\{{E_{\rm r}}(q,p)\}$, not only the
principal symbol $h_{0}(q,p)={E_{\rm r}}(q,p){\bf 1}_{\Hi _{\mathrm{f}}}$, but also the
subprincipal symbol $h_{1}(q,p)$ as given by (\ref{h1d}) is well defined
regardless of the gap condition, provided that the spectral projection  $\pi_0(q,p)$ 
is sufficiently regular. Indeed, it can be
shown, at least in some special cases, that there is still adiabatic decoupling to
leading order and an effective dynamics generated by
$\widehat{h}_{0}+\varepsilon \,\widehat{h}_{1}$ without a gap condition 
\cite{general}, \cite{note}.}
\end{remark}

To get even more explicit formulas for $h_{1}$ and $h_{2}$, note that in
most applications one has no naturally given transformation $u_{0}$. Instead
one chooses a suitable basis $\{\psi_{\alpha }(q,p)\}_{\alpha \in I}$ of Ran$\pi_{0}(q,p)$ 
and defines $u_{0}(q,p)=\sum_{\alpha \in I}
|\chi_{\alpha }\rangle \langle \psi_{\alpha
}(q,p)|+r(q,p)$, where the vectors $\chi
_{\alpha }$ form a basis for Ran$\pi_{\mathrm{r}}$ and $r(q,p)$ is some
arbitrary unitary intertwining  Ran$\pi
_{0}(q,p)^{\perp }$ and Ran$\pi_{\mathrm{r}}^{\perp }$.  $\pi_{\mathrm{r}}\,h_{j}(q,p)\,\pi_{\mathrm{r}}$ is 
independent of the choice of the unitary $r(q,p)$ for all $j\in \mathbb{
N}$.

We remark that such a basis $\{\psi_{\alpha }(q,p)\}_{\alpha \in I}$ of
global smooth sections of the bundle over $\mathbb{R}^{2d}$ defined by $\pi
_{0}(q,p)$ always exists, since $\mathbb{R}^{2d}$ is contractible (see Remark
\ref{Rem bundle}). However, we are not aware of a proof which insures $u_{0}\in S_{\rho
}^{0}$ . The situation changes completely, once one considers local domains
in the base space which are not contractible. Then it might become necessary
to chose as reference space the space of sections of
a globally nontrivial bundle.

Assuming that ${\sigma_{\rm r}}(q,p)=\{{E_{\rm r}}(q,p)\}$ consists of a single
eigenvalue of $H_{0}(q,p)$ of multiplicity $\ell$ (including $\ell=\infty $), we
obtain  the $\ell\times \ell$-matrix $\pi_{\mathrm{r}}\, h^{(1)}(q,p)\,\pi_{\mathrm{r}}$ as
\begin{equation}
h_{\alpha \beta }^{(1)}=\langle \chi_{\alpha },h^{(1)}\chi_{\beta }\rangle
={E_{\rm r}}\,\delta_{\alpha \beta }+\varepsilon \,h_{1\,\alpha \beta }\,,
\label{hs}
\end{equation}
with
\begin{eqnarray}
h_{1\,\alpha \beta } &=& \langle\chi_\alpha, h_1\chi_\beta\rangle  =  
\langle \psi_{\alpha },H_{1}\psi_{\beta }\rangle -
\frac{i}{2}\langle \psi_{\alpha },\{(H_{0}+{E_{\rm r}}),\psi_{\beta }\}\rangle
\nonumber \\
&=&\langle \psi_{\alpha },H_{1}\psi_{\beta }\rangle -i\langle \psi
_{\alpha },\{{E_{\rm r}},\psi_{\beta }\}\rangle -\frac{i}{2}\langle \psi_{\alpha
},\{(H_{0}-{E_{\rm r}}),\psi_{\beta }\}\rangle \,.  \label{h1w}
\end{eqnarray}
The indices $\alpha $ and $\beta $ are matrix-indices, both running from $1$
to $\ell$.
Equations (\ref{hs}) and (\ref{h1w}) are one of our central results. They are
still of a simple form and mostly suffice to compute the basic physics.
The first term in (\ref{hs}) is referred to as Peierls substitution and
the first order correction carries
information on the intraband spinor evolution. E.g., as will be discussed
in Section~\ref{SDIRAC}, for the Dirac equation $h_1$ governs the spin
precession.
The reason for the particular splitting of the terms in (\ref
{h1w}) will be discussed in Section~\ref{SSEMI}. Here we only remark that
the second term in (\ref{h1w}) is related to a ``generalized'' Berry
connection. We omit the analogous formula for $h_{2\,\alpha \beta }$, since
it is too complicated to be helpful.

\subsection{Born-Oppenheimer type Hamiltonians} \label{SBO}

An instructive example to which formula (\ref{h1w})  applies are
Born-Oppenheimer type Hamiltonians of the form
\begin{equation}  \label{BOHamiltonian}
H_{\mathrm{BO}}(q,p) = \frac{1}{2}p^2\mathbf{1}_{\Hi _{\mathrm{f}}} +
V(q)\,,
\end{equation}
$V\in S^0(\mathcal{B}(\Hi _{\mathrm{f}}))$, with an electronic
energy band ${e_{\rm r}}(q)$ of constant multiplicity $\ell$, i.e.\ $V(q)\pi_0(q) =
{e_{\rm r}}(q)\pi_0(q)$.
  Adiabatic decoupling for Born-Oppenheimer type Hamiltonians
is established with exponentially small errors by Martinez and Sordoni \cite{MS},
see also \cite{Sordoni}. Their result partly triggered our interest
to develop a  general theory. Exponentially accurate coherent state solutions
for Born-Oppenheimer type Hamiltonians have been constructed by Hagedorn and Joye in
\cite{JH}.

Note that the quadratic growth of $H_{\mathrm{BO}}(q,p)$ as
a function of $p$ prevents  from applying the general results directly. As
to be discussed  in  Section~\ref{SCUT}, energy cutoffs need to be
introduced.
For the moment we ignore this problem and proceed by working
out the perturbative scheme formally.

We fix arbitrarily
an orthonormal basis $\{\psi_{\alpha }(q)\}_{\alpha =1}^{\ell}$ of Ran$\pi
_{0}(q)$  depending  smoothly on $q$ which then satisfies
$H_{\mathrm{BO}}(q,p)\psi_{\alpha }(q)={E_{\rm r}}(q,p)\psi_{\alpha }(q)$ with 
${E_{\rm r}}(q,p)=\frac{1}{2}p^{2}+{e_{\rm r}}(q)$ for $1\leq \alpha \leq \ell$. 
Only the second term
of our formula (\ref{h1w}) contributes and yields
\[
h_{1\,\alpha \beta }(q,p)=-i\,p\cdot \langle \psi_{\alpha }(q),\nabla
_{q}\psi_{\beta }(q)\rangle =:-\,p\cdot A_{\alpha \beta }(q)\,,
\]
which is well known in the case of a nondegenerate eigenvalue, \cite{BerryBO,LW,TS}.
 $A_{\alpha \beta }(q)$ has the geometrical meaning
of a gauge potential, i.e.\ coefficients of a connection on the trivial
bundle $\mathbb{R}^{d}\times \mathbb{C}^{\ell}$, the so called Berry connection. As
mentioned already, a more detailed discussion of the origin of the Berry
connection will be given in Section~\ref{SSEMI}.

For the Born-Oppenheimer Hamiltonian  the calculation of
$h_{2\,\alpha \beta }$ is still feasible without much effort and the result is
\begin{eqnarray}
h_{2\,\alpha \beta }&=&\frac{1}{2}\sum_{\mu =1}^{\ell}A_{\alpha \mu }\cdot A_{\mu\beta }+
\frac{1}{2}\langle \nabla_q\psi_{\alpha },(1-\pi_0)\cdot
\nabla_{q}\psi_{\beta }\rangle \nonumber\\ &&-\,\langle \,p\cdot \nabla
_{q}\psi_{\alpha },\,R_{0}({E_{\rm r}})\,\,p\cdot \nabla_{q}\psi_{\beta
}\rangle \,.  \label{h2BO}
\end{eqnarray}
Recall the definition of $R_{0}({E_{\rm r}})=(H_{0}-{E_{\rm r}})^{-1}(1-\pi_{0})$,
which reduces to $R_{0}({E_{\rm r}})(q)=(V(q)-{e_{\rm r}}(q))^{-1}(1-\pi_{0}(q))$ in
the present case. Although we omit the details of the computation leading to (\ref{h2BO}),
we shortly describe how (\ref{h21}) relates to (\ref{h2BO}). Since $H_{1}=0$
and $H_{2}=0$ the corresponding terms in (\ref{h21}) do not contribute.
Since $u_{0}$ and $\pi_{0}$ are functions of $q$ only, the second term in
(\ref{u1}) is the only one contributing to $u_{1}$, and thus the third term
in (\ref{h21}) also vanishes after projecting with the $\pi_{\mathrm{r}}$'s
from outside the brackets. The last two terms in (\ref{h21}) cancel each
other. The seventh term in (\ref{h21}) yields the first term in (\ref{h2BO})
and the fourth and sixth term in (\ref{h21}) combine to the second and third
term in (\ref{h2BO}). In particular the calculation yields for the symbol of
the unitary
\[
u_{\mathrm{BO}}^*(q,p)\pi_{\mathrm{r}}=\sum_{\alpha =1}^{\ell}\Big(|\psi
_{\alpha }(q)\rangle +i\varepsilon \,R_{0}({E_{\rm r}})(q)\,\,|\,p\cdot \nabla
_{q}\psi_{\alpha }(q)\rangle \Big)
\langle \chi_{\alpha }|+\mathcal{O}(\varepsilon^{2})\,.
\]
Thus the symbol of the second order effective Born-Oppenheimer Hamiltonian reads
\begin{eqnarray}  \label{h3BO}
h_{\mathrm{BO}\,\alpha\beta}(q,p) & = & \frac{1}{2}\Big( p -
\varepsilon\,A(q) \Big)^2_{\alpha\beta} +\, {e_{\rm r}}(q)\delta_{\alpha\beta}\nonumber \\&& +
\frac{\varepsilon^2}{2}\langle\nabla_q \psi_\alpha(q),(1-\pi_0(q))\cdot
\nabla_q\psi_\beta(q)\rangle  \\
&& -\,\varepsilon^2\,\langle\,
p\cdot\nabla_q\psi_\alpha(q),\,R_0({E_{\rm r}})(q)\,\,p\cdot
\nabla_q\psi_\beta(q)\rangle + \mathcal{O}(\varepsilon^3)\,,\nonumber
\end{eqnarray}
where the first term from (\ref{h2BO}) nicely completes the square to the
first term in (\ref{h3BO}). Note that the third term on the right side of 
(\ref{h3BO}) depends on $q$ only and was interpreted in \cite{BerryBO} as a
geometric electric potential in analogy to the geometric vector potential $A(q)$.

In the special case of a nondegenerate eigenvalue ${e_{\rm r}}$ and a matrix-valued
Hamiltonian $H$, (\ref{h3BO}) reduces to the expression obtained by
Littlejohn and Weigert \cite{LW}. They also remark that the previous
studies \cite{BerryBO,AS} of the expansion of the effective
Born-Oppenheimer Hamiltonian missed the last term in (\ref{h3BO}). This
strengthens our point of the usefulness of a general and systematic
space-adiabatic perturbation theory.

The full power of our scheme is in force in cases where Ran$\pi_0$ is
degenerate and depends both on $q$ and $p$, since then the known techniques
\cite{LF,LW,NS,MS} cannot be applied. The simplest example of this kind is the
one-particle Dirac equation with slowly varying electric and magnetic
potentials, which will be discussed in Section~\ref{SDIRAC}.

\subsection{The time-adiabatic theory revisited} \label{STA}

With little additional effort our scheme can be applied even
 to the time-adiabatic setup. As for notation, we replace the
phase space $\mathbb{R}^d_q\times \mathbb{R}^d_p$ by
$\mathbb{R}_t\times \mathbb{R}_\eta$ in the following.
 Given a Hilbert space $\Hi $ and family $H^\epsi(t)$, $t\in\mathbb{R}$ of
self-adjoint operators such that $H^\epsi(t) =: H(t,\eta,\epsi)
\in S^0(\epsi, \mathcal{B(H)})$,
the solutions of the equations
\begin{equation}
i\varepsilon \partial_{t}U^\epsi(t,s)=H^\epsi(t)U^\epsi(t,s)\,,\qquad s\in \mathbb{R}\,,
\label{Unit propag}
\end{equation}
define a \emph{unitary propagator}. A unitary propagator is
a unitary operator-valued map
$U(t,s)$ strongly continuous in $t$ and $s$ jointly, such that
\[
U(t,t)={\bf 1}_{\Hi }\qquad \mbox{and\qquad }U(t,r)U(r,s)=U(t,s)
\]
for any $r,s,t\in \mathbb{R}$. In particular we have that $U^\epsi(t,0)\psi_0$
solves the time-dependent Schr\"odinger equation 
\begin{equation}\label{b}
i\epsi\frac{\partial}{\partial t}\psi(t)=H^\epsi(t)\psi(t)\,.
\end{equation} 
for any $\psi_0\in \Hi $.

It is assumed in addition that $H_0(t)$, the principal symbol of $H^\epsi(t)$,
has a relevant part ${\sigma_{\rm r}}(t)$ of its spectrum,
which is separated by a gap from the remainder uniformly for $t\in \mathbb{R}$. As before
we denote the spectral projection on  ${\sigma_{\rm r}}(t)$ by $\pi_0(t)$.

The following theorem is a variant  of the
{\em time-adiabatic theorem of quantum mechanics} \cite{Kato,ASY,JP,NenciuTA},
however formulated in the language of adiabatic perturbation theory.
 Sj\"ostrand first recognized the usefulness of pseudodifferential calculus
in this context \cite{Sjoestrand} and we are grateful to G.\ Nenciu for
pointing this out to us.
We remark that the proof below can be adapted to the case of
a time-dependent operator-valued classical symbol
$H(q,p,t)$, as -- for example -- the Dirac Hamiltonian or the Pauli-Fierz
Hamiltonian with \emph{slowly} varying time-dependent external potentials.

\begin{theorem}[Time-adiabatic theorem] 
Let $H(t)$ and ${\sigma_{\rm r}}(t)$ be as above.
\begin{enumerate}
\item {\rm Decoupled subspace.}
There exists a family of orthogonal projectors
$\Pi(t)$ such that $\Pi(\cdot)\in S^0(\epsi , \mathcal{B(H)})$,
$\Pi(t)-\pi_0(t) = \mathcal{O}_0(\epsi)$ and
\begin{equation}\label{TAsubspace}
U(t,s)^*\,\Pi(t)\,U(t,s) = \Pi(s) + \mathcal{O}_0(\epsi^\infty|t-s|)\
\end{equation}
uniformly for $s,t\in\mathbb{R}$. Whenever $\partial_t^\alpha H(t)=0$ for
some $t\in\mathbb{R}$ and all $\alpha\in\mathbb{N}$,
then $\Pi(t) = \pi_0(t)$.
\item {\rm Intertwining unitaries.} \hspace{-1.7pt} There exists a family of unitaries 
$u_0(\cdot)\in C^\infty_{\rm b}(\mathbb{R}, 
 \mathcal{B(H)})$ with $u_0(t)\,\pi_0(t)\,u_0^*(t) = \pi_0(0) =: \pi_{\rm r}$
and a family of unitaries $\mathcal{U}(\cdot) \in S^0(\epsi,
 \mathcal{B(H)})$ such that
\[
\mathcal{U}(t) \,\Pi(t)\, \mathcal{U}^*(t) = \pi_{\rm r}
\quad \mbox{and}\quad  \mathcal{U}(t) - u_0(t) = \mathcal{O}_0(\epsi)
\,.
\]
\item {\rm Effective dynamics.} There exists a family of self-adjoint operators $h(t)$,
$h(\cdot)\in S^0(\epsi, \mathcal{B(H)})$, such that
\begin{equation}
[\, h(t),\, \pi_{\rm r}\,] = 0\qquad \mbox{for all}\qquad t\in\mathbb{R}
\end{equation}
and the solution of the initial value problem
\[%\begin{equation}
i\varepsilon \partial_{t}U_{\rm eff}(t,s)=h(t)U_{\rm eff}
(t,s)\,,\qquad s\in \mathbb{R}\,,\qquad U_{\rm eff}(t,t) = {\bf 1}_{\Hi }
%\label{Unit eff propag}
\]%\end{equation}
satisfies
\begin{equation}\label{Kcompare}
 U(t,s) =
\mathcal{U}^*(t) \,U_{\rm eff}(t,s)\, \mathcal{U}(s)  +
\mathcal{O}_0(\epsi^\infty|t-s|)\,.
\end{equation}
The asymptotic expansion of $h(t)$ in $\mathcal{B(H)}$ reads
\begin{eqnarray} \label{kexpansion}
h(t) &\asymp& \sum_{n=0}^\infty\epsi^n \left(\sum_{j+k+l=n} u_j(t)\,H_k(t)\,u_l^*(t)\right.\\&&\left.
\qquad+\, \frac{i}{2}\sum_{j+k+1=n}  \left( u_j(t)\,\dot u_k^*(t) - \dot u_j(t)\,u_k^*(t)
\right)\right)\,,\nonumber
\end{eqnarray}
where $\sum_n\epsi^n H_n(t)$ is the asymptotic expansion of $H(t)$ in
 $\mathcal{B(H)}$ and $\sum_n \epsi^n u_n$ is the asymptotic expansion of
$\mathcal{U}(t)$ in  $\mathcal{B(H)}$.
\end{enumerate}
\end{theorem}

Before we turn to the proof we remark that, for ${\sigma_{\rm r}}(t) = \{{e_{\rm r}}(t)\}$
and $\{\varphi_\alpha(t)\}_{\alpha=1}^\ell$ an orthonormal basis of Ran$\pi_0(t)$,
 the effective Hamiltonian including second order
reads
\[
h_{\alpha \beta }(t)={e_{\rm r}}(t) \delta_{\alpha
\beta }\,-\,i\,\epsi\,\langle \varphi_\alpha(t),\dot{\varphi}_{\beta }(t)\rangle\,+\, 
\frac{\varepsilon^{2}}{2}\langle \dot{\varphi}_{\alpha
}(t),R_{0}({e_{\rm r}})\,\dot{\varphi}_{\beta }(t)\rangle \,+\,\mathcal{O}
(\varepsilon^{3})\,,
\]
where $R_{0}({e_{\rm r}})=(H(t)-{e_{\rm r}}(t))^{-1}\,(1-\pi_{0}(t))$. 
For the unitary $\mathcal{U}(t)$
one finds
\[
\mathcal{U}^*(t)\pi_{\mathrm{r}}=\sum_{\alpha =1}^{\ell}\Big(|\varphi_{\alpha }(t)\rangle
+i\varepsilon \,R_{0}({e_{\rm r}})(t)\,\,|\dot{\varphi}_{\alpha }(t)\rangle \Big)
\langle \varphi_{\alpha }(0)|+\mathcal{O}(\varepsilon^{2})\,.
\]

\noindent {\bf Proof.}\quad
In order to apply the general scheme developed in the previous
sections  it is convenient -- in analogy with
the extended configuration space in classical mechanics -- to introduce the
extended space $\mathcal{K}=L^{2}(\mathbb{R},\Hi )=\int_{\mathbb{R}
}^{\oplus }\mathcal{H\ }dt$  and to define the extended Hamiltonian
\[
\widehat{K}=-i\varepsilon \partial_{t}+H(t)\,
\]
which is self-adjoint on the domain
$\mathcal{D}(\widehat{K})=H^{1}(\mathbb{R},\Hi 
)\subseteq \mathcal{K}$.
By following Howland \cite{Howland}, we notice that
the unitary group $e^{-i\widehat{K}
\sigma }$, $\sigma \in \mathbb{R}$, is related to the unitary propagator 
(\ref{Unit propag})  through
 \begin{equation}\label{HowlandU}
\left( e^{-i\widehat{K}\sigma }\psi \right) (t)=U(t,t-\sigma )\psi (t-\sigma )\,.
\end{equation}
Moreover, the unitary group $e^{-i\widehat{K}\sigma }$ can now be studied by
means of the techniques developed in the previous sections, since $\widehat{K}$
is nothing but the Weyl quantization of the operator-valued function
$K(t,\eta )=\eta +H(t)$, and $K$ belongs to $S_{1}^{1}(\mathcal{B}(\Hi ))$.

By assumption
 $K\in S_{1}^{1}$ satisfies assumption (Gap)$_\sigma $ with
$\sigma =0$.  However, because of the simple dependence of $K(t,\eta )$ on
$\eta $, the conclusion of Theorem  \ref{Th Invariant subspace} and \ref{Th
unitaries} hold still  true in a sense to be made precise.

\noindent Indeed, by following the proof of Lemma \ref{Prop Moyal proj} \
one obtains a semiclassical symbol $\pi \in S_{0}^{0}(\varepsilon ,
\mathcal{B}(\Hi ))$, depending on $t$ only, such that
$\lbrack K,\pi ]_{\tilde{\#}}\asymp 0$ in $S_{0}^{1}(\varepsilon )$. On the other hand,
\[%\begin{equation}
\lbrack K,\pi ]_{\tilde{\#}}=[H,\pi ]_{\tilde{\#}}+[\eta ,\pi ]_{\tilde{\#}
}=[H,\pi ]_{\tilde{\#}}-i\varepsilon \left( \partial_{t}\pi \right)
%\label{Commutator}
\]%\end{equation}
where the last equality follows from the fact that $[\eta ,\pi ]_{\tilde{\#}}$ 
is the   symbol of $[-i\varepsilon \partial_{t},\pi
(t)]$ $ =-i\varepsilon \left( \partial_{t}\pi \right) (t)$.  Since both 
$[H,\pi ]_{\tilde{\#}}$ and $\partial_{t}\pi $ belong to $S_{0}^{0}(\varepsilon)$,
one concludes that the asymptotic expansion $[K,\pi
]_{\tilde{\#}}\asymp 0\ $ holds true in $S_{0}^{0}(\varepsilon )$,
and hence $[\widehat{K},\widehat{\pi }]=\mathcal{O}_{0 }(\varepsilon
^{\infty })$. Finally one defines
\[
\Pi(t) = \frac{i}{2\pi}\int_{|\zeta-1|=\frac{1}{2}} (\pi(t)-\zeta)^{-1}d\zeta
\]
and finds $\Pi(\cdot)\in S^0(\epsi,\mathcal{B(H)})$, 
$\Pi(t) - \pi(t) = \mathcal{O}_0(\epsi^\infty)$
and $[e^{-i\widehat K\sigma},\Pi]=\mathcal{O}_0(\epsi^\infty|\sigma|)$ 
as in Section~\ref{SPRO}.
Together with (\ref{HowlandU}) this implies
\[
{\rm ess}\sup_{\hspace{-0.4cm}t\in\mathbb{R}} \left\|
U(t,t-\sigma)^*\,\Pi(t)\,U(t,t-\sigma) - \Pi(t-\sigma)
\right\|_{\mathcal{B(H)}}
 = \mathcal{O}_0(\epsi^\infty|\sigma|)\,.
\]
However, since $\Pi(t)$ and $U(t,s)$ are continuous functions of $t$, 
the pointwise statement (\ref{TAsubspace}) follows.

For $u_0(t)$ one can use for example  Kato's construction \cite{Kato}
and define $u_0(t)$ as the solution of the initial value problem 
\[
\frac{d}{d t} u_0^*(t)  =  [\dot \pi_0(t),\pi_0(t)] \,u_0^*(t)\,,
\quad u_0^*(0) = {\bf 1}\,.
\]
Clearly $u_0(t)$ belongs to $S^0(\mathcal{B(H)})$. Notice that the same construction does
not work in the multidimensional case, since the evolutions in different directions do not 
commute.
 $\mathcal{U}$ can be obtained as in Section~\ref{SUNI},
where the fact that $\pi(t)$ and $u_0(t)$ both depend on $t$ only and not on $\eta$
simplifies the construction considerably and yields, in particular, a fibered
unitary $\mathcal{U}(t)$.

As in the general setting let the effective Hamiltonian be defined as
a resummation of
\[
k(\eta,t,\epsi) = \big(u\,\#\,K\,\#\,u^*\big)(\eta,t,\epsi) =: \eta + h(t,\epsi)\,,
\]
with the explicit expansion (\ref{kexpansion}).
According to Theorem \ref{hThm} we then have
\[%\begin{equation}
e^{-i\widehat K \sigma} - \mathcal{U}^*\,e^{-i\widehat k\sigma}\,\mathcal{U} = 
\mathcal{O}(\epsi^\infty
|\sigma|)\,,
\]%\end{equation}
which implies according to (\ref{HowlandU}) that
\[
{\rm ess}\sup_{\hspace{-0.4cm}t\in\mathbb{R}} \left\|
U(t,t-\sigma) -  \mathcal{U}^*(t)\,U_{\rm eff}(t,t-\sigma)\,\mathcal{U}(t-\sigma)
\right\|_{\mathcal{B(H)}}
 = \mathcal{O}_0(\epsi^\infty|\sigma|)\,.
\]
The pointwise statement (\ref{Kcompare}) follows again from the continuous
dependence on $t$ of all involved expressions.
\hfill$\qed$

\subsection{Energy cutoff} \label{SCUT}

The Born-Oppenheimer type Hamiltonians as well as many
other physically relevant Hamiltonians do not satisfy the general
assumptions we imposed in Sections~\ref{SPRO} and \ref{SUNI}. This is so for
two reasons. First of all they are quantizations of symbols taking values in
the unbounded operators. Secondly, the gap does not increase as fast as the
Hamiltonian for large momenta, e.g.\ quadratically in the Born-Oppenheimer
setting. The first problem is purely technical and the domain questions which arise
have to be dealt with case by case.
The second problem causes a qualitative change in the sense that the
adiabatic decoupling is no longer uniform, as can be seen from the
construction of the almost invariant subspace in Section~\ref{SPRO}.
To deal with the second problem one therefore needs a cutoff for large
momenta. There are basically two ways to implement such a cutoff. One
possibility is to directly cut off large momenta as was done in \cite{TS,ST},
but then one needs to control the times for which no momenta exceeding the
cutoff are produced under the dynamics. However, for a large class of Hamiltonians
including the Born-Oppenheimer type Hamiltonian (\ref{BOHamiltonian}),
cutting off high energies is  equivalent to cutting high momenta.
Then conservation of energy immediately ensures that no momenta exceeding
the cutoff are produced over time. This idea was developped in \cite{Sordoni}
and also  used in \cite{MS}.
We will briefly indicate an alternative way on how to implement
such an energy cutoff in order to fit the Born-Oppenheimer and similar settings
into our general assumptions.

Let $H_{0}\in S_{0}^{m}$ be elliptic and positive, i.e.\ there is a constant 
$C>0$ such that $H_{0}(q,p)\geq C\,\langle p\rangle^{m}$. For example the Born-Oppenheimer
Hamiltonian as defined in (\ref{BOHamiltonian}) satisfies $H_0\in S^2_0$ and it is elliptic
provided that $V$ is positive (otherwise just add a constant to $H_0$ since
$V\in S^0$).
Then we can prove
adiabatic decoupling uniformly for energies below any $\lambda \in \mathbb{R}$,
i.e.\ on Ran$\1I_{(-\infty ,\lambda ]}(\widehat{H}_{0})$.

Let $\Lambda = \{(q,p): H_0(q,p) <\lambda \}$, then
bounding the total energy by $\lambda$ essentially corresponds to confining
the slow degrees of freedom to the region $\Lambda$ in phase space. More
precisely, let $\chi_\lambda\in C_0^\infty(\mathbb{R})$ such that 
$\chi_\lambda|_{[0,\lambda]} =1$ and $\chi_\lambda|_{[\lambda+\delta,\infty)}
=0$ for some $\delta>0$, then $\chi_\lambda (\widehat H_0)\in \mathrm{OP}
S^{-\infty}_0$ is a semiclassical operator. Furthermore, its symbol $\chi :=
{\rm Symb}(\chi_\lambda (\widehat H_0))$ has an asymptotic expansion which is
identically equal to $1$ on $\Lambda$, i.e.\ $\chi|_\Lambda\asymp 1$ and
identically equal to $0$ on the set where $H_0(q,p) \geq \lambda +\delta$.
The statements about $\chi_\lambda (\widehat H_0)$ and its symbol follow
from the functional calculus for semiclassical operators as developped e.g.\
in \cite{DiSj}, Theorem 8.7.

Next we assume that one can define an auxiliary Hamiltonian 
$H_{\mathrm{aux}}(q,p) \in S^0_0$ such that
\begin{enumerate}
\item
$H_{\mathrm{aux}}(q,p) =
H_0(q,p)$ for all $(q,p) \in \Lambda + \delta := \{(q,p): H_0(q,p)<\lambda+\delta\}$,
\item
 $H_{\mathrm{aux}}(q,p) >
  H_0(q',p')$ for all  $(q,p) \not\in \Lambda + \delta$ and  
$(q',p') \in \Lambda + \delta$
\item
and $H_{\mathrm{aux}}(q,p)$
satisfies the global gap condition (Gap)$_0$.
\end{enumerate}
This can be easily achieved e.g.\ in the
Born-Oppenheimer setting by replacing $p^2$ by an appropriate bounded function.

It follows from the previous discussion that $( \, \widehat H_0 - \widehat
H_{\mathrm{aux}}\,) \,\chi_\lambda (\widehat H_0) = \mathcal{O}_{-\infty}
(\varepsilon^\infty)$ and
that $\chi_\lambda (\widehat H_0) -\chi_\lambda (\widehat H_{\mathrm{aux}})=
\mathcal{O}_{-\infty}(\varepsilon^\infty)$. Using 
\[
\chi_\lambda (\widehat H_0)\1I_{(-\infty,\lambda]} (\widehat H_0) = 
\1I_{(-\infty,\lambda]} (\widehat H_0)\,,
\] one finds, in particular, that
\begin{equation}  \label{H0Haux}
\left( \, \widehat H_0 - \widehat H_{\mathrm{aux}}\,\right) \,
\1I_{(-\infty,\lambda]} (\widehat H_0) = \mathcal{O}_0(\varepsilon^\infty)
\end{equation}
in the norm of bounded operators and thus also
\begin{eqnarray}  \label{H0Haux2} \lefteqn{
\left( e^{-i\widehat H_{\rm aux}t} -  e^{-i\widehat H_{0}t}\right)
\1I_{(-\infty,\lambda]} (\widehat H_0) }\\ &
&  \hspace{-10pt}=\,-\,  i  e^{-i\widehat H_{\rm aux}t}   \int_0^t \,ds\,  e^{i\widehat H_{\rm aux}s}
\left(   \widehat H_{\rm aux} - \widehat H_{0} \right) e^{-i\widehat H_{0}s} \,
\1I_{(-\infty,\lambda]} (\widehat H_0) = \mathcal{O}_0(\varepsilon^\infty|t|)\,.\nonumber
\end{eqnarray}

Now the scheme of Sections~\ref{SPRO}, \ref{SUNI} and \ref{SLOP} can be
applied to $H_{\mathrm{aux}}$ and by virtue of (\ref{H0Haux}) and  
(\ref{H0Haux2}) all results
are valid for $H_0$ up to $\mathcal{O}(\epsi^\infty)$ if one restricts to 
energies below $\lambda$. In
particular one finds that for $(q,p) \in \Lambda$ the leading order symbols
of $h_{\mathrm{aux}} = U^* \widehat H_{\mathrm{aux}} U$ are given by the
formulas obtained in Section~\ref{SFOE} using the symbol $H_0(q,p)$.

%%%%%%%%%%%%%   SEMICLASSICAL ANALYSIS  %%%%%%%%%%%%%%%%%%%%%%%%%%%%%%

\section{Semiclassical analysis for effective Hamiltonians \label{SSEMI}}

The results of the previous sections are genuine
quantum mechanical: semiclassical symbols have been used \emph{only
as a tool }in order to construct (and, eventually, to approximate)\ $\Pi $
and $U$, but no semiclassical limit has been performed. Indeed, the
adiabatic decoupling of energy bands is a purely quantum
phenomenon, which is, in general, independent from the semiclassical limit.

However, under the assumption that ${\sigma_{\rm r}}(q,p)=\{{E_{\rm r}}(q,p)\}$
consists of a single eigenvalue of $H_{0}(q,p)$ of necessarily constant  multiplicity $\ell$, the
principal symbol of $\widehat{h}$ is a scalar multiple of the identity,
i.e.\ $h_{0}(q,p)\pi_{\mathrm{r}}={E_{\rm r}}(q,p){\bf 1}_{\mathcal{K}_{\mathrm{f}}}$,
and a semiclassical analysis of $\widehat h$ can be done in a standard way. In
particular, the dynamics of quantum observables can be approximated by
quantities constructed using only the classical flow $\Phi^{t}$
generated by the (classical, scalar) Hamiltonian ${E_{\rm r}}(q,p)$. This results
 in a generalized Egorov's theorem, see Theorem  \ref{EgorovT}. We
emphasize that for more general energy bands ${\sigma_{\rm r}}(q,p)$
one cannot expect a simple semiclassical limit, at least not in the usual sense.

\subsection{Semiclassical analysis for matrix-valued symbols}

\noindent{\bf Egorov's Theorem.}
For the moment, we identify $\mathcal{K}_{\mathrm{f}}$ with $\mathbb{C}^{\ell}$
and $h$ with $\pi_{\mathrm{r}}\,h\,\pi_{\mathrm{r}}$, an $\ell\times \ell$-matrix-valued 
formal symbol. At least formally, Egorov's theorem is
obtained through an expansion of the Heisenberg equations of motion for
semiclassical observables: Let $a(q,p,\varepsilon )\in S_{1}^{0}(\varepsilon
,\mathcal{B}(\mathbb{C}^{\ell}))$, then the quantum mechanical time evolution of 
$\widehat{a}$ is given by
\[
\widehat{a}(t)=e^{i\widehat{h}t/\varepsilon }\,\widehat{a}\,e^{-i\widehat{h}t/\varepsilon }
\]
and satisfies
\begin{equation}
\frac{d\,\widehat{a}(t)}{dt}=\frac{i}{\varepsilon }[\,\widehat{h},\,\widehat{a}(t)]\,.  
\label{Heis1}
\end{equation}
Expanding both sides of (\ref{Heis1}) on the level of symbols and using
$[{E_{\rm r}}\mathbf{1},a_{n}(t)]\equiv 0$, $\mathbf{1}=\mathbf{1}_{\mathbb{C}^{\ell}}$,
one obtains the following hierarchy of equations:
\begin{eqnarray}
\frac{d\,a_{0}(t)}{dt} &=&\{{E_{\rm r}}\mathbf{1},a_{0}(t)\}+i[h_{1},a_{0}(t)]
\label{Eg1} \\
\frac{d\,a_{1}(t)}{dt} &=&\{{E_{\rm r}}\mathbf{1},a_{1}(t)\}+i[h_{1},a_{1}(t)]-
\frac{1}{2}\big( \{h_{1},a_{0}(t)\}-\{a_{0}(t),h_{1}\}\big)  \nonumber \\
&&+\,i[h_{2},a_{0}(t)]\label{Eg15} \\
\frac{d\,a_{2}(t)}{dt} &=&\{{E_{\rm r}}\mathbf{1},a_{2}(t)\}+i[h_{1},a_{2}(t)]+
\ldots \,.  \label{Egg2}
\end{eqnarray}
Since $da_{n}(t)/dt$ does not depend on higher orders, the equations can be
solved iteratively. The solution of (\ref{Eg1}) with initial condition 
$a_{0}(q,p,0)=a_{0}(q,p)$ is given through
\begin{equation}
a_{0}(q,p,t)=D^{*}(q,p,t)\,a_{0}(\Phi^{t}(q,p))\,D(q,p,t)\,,  \label{A0}
\end{equation}
where $\Phi^{t}:\mathbb{R}^{2d}\to \mathbb{R}^{2d}$ is the solution flow
corresponding to the scalar Hamiltonian ${E_{\rm r}}(q,p)$. More precisely, $\Phi
^{t}(q_{0},p_{0})=(q(t),p(t))$, where $(q(t),p(t))$ is the solution of the
classical equations of motion
\[
\dot{q}= \nabla_p {E_{\rm r}}\,,\qquad \dot{p}=-\nabla_q {E_{\rm r}}
\]
with initial condition $(q_{0},p_{0})$. $D(q,p,t)$ is the solution of
\begin{equation}
\frac{\partial }{\partial t}\,D(q,p,t)=-\,i\,h_{1}(\Phi
^{t}(q,p))\,D(q,p,t)\,.  \label{Deq}
\end{equation}
with initial condition $D(q,p,0)=\mathbf{1}$. One can think of (\ref{Deq})
for fixed $(q,p)\in \mathbb{R}^{2d}$ as an equation for the Schr\"{o}dinger-like 
unitary evolution induced by the time-dependent Hamiltonian 
$h_{1}(\Phi^{t}(q,p))$ on the Hilbert space $\mathbb{C}^{\ell}$. Since $h_{1}(q,p)$
is self-adjoint for all $(q,p)\in \mathbb{R}^{2d}$, the solution $D(q,p,t)$ of 
(\ref{Deq}) is unitary for all $(q,p,t)\in \mathbb{R}^{2d}\times \mathbb{R}$.

To see that (\ref{A0}) is indeed the solution of (\ref{Eg1}), note that the
mappings
\[
\mathcal{U}(t): C_{\mathrm{b}}(\mathbb{R}^{2d},\mathcal{B}(\mathbb{C}^\ell))\to 
C_{\mathrm{b}}(\mathbb{R}^{2d}, \mathcal{B}(\mathbb{C}^\ell))
\]
defined through (\ref{A0}) for $t\in\mathbb{R}$, i.e.\
\begin{equation}  \label{UDDEF}
\big( \mathcal{U}(t)\, a_0\big)(q,p) = D^*(q,p,t)\,a_0(\Phi^t(q,p))\,D(q,p,t)\,,
\end{equation}
form a one-parameter group of linear automorphisms on the Banach space 
$C_{\mathrm{b}}(\mathbb{R}^{2d},$ $\mathcal{B}(\mathbb{C}^\ell))$, since
\begin{eqnarray*}\lefteqn{\hspace{-5mm}
\big( \mathcal{U}(s)\,\mathcal{U}(t)\,a_0 \big)(q,p) =}\\&=& D^*(q,p,s)\,
D^*(\Phi^s(q,p),t)\,a_0(\Phi^s\circ\Phi^t(q,p))\, D(\Phi^s(q,p),t)\,D(q,p,s)
\\
&=& D^*(q,p,t+s)\, a_0(\Phi^{t+s}(q,p))\,D(q,p,t+s) \\
& =&\big( \mathcal{U}(t+s)a_0\big)(q,p)\,.
\end{eqnarray*}
Here the group structures of $\Phi^t$ and of the solutions of (\ref{Deq}) are used.
Hence $\mathcal{U}(t)$ is a group
and it suffices to check that (\ref{A0}) solves (\ref{Eg1}) at time $t=0$,
which is easy to see.

The physical interpretation becomes simpler when translated to   states: a
``classical'' particle which started at time $0$ at the phase space point $(q,p)$ 
with spinor $\varphi_{0}\in \mathbb{C}^{\ell}$, is at time $t$ located at
the phase space point $\Phi^{t}(q,p)$ with spinor $\varphi
_{t}=D(q,p,t)\varphi_{0}$. Hence (\ref{Deq}) implies that
\begin{equation}
\frac{d\varphi_{t}}{dt}=-\,i\,h_{1}(\Phi^{t}(q,p))\,\varphi_{t}\,.
\label{spintransport}
\end{equation}
One can also think of $\mathcal{U}(t)$ as being the action on observables of a
``classical'' flow $\Phi_{\ell}^{t}$ on phase space 
$\mathbb{R}^{2d}\times \mbox{SU}(\ell)$ defined as
\[
\Phi_{\ell}^{t}(q,p,U)=(\Phi^{t}(q,p),D(q,p,t)\,U)\,.
\]

Turning to the  higher order
corrections (\ref{Eg15}), (\ref{Egg2}) etc., they are of the form
\[
\frac{d\, a_{n}(t)}{dt} = \{{E_{\rm r}}\mathbf{1}, a_{n}(t)\} + i [h_1,
a_{n}(t)] + I_{n}(a_0(t), \ldots,a_{n-1}(t))
\]
with an inhomogeneity $I_{n}(t)$ depending only on the known functions
$a_0(t),$ $\ldots , a_{n-1}(t)$. Thus, assuming $a_{n}(0)=0$, one finds
\begin{equation}  \label{HOE}
a_{n}(t) = \int_0^t\,ds\, \mathcal{U}(t-s)\,I_{n}(s)\,.
\end{equation}

In order to solve Equation (\ref{Eg15}) for the subprincipal symbol one
needs to know $h_{2}$. However, if one is interested in semiclassical
observables with a principal symbol which is a scalar multiple of the
identity, e.g.\ in the position $a_{0}(0)=q\,\mathbf{1}$, the last term in (
\ref{Eg15}) vanishes at all times, since, according to (\ref{A0}), $a_{0}(t)$
is a scalar multiple of the identity for all times. In Section~\ref{SDIRAC} the back reaction of the spin of an electron on its translational
motion will be discussed on the basis of (\ref{Eg15}).

We summarize the preceding discussion on Egorov's theorem.

\begin{theorem}[Egorov]
\label{EgorovT} \, Let $H$ satisfy either (IG)$_{m}$ for $m\leq 1$ and $\rho=1$ or
(CG) with $\rho=0$. Let ${\sigma_{\rm r}}(q,p)=\{{E_{\rm r}}(q,p)\}$  be an eigenvalue of
$H_{0}(q,p)$ of finite multiplicity $\ell$.

\noindent Then the classical flow $\Phi^{t}$ generated by ${E_{\rm r}}(q,p)$ and
the solution of (\ref{Deq}) with initial condition $D(q,p,0)=\mathbf{1}$
exist globally in time. For $a_{0}\in S_\rho^{0}(\mathcal{B}(\mathbb{C}^{\ell}))$,
$a_{0}(t)$ given by (\ref{A0}) is a solution of (\ref{Eg1}) and $a_{0}(t)\in
S_\rho^{0}(\mathcal{B}(\mathbb{C}^{\ell}))$ for all $t$.

\noindent For each $T<\infty $ there is a constant $C_{T}<\infty $ such that
for all $t\in [-T,T]$
\begin{equation}
\left\| a(t)\,-\,\mathcal{W}_\epsi\left(a_{0}(t)\right)\right\| \,\leq \,\varepsilon \,C_{T}\,,
\label{EgEq}
\end{equation}
where $a(t)=e^{i\widehat{h}t/\varepsilon }\,\widehat{a}_{0}\,e^{-i\widehat{h}t/\varepsilon }$.
\end{theorem}

\noindent \textbf{Proof.} Up to the
modifications discussed before, the proof follows easily along the lines of
Egorov's theorem for scalar valued observables (cf.\ \cite{Robert,RobertPaper}): To make
the expansion of the Heisenberg equation (\ref{Heis1}) rigorous, note that
${E_{\rm r}}=\pi_{\mathrm{r}}h_{0}\pi_{\mathrm{r}}\in S_\rho^{m}(\mathbb{R})$ with $m\leq 1$ and
thus the corresponding Hamiltonian vector field is smooth and bounded.
It follows by standard ODE techniques  \cite{Robert} that
$\partial_t a_0(\Phi^t)\in S^0_1$ and hence also
$\partial_t a_0(t)\in S^0_1$, where $a_0(t)$ is given by  (\ref{A0}).
 Thus one can interchange quantization and differentiation with
respect to time and obtains
\begin{eqnarray*}\lefteqn{ 
a(t)-\mathcal{W}_\epsi(a_{0}(t)) =\int_{0}^{t}\,ds\,\frac{d}{ds}\left( e^{i\widehat{
h}s/\varepsilon }\,\mathcal{W}_\epsi( a_{0}(t-s))\,e^{-i\widehat{h}s/\varepsilon
}\right) } \\
&=&\int_{0}^{t}\,ds\,e^{i\widehat{h}s/\varepsilon }\,
\left( \frac{i}{\varepsilon }\left[ \,\widehat{h},
\mathcal{W}_\epsi( a_{0}(t-s))\,\right] -\mathcal{W}_{\varepsilon }
\left( \frac{da_{0}}{dt}(t-s)\right) \right) e^{-i\widehat{h}s/\varepsilon }\,.
\end{eqnarray*}
Now, by construction, $\frac{i}{\varepsilon }\left[ \widehat{h},
\widehat{a_{0}(t-s)}\right] -\mathcal{W}_{\varepsilon }
\left( \frac{da_{0}}{dt}(t-s)\right) $ is a semiclassical 
operator in OP$S_{1}^{1}(\varepsilon )$
with vanishing principal symbol. Hence the integrand is really 
$\mathcal{O}(\varepsilon )$ as a bounded operator and (\ref{EgEq}) follows. 
\hfill $\qed$\newline

This matrix-valued version of Egorov's theorem has been discussed several
times in the literature \cite{Ivrii,BN}.\bigskip

\noindent \textbf{Berry connection.} 
 With this preparation we explain the
motivation behind the particular splitting of the terms in (\ref{h1w}).
It is of geometrical origin and related to the Berry connection. Recall
that in the Born-Oppenheimer setting $h_{1\,\alpha \beta }(q,p)=-p\cdot
A_{\alpha \beta }(q)$ and thus $A_{\alpha \beta }(q)$ acts as a gauge
potential of a connection on the trivial bundle $\mathbb{R}^{d}\times \mathbb{C}^{\ell}$. 
Its origin is purely geometrical, since it comes from the connection
which the trivial connection on the trivial bundle $\mathbb{R}^{d}\times
\Hi _{\mathrm{f}}$ induces on the subbundle defined by $\pi_{0}(q)$.
If one assumes that Ran$\pi_{0}(q)$ is $1$-dimensional, the internal
rotations along classical trajectories are just phase changes, the so called
Berry phases, and are due to parallel transport with respect to the Berry
connection \cite{BerryPhase,BerryBO,Simon}.

In the general case the second term of $h_{1\,\alpha \beta }(q,p)$ in (\ref{h1w}),
which we denote by
\[
h_{\mathrm{Be\,\alpha \beta }}(q,p)=-i\langle \psi_{\alpha
}(q,p),\{{E_{\rm r}},\psi_{\beta }\}(q,p)\rangle ,
\]
corresponds exactly to this parallel transport along the generalized Berry
connection. More precisely, the trivial connection on the trivial bundle 
$\mathbb{R}^{2d}\times \Hi _{\mathrm{f}}$ induces a $U(\ell)$-connection on
the subbundle defined by $\pi_{0}(q,p)$. After unitary rotation $u_{0}(q,p)$
the coefficients of this connection on the bundle 
$\mathbb{R}^{2d}\times \mathbb{C}^{\ell}$ are
\[
A_{\alpha \beta }(q,p)=\,i\,\left(
\begin{array}{c}
\langle \psi_{\alpha }(q,p),\nabla_{q}\psi_{\beta }(q,p)\rangle \\
\langle \psi_{\alpha }(q,p),\nabla_{p}\psi_{\beta }(q,p)\rangle
\end{array}
\right) \,,
\]
in the sense that a section $s(q,p)$ is parallel if $(\nabla -iA)s=0$. It is
parallel along some curve $c(\tau )=\big(q(\tau ),p(\tau )\big)$ in $\mathbb{R}^{2d}$ if
\[
\Big( \partial_{\tau }-\dot{c}(\tau )\cdot iA\big(q(\tau ),p(\tau )\big)
\Big)
s\big(q(\tau ),p(\tau )\big)=0\,.
\]
For classical trajectories, where $\dot{c}(t)=(\nabla_{p}{E_{\rm r}},-\nabla
_{q}{E_{\rm r}})^{\mathrm{T}}$, this condition becomes
\begin{equation}
\Big( \partial_{t}+i\,h_{\mathrm{Be}}\big(q(t),p(t)\big) \Big)s\big(q(t),p(t)\big)=0\,.  
\label{parallel}
\end{equation}
If $h_{1}=h_{\mathrm{Be}}$, (\ref{parallel}) is exactly Equation (\ref
{spintransport}) for the rotation of the spinor 
$\varphi_{t}\big(q(t),p(t)\big) =D\big(q,p,t\big)\varphi_{0}$ along the trajectory 
of the particle.
This means if $h_{1}=h_{\mathrm{Be}}$, the spin dynamics corresponds to parallel
transport with respect to the Berry connection along classical trajectories.

Emmrich and Weinstein \cite{EW} give a geometric meaning also to the
remaining terms in their analog of $h_{1}$. While this is a natural venture
in the context of geometric WKB approximation, it seems to be less natural
in our approach, since we work in a fixed basis in order to obtain simple
analytic expressions.\bigskip

\noindent \textbf{Wigner function approach.}
The previous results on the time-evolution of semiclassical observables
translate, by the duality expressed through
\[
\langle \psi, \,\widehat a_0 \,\psi\rangle
 =\int_{\mathbb{R}^{2d}}\,{\rm Tr}_{\mathbb{C}^\ell}
\left(a_{0}(q,p)W^\psi(q,p)\right)\,dq\,dp \,,
\]
to the time-evolution of the Wigner transform 
\[
W^\psi(q,p) := {\rm Symb}(P_\psi)(q,p)=(2\pi)^{-d} \int_ {\mathbb{R}^{d}}\,d\xi\,
e^{i\xi\cdot p}\,
\psi(q+\epsi\xi/2)\otimes \psi^* (q-\epsi\xi/2)
\]
as
\begin{eqnarray*}\lefteqn{
\langle \psi, \,\widehat a_0(t) \,\psi\rangle
 =}\\&&\hspace{-13pt}=\,\int_{\mathbb{R}^{2d}}\hspace{-2pt}{\rm Tr}_{\mathbb{C}^\ell} \left(
a_{0}(q,p)\,D^*(q,p,-t)\,W^\psi(\Phi^{-t}(q,p))\,D(q,p,-t) \right)\,dq\,dp 
+\mathcal{O}(\epsi) \,.
\end{eqnarray*}
Transport equations for  matrix-valued Wigner measures were derived
in \cite{GMMP} and applied to the Dirac equation in \cite{Spohn}. \bigskip

\noindent \textbf{Semiclassical propagator.}
Often one is not only interested in the semiclassical propagation of
observables, but more directly in a semiclassical expansion of the kernel
$K(x,y,t)$ of the unitary group
\begin{equation}
\big(e^{-i\widehat{h}t/\varepsilon }\psi \big)(x)=
\int_{\mathbb{R}^{d}}\,dy\,K^{\varepsilon }(x,y,t)\,\psi (y)\,.  \label{KDef}
\end{equation}
As in the case of Egorov's theorem, generalizing the known results for
Hamiltonians with scalar symbols to the case of operator-valued symbols is
straightforward, whenever the principal symbol $h_{0}$ of $h$ is a scalar multiple of
the identity. As in the scalar case, see \cite{Robert}, one makes an ansatz
of the form
\[%\begin{equation}
K^{\varepsilon }(x,y,t)=\frac{1}{(2\pi \varepsilon )^{d}}
\int_{\mathbb{R}^{d}}\,dp\,e^{\frac{i}{\varepsilon }\big(S(x,p,t)-y\cdot p\big)}
\Big(\sum_{j=0}^{\infty }\varepsilon^{j}a_{j}(x,p,t)\Big)\,,  %\label{KAnsatz}
\]%\end{equation}
where $S(x,p,t)$ is real valued and the $a_{j}$'s take values in the bounded
linear operators on $\mathbb{C}^{\ell}$. Demanding (\ref{KDef}) at time $t=0$,
i.e.\ $K^{\varepsilon }(x,y,0)=\delta (x-y)$, imposes the following initial
conditions on $S$ and $\{a_{j}\}_{j\geq 0}$:
\[
S(x,p,0)=x\cdot p\,,\qquad a_{0}(x,p,0)=\mathbf{1}\quad \mbox{and}\quad
a_{j}(x,p,0)=0\quad \mbox{for}\,j\geq 1\,.
\]
For later times the coefficients are determined by formally expanding the
Schr\"{o}dinger equation for $K^{\varepsilon }(x,y,t)$
\[%\begin{equation}
i\,\varepsilon \frac{\partial }{\partial t}K^{\varepsilon }(\cdot ,y,t)=
\widehat{h}\,K^{\varepsilon }(\cdot ,y,t)
\]%\end{equation}
in orders of $\varepsilon $. At leading order only $\widehat{h}_{0}=\widehat{
E}_{0}$ contributes and one obtains as in the scalar case
\begin{equation}
\partial_{t}\,S(x,p,t)+{E_{\rm r}}\big(x,\nabla_{x}S(x,p,t)\big) =0\,,
\label{HJ}
\end{equation}
the Hamilton-Jacobi equation for the symbol $h_{0}$. The next to leading
order equation is the so called transport equation for $a_{0}$:
\begin{equation}
i\partial_{t}a_{0}(x,p,t)=\mathcal{L}(x,p,t)\,a_{0}(x,p,t)\,+h_{1}
\big(x,\nabla_{x}S(x,p,t)\big) \,a_{0}(x,p,t)\,.  \label{transport}
\end{equation}
The differential operator $\mathcal{L}(x,p,t)$ is the same as in the scalar
case, see \cite{Robert} for an explicit formula. Here we just want to
point out that the known techniques from the scalar case apply with one
modification: as in (\ref{Eg1}), also in (\ref{transport}) $h_{1}$
contributes as an additional rotation in the transport equation for the
leading order term. Since the solution of (\ref{HJ}) exists only until a
caustic is reached, the approximation (\ref{HJ}), (\ref{transport}) to the
propagator is a short time result only. The extension to arbitrary times is
a complicated task, in general \cite{Maslov}.

\subsection{An Egorov theorem}

Ultimately  the goal is  to
approximate expectation values of  observables in in the original Hilbert space
$\Hi =L^{2}(\mathbb{R}^{d},\Hi _{\mathrm{f}})$ rather than in
$\Hi =L^{2}(\mathbb{R}^{d},\mathcal{K}_{\mathrm{f}})$.
Before stating a theorem an obvious, but important observation should
be made, which seems to have been overlooked, or at least not stressed
sufficiently, in related discussions, e.g., \cite{LF,LW,BK,MS}: We proved
that in the case ${\sigma_{\rm r}}(q,p)=\{{E_{\rm r}}(q,p)\}$ the effective Hamiltonian
$\widehat{h}$
projected on the subspace $\mathcal{K}=\mbox{Ran}{\Pi_{\rm r}}$
has a semiclassical limit in the sense of a generalized Egorov
theorem, in principle, to any order in $\varepsilon $. However, the
variables $q$ and $p$ in the rotated representation
 are \emph{not} the canonical variables of the slow
degrees of freedom in the original problem. More precisely, let $\widehat{q}
_{\Hi }=x\otimes \mathbf{1}_{\Hi _{\mathrm{f}}}$ and $\widehat{
p}_{\Hi }=-i\varepsilon \nabla_{x}\otimes \mathbf{1}_{\Hi _{
\mathrm{f}}}$ be the position and momentum operators of the slow degrees of
freedom acting on $\Hi $ and let $\widehat{q}_{\mathcal{K}}=x\otimes
\mathbf{1}_{\mathcal{K}_{\mathrm{f}}}$ and $\widehat{p}_{\mathcal{K}
}=-i\varepsilon \nabla_{x}\otimes \mathbf{1}_{\mathcal{K}_{\mathrm{f}}}$ be
the same operators acting on $\mathcal{K}$. Then $\widehat{q}_{\mathcal{K}}=
{\Pi_{\rm r}}U\,\widehat{q}_{\Hi }\,U^{*}\,
{\Pi}_{\mathrm{r}}+\mathcal{O}(\varepsilon )$ and $\widehat{p}_{\mathcal{K}}=
{\Pi_{\rm r}}\,U\,\widehat{p}_{\Hi }\,U^{*}\,
{\Pi }_{\mathrm{r}}+\mathcal{O}(\varepsilon )$, with a, in general,
nonvanishing $\varepsilon $-correction. Physically this means that the
quantities which behave like position and momentum in the semiclassical
limit are only close to the position and momentum of the slow degrees of
freedom, but not equal.  This phenomenon is well known
in the case of the nonrelativistic limit of the
Dirac equation. The
Newton-Wigner position operator and not the standard position operator
goes over to the position operator in the Pauli equation. The standard
position operator has neither a nice nonrelativistic limit nor, as we will
see, a nice semiclassical limit, because of the Zitterbewegung. Switching to the
Newton-Wigner position operator corresponds to averaging over the
Zitterbewegung, or, in our language, to use the position operator 
$\widehat{q}_{\mathcal{K}}$ in the rotated representation. We remark that in the
Born-Oppenheimer case, and more generally whenever $\pi_{0}$ depends on $q$
only, one has $\widehat{q}_{\mathcal{K}}={\Pi_{\rm r}}U\,
\widehat{q}_{\Hi }\,U^{*}\,{\Pi_{\rm r}}+\mathcal{O}(\varepsilon^{2})$.

With this warning we exploit that semiclassical observables do not change after
unitary rotation in leading order and state the Egorov theorem for the
observables in the original representation.

\begin{corollary}
\label{Egorov2}Let $H$ satisfy either (IG)$_{m}$ with $m\leq 1$ and $\rho =1$ or (CG) with 
$\rho=0$ and
let ${\sigma_{\rm r}}(q,p)=\{{E_{\rm r}}(q,p)\}$ consist of a single eigenvalue of 
$H_{0}(q,p)$ of finite multiplicity $\ell$. Let $b_{0}\in S_{1}^{0}
(\mathcal{B(H}_{\mathrm{f}}))$ such that $[b_{0},\pi_{0}]=0$ and 
$B(t):=e^{i\widehat{H}t/\varepsilon }\,\widehat{b}_{0}\,e^{-i\widehat{H}t/\varepsilon }$.
Let $a_{0}:=\pi_{\mathrm{r}}\,u_{0}\,b_{0}\,u_{0}^{*}\,\pi_{\mathrm{r}}$ and define
$a_{0}(t)$ is in (\ref{A0}). 
Then for each $T<\infty $ there is a constant $C_{T}<\infty$
such that for all $t\in [-T,T]$
\begin{equation}
\left\|\big( B(t)\,-\,\mathcal{W}_{\varepsilon }(u_{0}^{*}\,a_{0}(t)\,u_{0})\big)\Pi 
\right\| \,\leq \,\varepsilon \,C_{T}\,.  \label{Eg2}
\end{equation}
For $b_{0}=f\,\mathbf{1}_{\Hi _{\mathrm{f}}}$, with $f\in
S_{1}^{0}(\mathbb{R})$, one obtains as a special case of (\ref{Eg2}) that
\[%\begin{equation}
\left\|\big( B(t)\,-\,\widehat{b_{0}(\Phi^{t})}\big)\Pi \right\| \,\leq
\,\varepsilon \,C_{T}\,.
\]%\end{equation}
\end{corollary}

Corollary \ref{Egorov2} follows from Theorem \ref{EgorovT} and a
straightforward expansion in $\varepsilon$ of the terms to be estimated
after rotation with $U$.

\section{The Dirac equation}

\label{SDIRAC}

\subsection{Adiabatic decoupling of electrons and positrons}

We apply the  adiabatic perturbation theory to the one-particle Dirac
equation with slowly varying external potentials, i.e.\ to
\[%\begin{equation}
\widetilde{H}_{\mathrm{D}}=c\alpha \cdot \left( -i\hbar \nabla_{y}-
\frac{e}{c}A(\varepsilon y)\right) +\beta mc^{2}+e{\phi}(\varepsilon y)\,
\]%\end{equation}
acting on $L^{2}(\mathbb{R}^{3},\mathbb{C}^{4})$. Here 
$A:\mathbb{R}^{3}\to \mathbb{R}^{3}$ is the vector potential of an external 
magnetic field $B=\nabla \wedge
A$ and ${\phi}:\mathbb{R}^{3}\to \mathbb{R}$ the potential of an external electric
field $E=-\nabla {\phi}$. For the Dirac matrices $\alpha $, $\beta $ we make the
standard choice
\[%\begin{equation}
\alpha =\left(
\begin{array}{cc}
0 & \sigma \\
\sigma & 0
\end{array}
\right) \,,\qquad \beta =\left(
\begin{array}{cc}
\mathbf{1}_{\mathbb{C}^{2}} & 0 \\
0 & -\mathbf{1}_{\mathbb{C}^{2}}
\end{array}
\right) \,,
\]%\end{equation}
where $\sigma =(\sigma_{1},\sigma_{2},\sigma_{3})$ denotes the vector of
the Pauli spin matrices. The small parameter $\varepsilon >0$ controls the
variation of the external potentials. 
To keep track of the size of the error terms, in this section
all physical constants, including $\hbar $, are displayed.

Transforming to the macroscopic space-scale $x=\varepsilon y$ one obtains the Dirac
Hamiltonian
\begin{equation}  \label{HDdef}
\widehat H_{\mathrm{D}} = c \alpha\cdot\left(-i\varepsilon\hbar\nabla_x -
\frac{e}{c}A(x)\right) + \beta m c^2 + e{\phi}( x)\,
\end{equation}
and we are interested in the solution of the time-dependent Dirac equation
for  times of order $\varepsilon^{-1}$, i.e.\ in solutions of
\begin{equation}  \label{TDDE}
i\,\varepsilon \hbar \frac{\partial}{\partial t} \psi_t =\widehat H_{\mathrm{
D}}\,\psi_t
\end{equation}
for $|t|=\mathcal{O}(1)$. The solutions of (\ref{TDDE}) for small
$\varepsilon$ approximately describe the dynamics of electrons, resp.\
positrons, in weak fields, as in storage rings, accelerators,
or cloud chambers, for example.

$\widehat{H}_{\mathrm{D}}$ is the Weyl quantization of the matrix-valued function
\[%\begin{equation}
H_{\mathrm{D}}(q,p)=c\alpha \cdot \left( p-\frac{e}{c}A(q)\right) +\beta
mc^{2}+e{\phi}(q)
\]%\end{equation}
on phase space $\mathbb{R}^{6}$, where now Weyl quantization is in the sense of
$p\mapsto -i\varepsilon \hbar \nabla_{x}$, i.e.\ on the right hand side of 
(\ref{Weyl def}) $\varepsilon $ must be replaced by $\varepsilon \hbar $.
 $\hbar $ appears here for dimensional reasons and is a fixed
physical constant. The small parameter of the
space-adiabatic expansion  is $\varepsilon $.
$H_{\mathrm{D}}(q,p)$ has two two-fold degenerate eigenvalues
\[
E_{\pm }(q,p)=\pm cp_{0}(q,p)+e{\phi}(q)
\]
with the corresponding eigenprojections
\[
P_{\pm }(q,p)=\frac{1}{2}\left( 1\pm \frac{1}{p_{0}(q,p)}\left( \alpha \cdot
\left( p-\frac{e}{c}A(q)\right) +\beta mc\right) \right) \,,
\]
where $p_{0}(q,p)=\sqrt{m^{2}c^{2}+(p-\frac{e}{c}A(q))^{2}}$. Obviously
\[
E_{+}(q,p)-E_{-}(q,p)=2cp_{0}(q,p)\geq C\langle p\rangle >0\,,
\]
whenever $A$ is uniformly bounded. Therefore the corresponding subspaces are
adiabatically decoupled and the effective dynamics on each of them can be
computed using our general scheme. Assuming $A\in C_{\mathrm{b}}^{\infty }
(\mathbb{R}^{3},\mathbb{R}^{3})$ and ${\phi}\in C_{\mathrm{b}}^{\infty }(\mathbb{R}^{3},
\mathbb{R})$, one finds that $H_{0}\in S_{1}^{1}$ and thus the
assumptions from Section~\ref{SPRO} are satisfied. In particular, 
$\widehat{H}_{\mathrm{D}}$ is  essentially self-adjoint on 
$\mathcal{S}(\mathbb{R}^{4},\mathbb{C}^{4})$ and $\widehat{E}_{\pm }$
 on $\mathcal{S}(\mathbb{R}^{4})$.

To be consistent with the notation from the previous sections, let 
$\pi_0(q,p)$ $ = P_+(q,p)$ be the projector on the electron band. The reference
subspace for the electrons is $\mathcal{K}=L^2(\mathbb{R}^3,\mathbb{C}^2)$ and it
is convenient to define it as the range of
\[
{\Pi_{\rm r}}:= \left(
\begin{array}{cc}
\mathbf{1}_{\mathbb{C}^2} & 0 \\
0 & 0
\end{array}
\right)
\]
in $L^2(\mathbb{R}^3,\mathbb{C}^4)$.

The only choice left is the one of $u_{0}(q,p)$ or, equivalently,  of a
basis $\{\psi_{\alpha }(q,p)\}_{\alpha =1,2}$ of Ran$\pi_{0}(q,p)$.
Since the degeneracy of Ran$\pi_{0}(q,p)$ is related to the spin
of the electron, a natural choice is the $\sigma_{z}$-representation with
respect to the ``mean''-spin $S(q,p)$ which commutes with 
$H_{\mathrm{D}}(q,p)$ \cite{FW,Thaller}. The eigenvectors $\psi_{\pm}(q,p)$ of the
operator $e_{3}\cdot S(q,p)$ in Ran$\pi_{0}(q,p)$ are
\[
\psi_{+}(q,p)=c {\textstyle\sqrt{\frac{p_{0}}{2(p_{0}+mc)}}}\left(
\begin{array}{c} \scriptstyle
\frac{(p_{0}+mc)}{p_0} \\
 \scriptstyle 0 \\
 \scriptstyle v_{3} \\
 \scriptstyle v_{1}+iv_{2}
\end{array}
\right) \,,\,\,\psi_{-}(q,p)=c {\textstyle\sqrt{\frac{p_{0}}{2(p_{0}+mc)}}}\left(
\begin{array}{c} \scriptstyle
0 \\
 \scriptstyle  \frac{(p_{0}+mc)}{p_0} \\
 \scriptstyle v_{1}-iv_{2} \\
 \scriptstyle -v_{3}
\end{array}
\right) .
\]
We abbreviated $v(q,p):=c\,(p-\frac{e}{c}A(q))/p_{0}(q,p)$ for the velocity.
The relevant part of $u_{0}$ for the analysis of the electron band is
thus given by $u_{0}^*(q,p) = \big(\psi_{+}(q,p),\psi_{-}(q,p),$ $*,*\big)$
with $u_{0}\in S_{1}^{0}$. Of course the positron part indicated by $*$'s would be
given through  charge conjugation. In our construction we want to emphasize, however, that
no specification is needed in
order to determine the expansion of the effective electron Hamiltonian
$\widehat{h}_{\mathrm{e}}:={\Pi_{\rm r}}\,\widehat{h}\,{\Pi_{\rm r}}$
up to arbitrary order.

An alternative way to arrive at the same $u_0(q,p)$ is to note that the
Foldy-Wouthuysen transformation $u_{\mathrm{FW}}(p)$, c.f.\ \cite{FW},
diagonalizes the free Dirac Hamiltonian $H_0(p)$, i.e.\ $H_{\mathrm{D}}$
with $A , {\phi}\equiv 0$. Including the fields  $u_0(q,p)
= u_{\mathrm{FW}}(p-\frac{e}{c}A(q))$ then diagonalizes $H_{\mathrm{D}}(q,p)$.

For the principal symbol of $h_{\mathrm{e}}$ one finds of course
\[%\begin{equation}
h_{\mathrm{e,0}}(q,p)=E_{+}(q,p)\mathbf{1}_{\mathbb{C}^{2}}\,.
\]%\end{equation}
For the subprincipal symbol after a lengthy but straightforward
calculation our basic formula (\ref{h1w}) yields
\begin{eqnarray}
h_{\mathrm{e,1}}(q,p)&=&-\frac{\hbar e}{2p_{0}(q,p)}\,\,\sigma \cdot \left(\hspace{-2pt}
B(q)-\frac{p_{0}}{c\,(p_{0}(q,p)+mc)}\,v(q,p)\wedge E(q)\hspace{-2pt}\right) \nonumber\\ &=:&-\,\frac{
\hbar }{2}\,\sigma \cdot \Omega (q,p)\,.  \label{h1D}
\end{eqnarray}
Note that the factor $\hbar$ comes from the fact that
 the $n$th term in the space-adiabatic expansion carries a
prefactor $\hbar^{n}$. Defining
\[
\gamma (q,p)=1/\sqrt{1-(v(q,p)/c)^{2}}=p_{0}(q,p)/(mc)
\]
 one concludes that
\begin{equation}
\Omega (q,p)=\frac{e}{mc}\,\left( \frac{1}{\gamma (q,p)}\,B(q)-\frac{1}{c\,
(1+\gamma (q,p))}\,v(q,p)\wedge E(q)\right) \,.  \label{ODef}
\end{equation}
We remark that the second term
in (\ref{h1w}), the ``Berry term'', does not coincide with any of the terms
in (\ref{ODef}).
Indeed, the compact expression (\ref{ODef}) is
obtained only through cancellations in more complicated expressions coming from
both terms contributing in (\ref{h1w}).

We summarize our results on the adiabatic decoupling and the effective
dynamics for the Dirac equation in the following

\begin{theorem}
\label{DiracTheorem} Let $A\in C_{\mathrm{b}}^{\infty }(\mathbb{R}^{3},\mathbb{R}^{3})$ 
and ${\phi}\in C_{\mathrm{b}}^{\infty }(\mathbb{R}^{3},\mathbb{R})$. Then there
exist orthogonal projectors $\Pi_{\pm }$ with $\Pi_{+}+\Pi_{-}=\mathbf{1}$
such that $[\widehat{H}_{\mathrm{D}},\Pi_{\pm }]=\mathcal{O}_{0}(\varepsilon^{\infty })$, 
and there exists a unitary $U$
and $\widehat h\in {\rm OP}S^1_1$ with
\begin{equation}
\widehat h = \left(
\begin{array}{cc}
\widehat{h}_{\mathrm{e}} & 0 \\
0 & \widehat{h}_{\mathrm{p}}
\end{array}
\right)\,,
\end{equation}
such that
\begin{equation}
e^{-i\widehat H_{\rm D} t} - U^*\,e^{-i\widehat h t}\,U = \mathcal{O}_0(\epsi^\infty|t|) \,.
  \label{HDexp}
\end{equation}
Here $\widehat{h}_{\mathrm{e}}$ and $\widehat{h}_{\mathrm{p}}$ are
semiclassical operators on $L^{2}(\mathbb{R}^{3},\mathbb{C}^{2})$ with 
\[
h_{\mathrm{
e}}(q,p,\varepsilon )\asymp E_{+}(q,p)\mathbf{1}_{\mathbb{C}^{2}}+
\sum_{j=1}^{\infty }\,\varepsilon^{j}\,h_{\mathrm{e},j}(q,p)
\]
 and $h_{
\mathrm{e},j}=\pi_{\mathrm{r}}(u\,\#\,H_{\mathrm{D}}\,\#\,u^{*})_{j}\pi_{
\mathrm{r}}\in S_{1}^{1-j}(\mathcal{B}(\mathbb{C}^{2}))$ for all $j\geq 0$,
where $u\in S_{1}^{0}(\varepsilon )$ is constructed as in Section~\ref{SUNI}.
In particular, $h_{\mathrm{e,1}}(q,p)$ is given by (\ref{h1D}) and thus
\begin{equation}
\widehat{h}_{\mathrm{e}}=\Big(c\sqrt{m^{2}c^{2}+(-i\varepsilon \hbar \nabla -
\frac{e}{c}A(x))^{2}}+e{\phi}(q)\Big)\mathbf{1}_{\mathbb{C}^{2}}\,-\,\varepsilon \,
\frac{\hbar }{2}\,\sigma \cdot \widehat{\Omega (q,p)}\,+\,\mathcal{O}_0
(\varepsilon^{2})\,.  \nonumber  \label{hel}
\end{equation}
Analogous results hold for $h_{\mathrm{p}}$. The errors in (\ref{HDexp})
and (\ref{hel}) are in the norm of bounded operators on $L^{2}(\mathbb{R}^{3},
\mathbb{C}^{4})$, resp.\ on $L^{2}(\mathbb{R}^{3},\mathbb{C}^{2})$.\emph{\ }
\end{theorem}

According to the effective Hamiltonian (\ref{hel}) the $g$-factor of the electron
equals $2$. There would be no problem to add to the Dirac Hamiltonian the standard
subprincipal symbol \cite{Thaller}, which accounts for the slightly larger $g$-factor
of real electrons.
Blount \cite{BlountDirac} computes the second order effective Hamiltonian $h_{\rm e,2}$,
which he finds to be proportional to ${\bf 1}_{\mathbb{C}^2}$.
 $h_{\rm e,2}$ is a sum of terms allowed by dimensional reasoning, i.e.\
proportional to $\nabla B$, $\nabla E$, $B^2$, $E^2$, $EB$. Second order corrections
seem to be of interest for the dynamics of electrons in storage rings. Ignoring the
contribution \cite{BlountDirac}, nonrigorous expansions are \cite{DK} and \cite{HB}.

\subsection{The semiclassical limit of the Dirac equation}

Equipped with $h_{\rm e,0}$ and  $h_{\rm e,1}$
we can apply the general results of Section~\ref{SSEMI} on the
semiclassical limit to the Dirac equation. Let $\Phi_{\pm }^{t}$ be the
Hamiltonian flows generated by $E_{\pm }(q,p)$ on phase space $\mathbb{R}^{6}$
and let $\widehat{B}=\widehat{b}\,\mathbf{1}$, $b\in S_{1}^{0}(\mathbb{R})$, be
a semiclassical observable in the unrotated Hilbert space which does not
depend on spin. From Corollary \ref{Egorov2} we conclude for each $T<\infty $
the existence of a constant $C_{T}$ such that for all $t\in [-T,T]$
\[
\Big\| \Big( B(t)-\mathcal{W}_\epsi \big(b(\Phi_{+}^{t})\big)\,\mathbf{1}\Big) \Pi_{+}\Big\|
\leq \,\varepsilon \,C_{T}\,,\,\, \Big\| \Big( B(t)-\mathcal{W}_\epsi\big(
b(\Phi_{-}^{t})\big)\,\mathbf{1}\Big) \Pi_{-}\Big\|
\leq \,\varepsilon \,C_{T}\,,
\]
where $B(t)=e^{i\widehat{H}_{\mathrm{D}}t/(\varepsilon \hbar )}\,
\widehat{B}\,e^{-i\widehat{H}_{\mathrm{D}}t/(\varepsilon \hbar )}$. Hence, to leading
order, states in the range of $\Pi_{+}$ behave like classical relativistic
electrons and states in the range of $\Pi_{-}$ like classical relativistic
positrons. We emphasize that, in general, $\Pi_{\pm }$ are not spectral
projections of $\widehat{H}_{\mathrm{D}}$, since the variation of ${\phi}$ can be
larger than the mass gap $2mc^{2}$. Hence in the limit of slowly varying
potentials a natural characterization of ``electronic'' and ``positronic''
subspaces is obtained which does \emph{not} come from spectral projections
of the free or full Dirac Hamiltonian.

Next we discuss the leading order spin dynamics, which in the first place
requires to figure out which operator represents the spin of the electron.
There has been a considerable discussion on this point, cf.\ \cite{Thaller},
with no general consensus reached. We suspect that the problem is
void. The wave function is  spinor valued and what is observed is the
spatial splitting of different spinor components in inhomogeneous magnetic
fields. Hence we should pick the ``spin observable'' $\Sigma$ such that the
splitting can nicely be attributed to it. E.g., 
in a magnetic field with gradient along the $z$-direction the
eigenvectors of $\Sigma_z$
should have the property that their spatial support goes either parallel to $+z$ 
or to $-z$, but should not split.
In view of (\ref{hel}) a natural choice is to take as spin operator
the vector of Pauli-matrices $\sigma$ in the rotated electronic subspace.
In the original Hilbert space this amounts to
\[
\Sigma = U^*\,\left(
\begin{array}{cc}
\sigma & 0 \\
0 & \sigma
\end{array}
\right)\,U = \frac{2}{\hbar}\,\widehat S + \mathcal{O}(\varepsilon)\,,
\]
where $S(q,p)$ is the ``mean'' spin defined before.

The leading order semiclassical approximation for 
\[
\sigma(t) = e^{i\widehat
h_{\mathrm{e}}t/(\varepsilon\hbar)}\,\sigma\, e^{-i\widehat h_{\mathrm{e}
}t/(\varepsilon\hbar)}
\]
 follows from Theorem \ref{EgorovT}.
For each $T<\infty$ there is a constant $C_T<\infty$ such that for $t\in [-T,T]$
\begin{equation}
\big\| \sigma(t) - \widehat{ \sigma_{0}(t)} \big\| \leq\,\varepsilon\,C_T\,,
\end{equation}
where $\sigma_{0k}(q,p,t)$, $k\in\{1,2,3\}$, is obtained as the solution of
\begin{equation}  \label{BMT0}
\frac{\partial\,\sigma_{0k}(q,p,t)}{\partial t} =-\, \frac{ i}{2}\, \Big[\,
\sigma\cdot\Omega(\Phi_+^{t}(q,p)),\, \sigma_{0k}(q,p,t)\,\Big]
\end{equation}
with initial condition $\sigma_{0k}(q,p,0)=\sigma_k$. This follows from the
Equations (\ref{A0}) and (\ref{Deq}) by setting $\sigma_{0k}(q,p,t)
= D^*(q,p,t)\,\sigma_k\, D(q,p,t)$.

To solve Equation (\ref{BMT0}) one makes an ansatz $\sigma_{0k}(q,p,t) =
s_k(q,p,t)\cdot \sigma$ with $s_k(q,p,0) = e_k$. Using 
$[\sigma_n,\sigma_m]=2\,i\,\varepsilon_{nmk}\,\sigma_k$, one finds that the spin- or
``magnetiza\-tion''-vector $s_k(q,p,t)$ is given as the solution of
\begin{equation}  \label{BMT1}
\frac{\partial\,s_k(q,p,t)}{\partial t} = - \, s_k(q,p,t)\,\wedge\,
\Omega(\Phi_+^{t}(q,p))\,.
\end{equation}
(\ref{BMT1}) is the BMT-equation \cite{BMT,Jackson} on the level of
observables. It was derived by Bargmann, Michel and Telegdi in 1959 on
purely classical grounds as the simplest Lorentz invariant equation for the
spin dynamics of a classical relativistic particle.

The semiclassical limit of the Dirac equation has been discussed repeatedly
and we mention only some recent work. Yajima \cite{Yajima} considers
time-dependent external fields and proves directly a semiclassical expansion
for the corresponding propagator. As mentioned already at the end of Section~\ref{SSEMI}, this program is mathematically rather involved, since one faces the
problem of caustics in the classical flow, and different expansions have to
be glued together in order to obtain results valid for all macroscopic times.  Based on the same
approach  Bolte and Keppeler
\cite{BK}  derive a Gutzwiller type trace formula. Since $\widehat{H}_{
\mathrm{D}}$ and $U^{*}\,\widehat{H}_{\mathrm{D}}\,U$ are isospectral
and since (\ref{HDexp}) holds, a
trace formula for the eigenvalue statistics of $\widehat{H}_{\mathrm{D}}$
could as well be derived from the semiclassical propagator of 
$\widehat h=\widehat{h}_{\mathrm{e}}\otimes \mathbf{1}+\mathbf{1}\otimes 
\widehat{h}_{\mathrm{p}}$.
As argued in Section~\ref{SSEMI},
the latter is somewhat easier to obtain. In \cite{GMMP,Spohn} the
semiclassical limit of the Dirac equation is discussed using matrix-valued
Wigner functions. Their results hold for an arbitrary macroscopic time interval,
but fuse, as does the WKB
approach, adiabatic and semiclassical limit. No higher order corrections
seem to be accessible and the results are weaker than ours in the sense that
the approximations do not hold uniformly in the states.

This leads us to the next natural question: What can be said about higher order
corrections?
While in general one would need $h_{\rm e,2}$, according to
(\ref{Eg15})
the  semiclassical limit of observables of the type
$\widehat b = \widehat b_0 \mathbf{1}_{\mathbb{C}^2}$, $b_0\in S^0_1(\mathbb{R})$,
can be determined without this explicit information.
For such a scalar symbol the principal symbol
$b_0(t) $, i.e.\ the solution to (\ref{Eg1}), will remain scalar and thus its
commutator with $h_{\mathrm{e,2}}$ in (\ref{Eg15}) vanishes identically for
all times. The solution $b_1(t)$ of (\ref{Eg15}) with initial condition $b_1(0)=0$, 
is not scalar, in general. Hence, at this order there is  back reaction
of the spin dynamics on the translational motion. We
illustrate this point for the position operator $x(q,p) =x_0(q,p) := q\,
\mathbf{1}_{\mathbb{C}^2}$ and refer to \cite{Te3} for a general analysis of the higher
order effects in the semiclassical dynamics of Dirac particles. Now $x_0(q,p,t) = x_0\big(\Phi^t_{+}(q,p)\big)$
and $x_1(t)$ is obtained, according to Equation (\ref{Eg15}), as the solution
of
\begin{equation}  \label{x1}
\frac{d\, x_1(t)}{dt} = \{E_+\mathbf{1}, x_1(t)\} + i [h_{\mathrm{e,1}},
x_1(t)] - \{h_{\mathrm{e,1}},x_0(t)\}
\end{equation}
with initial condition $x_1(0) = 0$. The homogeneous part of this equation
is just the classical translational and spin motion and the inhomogeneity is
\begin{equation}\label{84}
\{h_{\mathrm{e,1}},x_0(t)\} =-\frac{\hbar}{2}\, \sigma \cdot\{ \,\Omega,
x_0(t) \}\,,
\end{equation}
which is not scalar and thus responsible for the splitting of trajectories
of electrons with distinct spin orientation. Hence, as in (\ref{HOE}),
\[%\begin{equation}  \label{x1D}
x_1(t) = - \frac{\hbar}{2} \int_0^t\,ds\, \mathcal{U}(t-s)\,\sigma \cdot\{
\,\Omega,\, x_0(s) \}\,,
\]%\end{equation}
where $\mathcal{U}(t)$ is the ``classical flow'' defined through (\ref{UDDEF}).

Without claim of rigor, we observe in (\ref{ODef}) that for small
velocities $v(q,p)$ one has
\[
\Omega (q,p)\approx \frac{e}{mc}\,B(q)\,.
\]
Let us further assume that $B(q)=b\,q_z\,e_{z}$, then
\[
\frac{\hbar }{2}\,\sigma \cdot \{\Omega ,x_{0}(t)\}=
\frac{\hbar e}{2mc}\sigma_{z}\frac{\partial B}{\partial q_{z}}\,
\frac{\partial \Phi_{q}^{t}}{\partial p_{z}}=t\,\frac{\hbar e}{2m^{2}c}\left(
\begin{array}{cc}
b & 0 \\
0 & -b
\end{array}
\right)
\]
and thus according to (\ref{x1}), (\ref{84}) the correction to the velocity
 is proportional to $t$, corresponding to a constant force with absolute
value $\hbar e/(2mc)|\nabla B|$, as expected for a spin-$\frac{1}{2}$  particle.

\section{Conclusions}

The basic formulae (\ref{h1d}), (\ref{h21}) can be applied, in essence
in a mechanical fashion, to any concrete quantum problem with two provisos.
First of all the problem has to be cast into the general form  (\ref{a})
and secondly one must have sufficient information on the principal
symbol $H_0(q,p)$. Depending on $H_0$ considerable simplifications
of (\ref{h1d}), (\ref{h21}) may be in force, one example being the effective
Hamiltonian of the time-adiabatic theorem studied in Section~\ref{STA}.
As a net result, if the conditions of the space-adiabatic Theorem 
\ref{Th Invariant subspace} are satisfied, the full Schr\"odinger equation 
is approximated by an effective Schr\"odinger equation referring to a specific
relevant energy band. The errors are estimated and, in general, the time scale of validity
is much larger than the one which can be reached within a semiclassical
approximation.

We focused our interest on a single relevant energy band. No information
on the complement is needed except for global quantities like the resolvent
$(H_0(q,p) -{E_{\rm r}}(q,p))^{-1}(1-\pi_0(q,p))$. In previous investigations
\cite{BlountBloch,LF} all energy bands are treated simultaneously. An example which
would not fall under such a scheme is nonrelativistic QED, which governs
electrons coupled to the quantized radiation field. In this case the principal
symbol has a two-fold degenerate eigenvalue at the bottom of the spectrum separated
by a gap from the continuous spectrum, provided $|p|$ is sufficiently
small and there is a suitable infrared cutoff \cite{PST}.

The main restriction of our work is the gap condition of Section~\ref{SPRO}.
There are two standard mechanisms of how this condition is violated.
(1) There are two (or possibly more) locally isolated energy bands of constant
multiplicity which cross on a lower dimensional submanifold.
Away from the crossing region the wave function in one band is governed by
the effective Hamiltonian discussed before. If the wave function comes close
to the crossing manifold, there is a certain probability to make a transition to
the other band. In rather specific model systems such transitions have been
studied in considerable detail \cite{HagedornCross,JHcross,FeGe,FL}.
(2) $H_0$ has a smooth band of constant multiplicity bordering the continuous
spectrum {\em without} gap. This is the rule in models from nonrelativistic QED with
massless photons. Results  for the massless Nelson model \cite{general}
indicate that smoothness of $\pi_0(q,p)$ suffices also in general
for adiabatic decoupling at leading order with intraband dynamics
generated by $h_0+\epsi h_1$ as defined by  (\ref{hs}).
However, the expansion stops at this stage. Physically, the electron looses
energy through radiation, which means that the next order correction must be
dissipative.

>From the physics point of view the dynamics of molecules and the dynamics of
electrons in a solid are the two most prominent areas of application for the
space-adiabatic perturbation theory. The former has been discussed already in
Section~\ref{SBO}. Bloch electrons do not quite fall into our scheme, since the 
classical phase space is $\mathbb{R}^d\times\mathbb{T}^d$, $\mathbb{T}^d$ a flat
$d$-dimensional torus. This   requires substantial  changes which are discussed in \cite{PST2}.

\appendix

\section{Operator-valued Weyl calculus}

\label{SPREL}

Pseudodifferential operators with operator-valued symbols have been widely  discussed
 in the literature. 
The results presented in this Appendix can be found in  \cite{Hormander,Folland,Ivrii,GMS}.
We start with some notation.
  Let  $\mathcal{E}$  be a Banach space, then  $\mathcal{C}(\mathbb{R}^{d},\mathcal{E})$ denote    the space of $\mathcal{
E}$-valued continuous functions on $\mathbb{R}^{d}$. In the same spirit we
will employ the notation $\mathcal{S}(\mathbb{R}^{d},\mathcal{E})$, $L^{p}(\mathbb{
R}^{d},\mathcal{E})$,
with the obvious meaning. Note that, in the special case where $\mathcal{E=H}_{\rm f}
$ is an Hilbert space, one has $L^{2}(\mathbb{R}^{d},\Hi _{\rm f})\cong L^{2}(
\mathbb{R}^{d})\otimes \Hi _{\rm f}$. The space of the bounded operators on
$\mathcal{E}$ will be denoted as $\mathcal{B}(\mathcal{E})$.

\subsection{Weyl quantization}

Let $A$ be a $\mathcal{B}(\Hi _{\mathrm{f}})$-valued
rapidly decreasing smooth function on $\mathbb{R}^{2d}$, i.e.\ $A\in
\mathcal{S}(\mathbb{R}^{2d},\mathcal{B}(\Hi _{\mathrm{f}}))$. If we
denote by $\mathcal{F}A$ the Fourier transform of $A$ then, by Fourier
inversion formula,
\[
A(q,p)=\frac{1}{(2\pi )^{d}}\int_{\mathbb{R}^{2d}}(\mathcal{F}A)(\eta ,\xi )\
e^{i(\eta \cdot q+\xi \cdot p)}\ d\eta d\xi\,,
\]
where the integral is a Bochner integral for $\mathcal{B}(\Hi _{\rm f})$-valued
functions.
This suggest to define an operator $\widehat{A}\in \mathcal{B}(\Hi )$,
called the \textbf{Weyl quantization} of $A$, by substituting $e^{i(\eta
\cdot q+\xi \cdot p)}$ with $e^{i(\eta \cdot \widehat q+\xi \cdot\widehat p)}\otimes 1_{
\Hi _{\mathrm{f}}}$ where $\widehat q$ is multiplication by $x$ and $
\widehat p=-i\varepsilon \nabla_{x}$ in $L^{2}(\mathbb{R}^{d})$. The exponential is
defined by using the spectral theorem and it is explicitly given by
\begin{equation}
\left( e^{i(\eta \cdot \widehat q+\xi \cdot \widehat p)}\psi \right) (x)=e^{i\varepsilon (\eta
\cdot \xi )/2}e^{i\eta \cdot x}\psi (x+\varepsilon \xi )\qquad \mbox{for }
\psi \in L^{2}(\mathbb{R}^{d}).  \label{Weyl op}
\end{equation}
Thus
\begin{equation}
\widehat{A}=\frac{1}{(2\pi )^{d}}\int_{\mathbb{R}^{2d}}(\mathcal{F}A)(\eta ,\xi )\
\left( e^{i(\eta \cdot \widehat q+\xi \cdot \widehat p)}\otimes 1_{\Hi _{\mathrm{f}
}}\right) d\eta d\xi\,,   \label{Weyl1}
\end{equation}
and, in particular,
\[
\big\| \widehat{A} \big\|_{\mathcal{B}(\Hi )}\leq \frac{1}{(2\pi )^{d}}
\int_{\mathbb{R}^{2d}}\left\| (\mathcal{F}A)(\eta ,\xi )\right\|_{\mathcal{B}(
\Hi _{\mathrm{f}})}\ d\eta d\xi \,,
\]
which implies that $\widehat{A}$ belongs to $\mathcal{B}(\Hi )$ provided
the Fourier transform of $A$ belongs to $L^{1}(\mathbb{R}^{2d},$ $\mathcal{B}(
\Hi _{\mathrm{f}}))$.  We will also use the notation $\mathcal{W}
_{\varepsilon }(A)\equiv \widehat{A}$ in order to emphasize the $\varepsilon$
-dependence.

Substituting (\ref{Weyl op}) in (\ref{Weyl1}) one obtains that for every 
$\psi \in \mathcal{S}(\mathbb{R}^{d},\Hi _{\mathrm{f}})$
\begin{equation}
\big( \widehat{A}\psi \big) (x)=\frac{1}{(2\pi \varepsilon )^{d}}\int_{\mathbb{R}
^{2d}}A\big(\textstyle{\frac{1}{2}}(x+y),\xi \big)\, e^{i \xi \cdot (x-y)/\epsi}\,
\psi (y)\ d\xi dy,  \label{Weyl def}
\end{equation}
i.e.\ $\widehat{A}$ is an integral operator with kernel
\[
K_{A}(x,y)=\frac{1}{(2\pi \varepsilon )^{d}}\int_{\mathbb{R}^{d}}A\big({\textstyle\frac{1}{2}}
(x+y),\xi \big)\ e^{i\xi \cdot (x-y)/\varepsilon }\ d\xi\,.
\]

Taking (\ref{Weyl def}) as a definition, the Weyl quantization
 can be extended to much larger classes
of {\bf symbols} $A(q,p)$.

\begin{definition}
A function $A\in C^{\infty }(\mathbb{R}^{2d},\mathcal{B}(\Hi _{\mathrm{f}}))$ 
belongs to the  symbol class $S_{\rho }^{m}(\mathcal{B}(\Hi _{
\mathrm{f}}))$ $($with $m\in \mathbb{R}$ and $0\leq \rho \leq 1)$ if for every
$\alpha ,\beta \in \mathbb{N}^{d}$ there exists a positive constant
$C_{\alpha ,\beta }$ such that
\[
\sup_{q\in \mathbb{R}^{d}}\left\| (\partial_{q}^{\alpha }\partial_{p}^{\beta
}A)(q,p)\right\|_{\mathcal{B}(\Hi _{\mathrm{f}})}\leq C_{\alpha
,\beta }\ \left\langle p\right\rangle^{m-\rho |\beta |}
\]
for every $p\in \mathbb{R}^{d}$, where $\left\langle p\right\rangle
=(1+|p|^{2})^{1/2}$.
\end{definition}

\noindent The space $S_{\rho }^{m}(\mathcal{B}(\Hi _{\mathrm{f}}))$
is a Fr\'{e}chet space, whose topology can be defined by the (directed)
family of semi-norms
\begin{equation}
\left\| A\right\|_{k}^{(m)}=\sup_{|\alpha |+|\beta |\leq k}
 \sup_{q,p\in \mathbb{R}^{d}}\left\langle
p\right\rangle^{-m+\rho |\beta |}\left\| (\partial_{q}^{\alpha }\partial
_{p}^{\beta }A)(q,p)\right\|_{\mathcal{B}(\Hi _{\mathrm{f}})}\,,\qquad
k\in \mathbb{N}\,.  \label{Seminorms}
\end{equation}
The following result is proved exactly as in the scalar case, cf.\ also \cite{GMS}.
\begin{proposition}
Let $A\in S^m_\rho(\mathcal{B}(\Hi _{\rm f}))$, then $\widehat A$
given trough (\ref{Weyl def}) maps $\mathcal{S}(\mathbb{R}^d,\Hi _{\rm f})$
continuously into itself.
\end{proposition}
Since  $A\in S^m_\rho(\mathcal{B}(\Hi _{\rm f}))$ implies
$A^*\in S^m_\rho(\mathcal{B}(\Hi _{\rm f}))$, the previous result 
allows to extend $\widehat A$ to a continuous map on $\mathcal{S}'(\mathbb{R}^d,
\Hi _{\rm f})$.

It is convenient to introduce a special 
notation for such classes of operators acting on 
$\mathcal{S}(\mathbb{R}^{d},\Hi _{\mathrm{f}})$,  
called pseudodifferential operators,
\[
OPS_{\rho }^{m}:=\left\{ \mathcal{W}_{\varepsilon }(A):\, A\in S_{\rho
}^{m}(\mathcal{B}(\Hi _{\mathrm{f}}))\right\} .
\]
In the following we will sometimes denote 
$S_{\rho }^{m}(\mathcal{B}(\Hi _{\mathrm{f}}))$ simply as $S_{\rho
}^{m}$ and we will use the shorthand $S_{{}}^{m}:=S_{0}^{m}$. Notice
that $S_{\rho }^{m}\subseteq S_{\rho^{\prime }}^{m}$ for any $\rho \geq
\rho^{\prime }$.

 If $A$ belongs to $S_{{}}^{0}(\mathcal{B}(\Hi _{\mathrm{f}}))$
then the corresponding Weyl quantization is a \emph{bounded} operator on 
$\Hi =L^{2}(\mathbb{R}^{d},\Hi _{\mathrm{f}})$. The following
proposition sharpens this statement (see  \cite{Folland}, Theorem  2.73).

\noindent \textbf{Notation.} Denote by $C_{\mathrm{b}}^{k}(\mathbb{R}^{d},
\mathcal{E})$  the space of  $\mathcal{E}$-valued, $k$ times
continuously differentiable functions on $\mathbb{R}^{d},$ such that all the
derivatives up to the order $k$ are bounded. Equipped with the norm
\[
\left\| A\right\|_{C_{\mathrm{b}}^{k}}:=\sup_{|\alpha |\leq k}\
\sup_{x\in \mathbb{R}^{d}}\left\| (\partial_{x}^{\alpha }A)(x)
\right\|_{\mathcal{E}}
\]
it is a Banach space.

\begin{proposition}
\emph{(Calderon-Vaillancourt)} There exists a  constant 
$C_{d}<\infty$ such that for every 
$A\in C_{\mathrm{b}}^{2d+1}(\mathbb{R}^{2d},
\mathcal{B}(\Hi _{\mathrm{f}}))$ one has \label{Prop CaldVaill}
\[
\big\| \widehat{A}\big\|_{\mathcal{B}(\Hi )}\leq C_{d}\sup_{|\alpha
|+|\beta |\leq 2d+1}\ \sup_{q,p\in \mathbb{R}^{d}}\left\| (\partial
_{q}^{\alpha }\partial_{p}^{\beta }A)(q,p)\right\|_{\mathcal{B}(\Hi 
_{\mathrm{f}})}=C_{d}\left\| A\right\|_{C_{\mathrm{b}}^{2d+1}}\,.
\]
\end{proposition}

\noindent This implies, in particular, that the Weyl quantization, regarded
as a map $\mathcal{W}_{\varepsilon }:S^{0}(\mathcal{B}
(\Hi _{\mathrm{f}}))$ $\rightarrow \mathcal{B}(\Hi )$, 
is \emph{continuous}
with respect to the Fr\'{e}chet topology on $S^{0}(\mathcal{B}
(\Hi _{\mathrm{f}}))$.

\subsection{The Weyl-Moyal product}

Next we consider the composition of symbols.
The behavior of the symbol classes with respect to the
pointwise product is very simple, as can be proved by using the Leibniz
rule.

\begin{proposition}
If $A$ $\in S_{\rho }^{m_{1}}(\mathcal{B}(\Hi _{\mathrm{f}}))$ and 
$B\in S_{\rho }^{m_{2}}(\mathcal{B}(\Hi _{\mathrm{f}}))$, then $AB$
belongs to\\ $S_{\rho }^{m_{1}+m_{2}}(\mathcal{B}(\Hi _{\mathrm{f}}))$
for every $m_{1},m_{2}\in \mathbb{R}$.\ \label{Prop point product}
\end{proposition}

\noindent The behavior under pointwise inversion is described in the
following proposition. For every $T\in \mathcal{B}(\Hi _{\mathrm{f}})$
let the internal spectral radius be $\rho_{\mathrm{int}}(T):=\inf
\left\{ |\lambda |:\lambda \in \sigma (T)\right\} .$

\begin{proposition}
Assume that $A$ $\in S_{\rho }^{m}(\mathcal{B}(\Hi _{\mathrm{f}}))$
is a normal symbol which is elliptic,
in the sense that there exists a constant $C_{0}$ such that
\[
\rho_{\mathrm{int}}(A(q,p))\geq C_{0}\left\langle p\right\rangle^{m}\qquad
\mbox{for any }\,\,p\in \mathbb{R}^{d}\,.
\]
Then the pointwise inverse $A^{-1}$ exists and belongs to 
$S_{\rho }^{-m}(\mathcal{B}(\Hi _{\mathrm{f}}))$. \label{Prop point inversion}
\end{proposition}

\noindent \textbf{Proof.} As a consequence of the spectral theorem (for
bounded normal operators) one has 
\[
\left\| A^{-1}(q,p)\right\|_{\mathcal{B}(\Hi _{\mathrm{f}})}=\rho_{
\mathrm{int}}(A(q,p))^{-1}\leq C\left\langle p\right\rangle^{-m} \,.
\]
Similar bounds on derivatives can be obtained by noticing that
\[
\left\| \nabla_{p}(A^{-1})\right\|_{\mathcal{B}(\Hi _{\mathrm{f}
})}=\left\| -A^{-1}(\nabla_{p}A)\ A^{-1}\right\|_{\mathcal{B}(\Hi _{
\mathrm{f}})}\leq C^{\prime }\left\langle p\right\rangle^{-m-\rho }
\]
and applying the chain rule.\hfill$\qed$\bigskip

The crucial result for pseudodiferential calculus is the following. 
One can define an associative product in the space of classical symbols
which corresponds to the composition of the operators.
Given $A\in S_{\rho }^{m_1}(\mathcal{B}(\Hi _{\mathrm{f}}))$ 
and $B\in S_{\rho }^{m_2}(\mathcal{B}(\Hi _{\mathrm{f}}))$ 
we know
that $\widehat{A}$ and $\widehat{B}$ map $\mathcal{S}(\mathbb{R}^{d},\Hi _{\mathrm{f}
})$ into itself. Then  $\widehat{A}\widehat{B}$
is still an operator on $\mathcal{S}
(\mathbb{R}^{d},\Hi _{\mathrm{f}})$ and  one can show that there exists a unique
$\varepsilon $-dependent symbol ${\rm Symb}({\widehat{A}\widehat{B}})=:
A\,\tilde{\#}\,B\in S_{\rho }^{m_1+m_2}(\mathcal{B}(\Hi _{\mathrm{f}}))$
 such that
\[
\mathcal{W}_{\varepsilon }(A)\mathcal{W}_{\varepsilon }(B)=\mathcal{W}
_{\varepsilon }(A\ \tilde{\#}\ B)\mbox{. }  %\label{Weyl product}
\]
The symbol $A\, \tilde{\#}\, B$ is called the \textbf{Weyl product} (or the
twisted product) of the symbols $A$ and $B$. For the
proof of the following proposition in the operator valued case we refer
again to \cite{GMS}.

\begin{proposition} \label{Prop Moyal product}
Let  $A\in S_{\rho }^{m_1}(\mathcal{B}(\Hi _{\mathrm{f}}))$ 
and $B\in S_{\rho }^{m_2}(\mathcal{B}(\Hi _{\mathrm{f}}))$,
then $\widehat A\widehat B =\widehat C$ with 
 $C\in S_{\rho }^{m_1+m_2}(\mathcal{B}(\Hi _{\mathrm{f}}))$ 
given through 
\begin{eqnarray}\label{Weyl product}
C(q,p) &=& \exp\left(\frac{i\epsi}{2} (\nabla_p\cdot\nabla_x - \nabla_\xi\cdot\nabla_q)
\right) \big( A(q,p)\,B(x,\xi) \big)
\Big|_{x=q,\xi=p} \nonumber\\ &=:& \big(A\,\tilde\#\,B\big)(q,p)\,.
\end{eqnarray}
\end{proposition}

\noindent In particular, $S_{\rho }^{0}(\mathcal{B}(\Hi _{\mathrm{f}
}))$ and $S_{\rho }^{\infty }(\mathcal{B}(\Hi _{\mathrm{f}
})):=\bigcup_{m\in \mathbb{R}}S_{\rho }^{m}(\mathcal{B}(\Hi _{\mathrm{f}
}))$ are algebras with respect to the Weyl product $\tilde{\#}$. \

Since the product $A\ \tilde{\#}\ B$ depends on $\varepsilon $ by
construction, one can expand (\ref{Weyl product}) in orders of $\varepsilon$.
 To this end, it is convenient to define suitable classes of 
$\varepsilon $-dependent symbols, called \textbf{semiclassical symbols},
which -- roughly speaking -- are close to a power series in $\varepsilon $
of classical symbols with nicer and nicer behavior at infinity. Our
definition is a special case of the standard ones 
(see \cite{DiSj,Martinez,Folland,Hormander}).

\begin{definition}
A map $A:[0,\varepsilon_{0})\rightarrow S_{\rho }^{m},\varepsilon \mapsto
A_{\varepsilon }$ is called a semiclassical symbol of order $m$ and weight $
\rho $ if there exists a sequence $\{A_{j}\}_{j\in \mathbb{N}}$ with $A_{j}\in
S_{\rho }^{m-j\rho }$ such that for every $n\in \mathbb{N}$ one has that $
\left( A_{\varepsilon }-\sum_{j=0}^{n-1}\varepsilon^{j}A_{j}\right) $
belongs to $S_{\rho }^{m-n\rho }$ {\em uniformly} in $\varepsilon $, in the
following sense: for any $k\in \mathbb{N}$ there exists a constant $C_{n,k}$
such that for any $\varepsilon \in [0,\varepsilon_{0})$ one has
\begin{equation}
\Big\| A_{\varepsilon }-\sum_{j=0}^{n-1}\varepsilon^{j}A_{j}\Big\|
_{k}^{(m-n\rho )}\leq C_{n,k}\, \varepsilon^{n} \,, \label{Semi symb}
\end{equation}
where $\left\| \ldots \right\|_{k}^{(m)}$ is the $k$-th Fr\'{e}chet
semi-norm in $S_{\rho }^{m},$ introduced in (\ref{Seminorms}).
\end{definition}

\noindent One calls $A_{0}$ and $A_{1}$ the \textbf{
principal symbol} and the \textbf{subprincipal symbol} of ${A}$. The
space of semiclassical symbols of order $m$ and weight $\rho $ will be
denoted as $S_{\rho }^{m}(\varepsilon )$. If condition (\ref{Semi symb}) is
fulfilled, one writes
\[%\begin{equation}
A\asymp \sum_{j\geq 0}\varepsilon^{j}A_{j}\qquad \mbox{in }S_{\rho
}^{m}(\varepsilon )  %\label{Asympotic}
\]%\end{equation}
and one says that $A$ is asymptotically equivalent to the series $\sum_{j\geq
0}\varepsilon^{j}A_{j}$ in $S_{\rho }^{m}(\varepsilon )$. If $A$ is
asymptotically equivalent to the series in which $A_{j}=0$ for every $j\in
\mathbb{N}$, we write $A=\mathcal{O}(\varepsilon^{\infty })$. To be precise,
we should write $A=\mathcal{O}(\varepsilon^{\infty })$ in $S_{\rho
}^{m}(\varepsilon )$, but the latter specification is omitted whenever
it is unambiguous from the context.

In general a formal power series $\sum_{j\geq 0}\varepsilon^{j}A_{j}$
is not convergent, but it is always the asymptotic expansion of a (non
unique) semiclassical symbol (e.g.\ \cite{Martinez}).

\begin{proposition}
Let be $\{A_{j}\}_{j\in \mathbb{N}}$ an arbitrary sequence such that $A_{j}\in
S_{\rho }^{m-j\rho }.$ Then there exists $A\in S_{\rho }^{m}(\varepsilon )$
such that $A\asymp \sum_{j\geq 0}\varepsilon^{j}A_{j}$ in $S_{\rho
}^{m}(\varepsilon )$ and $A$ is unique up to $\mathcal{O}(\varepsilon
^{\infty })$, in the sense that the difference of two such symbols is 
$\mathcal{O}(\varepsilon^{\infty })$ in $S_{\rho }^{m}(\varepsilon )$. The
semiclassical symbol $A$ is called a \textbf{resummation} of the formal
symbol $\sum_{j\geq 0}\varepsilon^{j}A_{j}$.\label{Prop resummation}
\end{proposition}

\noindent The Weyl product of two semiclassical symbols is again a 
semiclassical symbol with
an explicit asymptotic expansion (see \cite{Folland}, Theorem 2.49).

\begin{proposition}
If $A\asymp \sum_{j\geq 0}\varepsilon^{j}A_{j}$ in $S_{\rho
}^{m_{1}}(\varepsilon )$ and $B\asymp \sum_{j\geq 0}\varepsilon^{j}B_{j}$
in $S_{\rho }^{m_{2}}(\varepsilon )$, then $A\ \tilde{\#}\ B\in S_{\rho
}^{m_{1}+m_{2}}(\varepsilon )$ has an asymptotic expansion given by
\begin{equation}
\left( A\ \tilde{\#}\ B\right)_{k}(q,p)= (2i)^{-k}
\sum_{|\alpha |+|\beta |+j+l=k}\frac{(-1)^{|\alpha |}}{|\alpha |!|\beta
|!}\left( (\partial_{q}^{\alpha }\partial_{p}^{\beta }A_{j})(\partial
_{p}^{\alpha }\partial_{q}^{\beta }B_{l})\right) (q,p)  \label{Composition}
\end{equation}
where it is understood that $k,j,l\in \mathbb{N}$ and $\alpha ,\beta \in 
\mathbb{N}^{d}$.\label{Prop composition}
\end{proposition}

\noindent For example\ $(A\ \tilde{\#}\ B)_{0}$ is simply given by the
pointwise product $A_{0}B_{0}$ and
\[
(A\ \tilde{\#}\ B)_{1}=A_{0}B_{1}+A_{1}B_{0}-\frac{i}{2}\{A_{0},B_{0}\}
\]
where $\{\cdot ,\cdot \}$ denotes the Poisson bracket on 
$S_{\rho }^{\infty }(\mathcal{B}(\Hi _{\mathrm{f}}))$, defined through
\begin{equation}
\{A,B\}=\sum_{j=1}^{d}\frac{\partial A}{\partial p_{j}}\frac{\partial B}{
\partial q_{j}}-\frac{\partial A}{\partial q_{j}}\frac{\partial B}{\partial
p_{j}}.  \label{Poisson}
\end{equation}
Notice that, in general, $\{A,B\}\neq -\{B,A\}$ since operator-valued
derivatives do not commute, in particular $\{A,A\}\neq 0$. The usual
Poisson algebra is recovered in the special case in which one of the two
arguments is a multiple of the identity, i.e.\ $A(z)=a(z){\bf 1}_{\Hi _{
\mathrm{f}}}$.

As a consequence of the previous result, it is convenient to introduce the
space of the formal power series with coefficients in 
$S_{\rho }^{\infty }(\mathcal{B}(\Hi _{\mathrm{f}}))$. This space, equipped with the
associative product given by (\ref{Composition}) and with the involution 
defined by taking the adjoint of every coefficient, will be called the
\textbf{algebra of formal symbols} over $\mathcal{B}(\Hi _{\mathrm{f}})$. 
In particular we will denote as $M_{\rho }^{m}(\varepsilon )$ the
subspace of the formal power series with a resummation in $S_{\rho
}^{m}(\varepsilon )$, i.e.\
\[%\begin{equation}
M_{\rho }^{m}(\varepsilon ):=\left\{ \sum_{j\geq 0}\varepsilon
^{j}A_{j}:A_{j}\in S_{\rho }^{m-j\rho }\right\}\, .  %\label{Moyal classes}
\]%\end{equation}
In the context of formal power series, the product defined by
(\ref{Composition}) will be called the \textbf{Moyal product }and denoted simply
as $\#$ . Notice that $\#$ defines a map from $M_{\rho
}^{m_{1}}(\varepsilon )\times M_{\rho }^{m_{2}}(\varepsilon )$ to $M_{\rho
}^{m_{1}+m_{2}}(\varepsilon )$.
The Moyal product can also be
regarded as a map from $M_{\rho }^{m_{1}}(\varepsilon ,\mathcal{B}(\mathcal{H
}_{\mathrm{f}}))\times M_{\rho }^{m_{2}}(\varepsilon ,\Hi _{\mathrm{f}
})$ to $M_{\rho }^{m_{1}+m_{2}}(\Hi _{\mathrm{f}})$, where
in (\ref{Composition}) the operator $A$ and its derivatives act on the vector
$B$ and its derivatives.

To sum up  the previous discussion, we wish to point out that one can
prove statements on three levels: formal symbols (i.e.\ formal power series),
semiclassical symbols, and operators on 
$\mathcal{S}(\mathbb{R}^{d},\Hi _{\mathrm{f}})\subseteq 
L^{2}(\mathbb{R}^{d},\Hi _{\mathrm{f}})$. A simple
example illustrates the interplay between these levels.
Suppose that two formal symbols $A\in M_{\rho
}^{m_{1}}(\varepsilon )$ and $B\in M_{\rho }^{m_{2}}(\varepsilon )$ Moyal
commute, i.e.\ $[A,B]_{\#}=A\#B-B\#A=0$.  Let  
$A_{\varepsilon }\in S_{\rho}^{m_{1}}(\varepsilon )$ and 
$B_{\varepsilon }\in S_{\rho}^{m_{2}}(\varepsilon )$ be any two 
resummations of $A$ and, respectively, $B$.
Since we know \emph{a priori} (by Prop. \ref{Prop composition}) that the
Weyl product $A_{\varepsilon }\ \tilde{\#}\ B_{\varepsilon }$ belongs to 
$S_{\rho }^{m_{1}+m_{2}}(\varepsilon )$ it follows that the Weyl commutator 
$[A_{\varepsilon },B_{\varepsilon }]_{\tilde{\#}}$ is asymptotically close to
zero in $S_{\rho }^{m_{1}+m_{2}}(\varepsilon )$, which can be rephrased in
the following way: for any $n,k\in \mathbb{N}$ there exists a constant $C_{n,k}$
such that for any $\varepsilon \in [0,\varepsilon_{0})$ one has
\[
\left\| \lbrack A_{\varepsilon },B_{\varepsilon }]_{\tilde{\#}}\right\|
_{k}^{(m_{1}+m_{2}-n\rho )}\leq C_{n,k}\ \varepsilon^{n}\,.
\]
If $\rho >0$  we obtain that definitely
$m_{1}+m_{2}-n\rho \leq 0$ for some $n\in \mathbb{N}$
and then Prop.\ \ref{Prop CaldVaill} assures that
the operator commutator $[\widehat{A}_{\varepsilon },\widehat{B}_{\varepsilon }]$
can be bounded in the $\mathcal{B}(\Hi )$-norm. Moreover, for $\rho
>0 $, we can conclude that $[\widehat{A}_{\varepsilon },\widehat{B}_{\varepsilon }]$
 is a smoothing operator (i.e. it belongs to $OPS_{\rho }^{-\infty }:=\cap
_{m\in \mathbb{R}}OPS_{\rho }^{m}$) and in particular one can prove that it is a
``small'' bounded operator between the Sobolev spaces $H^{q}$ and $H^{q+r}$
for any $q,r\in \mathbb{N}$. To be precise, for any $q,r,n\in \mathbb{N}$
there exist a constant $C_{n,q,r}$ such that
\[
\left\| \lbrack \widehat{A}_{\varepsilon },
\widehat{B}_{\varepsilon }]\right\|_{\mathcal{B}(H^{q},H^{q+r})}
\leq C_{n,q,r}\,\varepsilon^{n}
\]
for any $\varepsilon \in [0,\varepsilon_{0})$, where $H^{q}$ stands for 
$H^{q}(\mathbb{R}^{d},\Hi _{\mathrm{f}})$.
Notice that for $\rho =0$ and $m_{1}+m_{2}=:m>0$ it is not
possible to conclude from $[A,B]_\#=0$ that 
$[\widehat{A}_{\varepsilon },\widehat{B}_{\varepsilon }]$
\ is a bounded operator, since it could happen -- for example -- that 
$[A_{\varepsilon },B_{\varepsilon }]_{\tilde{\#}}=e^{-\frac{1}{\varepsilon }}p^{m},$ 
which is asymptotically close to zero in $S_{\rho }^{m}(\varepsilon
)$.
In the following we will use the same symbol for an element in 
$S^m_\rho(\epsi)$ and its
expansion in $M^m_\rho(\epsi)$. As suggested by the preceding discussion, we introduce
 the following
synthetic notation.\medskip

\noindent \textbf{Notation.} Let be $A$ and $B$ semiclassical symbols in
$S_{\rho }^{m}(\varepsilon )$. We will say that $B=A+\mathcal{O}_{-\infty
}(\varepsilon^{\infty })$ if $B-A$ is asymptotically close to zero in 
$S_{\rho }^{m}(\varepsilon )$ for $\rho >0$.\bigskip

\noindent With a little abuse, we will employ the same notation for
pseudodifferential operators too, i.e. we write 
$\widehat{B}=\widehat{A}+\mathcal{O}_{-\infty }(\varepsilon^{\infty })$ 
if $B=A+\mathcal{O}_{-\infty
}(\varepsilon^{\infty })$. As noticed above this is a strong concept of
closeness, since it implies that $\widehat{B}-\widehat{A}$ \ is a smoothing
operator. Compare with the following weaker concept. \bigskip

\noindent \textbf{Notation.} Let be $R$ and $S$ two ($\varepsilon$-dependent) 
operators on $\Hi $. We will say that 
$R=S+\mathcal{O}_{0}(\varepsilon^{\infty })$ if for every $n\in \mathbb{N}$ 
 there exists a constant $C_{n}$ such that
\[
\left\| R-S\right\|_{\mathcal{B}(\Hi )}\leq C_{n}\varepsilon^{n}
\]
for every $\varepsilon \in [0,\varepsilon_{0})$. In such a case we will say
that $R$ is $\mathcal{O}_{0}(\varepsilon^{\infty })$-close to $S$.

%%%%%%%%%%%%%%%%%   BIBLIOGRAPHY   %%%%%%%%%%%%%%%%%%%%%%%%%%%%%%%%%%%%%%%

\end{document}